\documentclass[11pt]{article}

\usepackage{physics} 
\usepackage{siunitx} 
\usepackage{enumerate} 
\usepackage{pgfplots}
\usepackage{pgfplotstable}
\usepackage{tikz,pgfplots}
\usepackage{cite}
\usepackage{amsmath}  
\usepackage{wasysym} 
\usepackage{amsfonts}
\usepackage{multirow}
\usepackage{float}  
\usepackage{placeins}  
\usepackage{longtable}
\usepackage{hyperref}
\usepackage{float}
\usepackage{hyperref}
\usepackage{float}
\usepackage{lipsum}
\usepackage{authblk}
\usepackage{soul} 

\usepackage{subcaption}
\usepackage[square,numbers,sort&compress]{natbib}
\usepackage{amsmath,amssymb}

\usepackage{geometry}
\def\l{\left(} 
\def\r{\right)} 
\def\d3{\mathrm{d}^3}
\def\d4{\mathrm{d}^4}
\def\pt{\partial} 

\def\rd{\mathrm{d}} 
\def\ri{\mathrm{i}} 
\def\re{\mathrm{e}} 
 
\begin{document}
\title{\bf Kerr black holes with synchronised Proca hair: \\ lensing, shadows and EHT  constraints} 

\author[1]{Ivo Sengo\thanks{sengo@ua.pt}}
\author[1]{Pedro V.P. Cunha\thanks{pvcunha@ua.pt}}
\author[1]{ Carlos A. R. Herdeiro\thanks{herdeiro@ua.pt}}
\author[1]{Eugen Radu\thanks{eugen.radu@ua.pt}}
\affil[1]{Departamento de Matemática da Universidade de Aveiro and Centre for Research and Development
	in Mathematics and Applications (CIDMA), Campus de Santiago, 3810-183 Aveiro, Portugal}
 
 \date{September 2022}
 
\maketitle
\begin{abstract}
We investigate the gravitational lensing by spinning Proca stars and the shadows and lensing by Kerr black holes (BHs) with synchronised Proca hair, discussing both theoretical aspects and observational constraints from the Event Horizon Telescope (EHT) M87* and Sgr A* data. On the theoretical side, this family of BHs interpolates between Kerr-like solutions -- exhibiting a similar optical appearance to that of Kerr BHs -- to very non-Kerr like solutions, exhibiting exotic features such as cuspy shadows, egg-like shadows and ghost shadows. We interpret these features in terms of the structure of the fundamental photon orbits, for which different  branches exist, containing both stable and unstable orbits, with some of the latter not being shadow related. On the observational side, we show that current EHT constraints are compatible with all such BHs that could form from the growth of the superradiant instability of Kerr BHs. Unexpectedly, given the (roughly) $10\%$ error bars in the EHT data -- and in contrast to their scalar cousin model --, some of the BHs with up to $40\%$ of their energy in their Proca hair are compatible with the current data. We estimate the necessary resolution of future observations to better constrain this model.
\end{abstract}

\newpage

\tableofcontents

\section{Introduction}
The present golden age of observational strong gravity invites us to test the Kerr hypothesis and, in particular, its universality~\cite{Herdeiro:2022yle}. It is therefore timely to consider well motivated non-Kerr models and compare their phenomenology with the current data. In parallel, considering non-Kerr models is a valuable theoretical arena, often enlightening in understanding how generic or special the General Relativity (GR) (electro-)vacuum black holes (BHs) are, since much of our intuition about the physics of BHs is constructed upon them. 

For either of these -- $i.e.$ observational or theoretical -- perspectives, light is a privileged probe. It has been so since the genesis of GR, when the bending of light was proposed as a test~\cite{Einstein:1915bz}, and its confirmation~\cite{Dyson:1920cwa} elevated GR above Newton's law. Einstein himself studied gravitational lensing and the possibility of multiple images, including what we now call \textit{Einstein rings}~\cite{Einstein:1936llh}. The deeper understanding of BHs that emerged in the 1960s, in particular the discovery of the Kerr metric~\cite{Kerr:1963ud}, led to the first (academic) discussions on the optical appearance of BHs~\cite{Bardeen:1973tla}. It was understood that such optical appearance is intimately connected to the bound orbits of light around BHs. When planar, these orbits are called \textit{light rings}, which have been well understood for the Kerr BH since the 1970s~\cite{Bardeen:1972fi}. The non-planar bound orbits of light around Kerr BHs are called \textit{spherical photon orbits} (SPOs) and have been studied in detailed only more recently~\cite{Teo2003}. For generic non-Kerr spacetimes, in particular where the geodesic motion may not be integrable, we shall call such bound orbits of light \textit{fundamental photon orbits} (FPOs), following~\cite{Cunha2017}.

These academic studies were accompanied by more astrophysical investigations on the optical appearence of BHs, most notably the \textit{tour de force} image of a Schwarzschild BH surrounded by an accretion disk, by Luminet~\cite{Luminet:1979nyg}. At the turn of the last century Falcke \textit{et al.} proposed that the optical appearence the supermassive BH at our galactic centre could actually be resolved, to produce its ``photography", observing in particular its silhouette or  \textit{shadow}~\cite{Falcke:1999pj}. This remarkable proposal led to a worldwide collaborative effort -- the \textit{Event Horizon Telescope} (EHT) -- that in 2019~\cite{EHT1,EHT2,EventHorizonTelescope:2019_6} and 2022~\cite{EventHorizonTelescope:2022_1,EventHorizonTelescope:2022_2,EventHorizonTelescope:2022_3,EventHorizonTelescope:2022_6} published the first images of the M87* and Sgr A* supermassive BHs, respectively.

BH imaging is still in its infancy, and the future promises to deliver higher accurate results with the next generation EHT observatories. It is therefore timely to study how much these observations can distinguish well motivated models of non-Kerr BHs, emerging as solutions of sound physical theories, without known pathologies and with a plausible formation mechanism. Such conditions are quite restrictive, but they can be met even in GR with simple matter contents obeying the most fundamental energy conditions. This is the case of Kerr BHs with (synchronised) Proca Hair  (KBHsPH)~\cite{Herdeiro2016,Santos2020} that are solution of Einstein's gravity minimally coupled to a free, complex Proca field ($cf.$ action~\ref{eq1} below). These BHs circumvent no-Proca hair theorems~\cite{Bekenstein:1972ny} by virtue of a harmonic time dependence of the bosonic field~\cite{Herdeiro2016}. The model is free of known pathologies, and the hairy BHs could emerge dynamically via the superradiant instability of Kerr~\cite{Brito:2015oca}, triggered by the bosonic field~\cite{East:2017ovw,Herdeiro_2017,Dolan:2017otg}. This family of hairy BHs interpolates between the bald Kerr solution and horizonless self-gravitating solitons known as \textit{Proca stars} (PSs)~\cite{Brito2016}. Spinning PSs, unlike their scalar cousins - boson stars (BSs) - have been shown to be dynamically robust~\cite{Sanchis-Gual:2019ljs}, lending further dynamical credibility to the model - see also~\cite{Liebling:2012fv}. These PSs, moreover, have been recently advocated to match gravitational wave data~\cite{Bustillo_2021,CalderonBustillo:2022cja}, further motivating the model from a phenomenological viewpoint. Finally, even in the spherical case, PSs exhibit qualitatively distinct features from their scalar cousins, and have been shown to be able to imitate imaging observations under some conditions~\cite{Herdeiro:2021lwl} (see also~\cite{Cunha2017-2}).

The goal of this paper is study the lensing, shadows of KBHsPH from both a theoretical and observational viewpoint. This study parallels previous studies for Kerr BHs with (synchronised) scalar hair~\cite{Cunha_2015,Cunha2016,Chaotic,Cunha2017,Cunha_2019}, and some of our findings are qualitatively similar to the scalar case; others, however, are different. We observe, for instance, that the shadows of these Proca hairy BHs can vary from Kerr-like to very non-Kerr like, as one scans the domain of existence of solutions. When the latter occur, qualitatively distinct features emerge, such as cuspy shadows, egg-like shadows and ghost shadows. We shall interpret these features considering the study of the FPOs of the corresponding solutions. One of the most unexpected aspects of our study emerges when considering the comparison with EHT data. As we shall show below, KBHsPH can be compatible with the current EHT observations for much hairier solutions than in the scalar case. We shall put forward a speculation to why this is the case in the discussion section.

This paper is organized as follows. Sections 2 and 3 discuss generalities about the solutions and the setup for studying lensing and shadows. Then, sections 4-6 discuss theoretical aspects, namely the lensing by PSs, the shadows and lensing obtained by KBHsPH from ray-tracing and the FPOs structure for some chosen solutions. In sections 7-9 we turn to observational aspects, developing a setup for comparing solutions in a region of interest of the domain of solutions and using the EHT M87* and Sgr A* data to constrain the parameter space. Finally, in section 10 we offer some conclusions and a discussion.

\section{The solutions}
KBHsPH~\cite{Herdeiro2016,Santos2020} are fully non-linear solutions of the Einstein-complex-Proca model, described by the action 

\begin{align} \label{eq1}
S = \int \rd^4 x \sqrt{-g} \l \frac{R}{16 \pi} + \mathcal{L}_M \r \, , 
\end{align}

\noindent where $g$ is the determinant of the metric, $R$ is the Ricci scalar and $\mathcal{L}_M$ is the Lagrangian density of a massive complex vector boson $A^\alpha$, which reads:

\begin{align} \label{eq2}
\mathcal{L}_M = - \frac{1}{4} F_{\alpha \beta} \bar{F}^{\alpha \beta} - \frac{1}{2} \mu^2 A_{\alpha} \bar{A}^\alpha \, .
\end{align}
In Eq.~\eqref{eq2}, $F_{\alpha \beta} = 2 A_{\left[\beta ;\alpha \right]}$ is the electromagnetic-field tensor, which is antisymmetric and gauge invariant, $\alpha , \beta = \{0, 1, 2, 3\}$, $\mu$ is the boson's mass and overbar denotes complex conjugation.

Varying the action (\ref{eq1}) one obtains the Einstein field equations sourced by the Proca energy-momentum tensor, as well as the Proca equations for the boson field -- see \cite{Herdeiro2016,Santos2020} for a detailed discussion of the Einstein-complex-Proca model. This system of equations can be solved numerically by considering the following metric ansatz for stationary and axially-symmetric BH solutions (as well as for horizonless, self-gravitating solitons):

\begin{align}
\begin{split}
\label{Eqansatz} ds^2 = & - \re^{2 F_0}N dt^2 +  \re^{2 F_1} \l \frac{dr^2}{N} + r^2 d \theta^2 \r +  \re^{2 F_2} r^2 \sin^2 \theta \l d \phi - W dt  \r^2 \, ,
\end{split}
\end{align}

\noindent where $N = 1 -r_H/r$ and $F_i, W$ ($i=0,1,2$) are functions of the spheroidal coordinates $\l r, \theta \r$. The parameter $r_H$ is the radial coordinate of the event horizon; when it is set to zero, the solutions describe spinning PSs. The metric possesses two Killing vector fields $\pt/\pt t$ and $\pt /\pt \phi$, connected respectively to stationarity and axial-symmetry. In addition, a $\mathbb{Z}_2$ reflection symmetry around the equatorial plane ($\theta=\pi/2$) will be assumed. For the Proca potential it is considered an ansatz of the form

\begin{align}
A= \re^{\ri \l m \phi - \omega t \r} \l \ri V dt +H_1 dr + H_2 d \theta + \ri H_3 \sin \theta d \phi \r \, ,
\end{align}

\noindent where the four functions $\l V, H_i \r$ all depend on $\l r, \theta \r$.  This ansatz has an harmonic time and azimuthal dependence, associated with the frequency $\omega > 0 $ and the azimuthal harmonic index $m \in \mathbb{Z} $, respectively.

\begin{figure}[H] 
	\centering
	\includegraphics[width=0.6\textwidth]{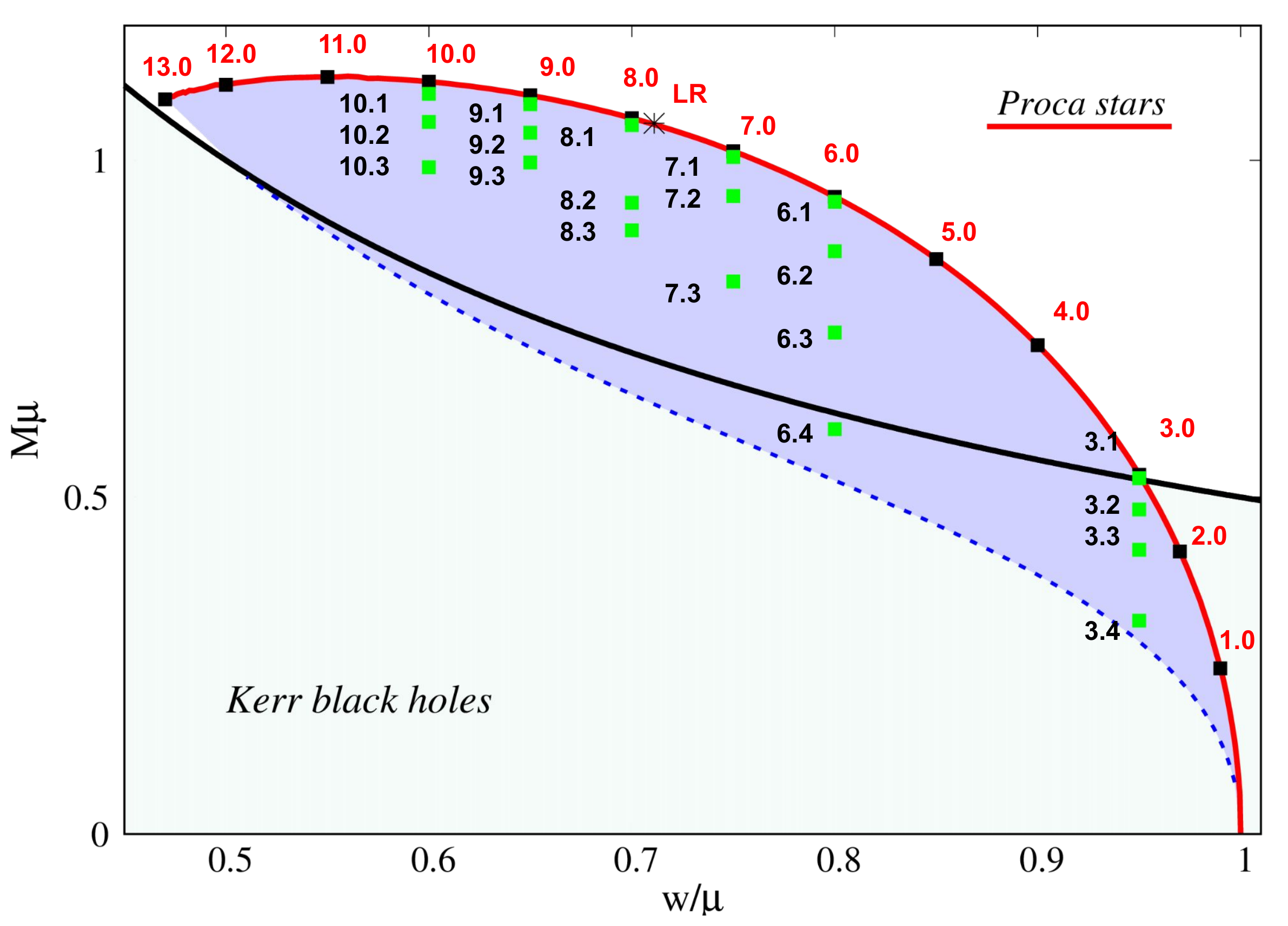}
	\caption{\small \label{fig1} The domain of existence of KBHsPH (blue shaded region) and the corresponding PSs (red line).  The  asterisk symbol on the PS line marks the first PS solution to feature a light ring (LR) orbit. The relevant physical quanties of these solutions can be found in Appendix \ref{App1}}
\end{figure}

We are going to focus on fundamental~\footnote{These are solutions with a nodeless Proca potential temporal component $V$, and $m=1$. These solutions are the ones that bifurcate from the linear bound state of a test Proca field around a Kerr BH, that corresponds to the lowest energy state.} KBHsPH solutions, which were first discussed in~\cite{Herdeiro_2017} and reported in detail in~\cite{Santos2020}. The domain of existence of these solutions is displayed in Fig.~\ref{fig1}, where hairy BHs solutions exist on a bound set (blue shaded region), bounded above by the existence line of PS solutions (solid red line) and below by a set of Kerr BHs (blue dashed line).

The solutions featured in this paper are identified by a symbol of the form $u.v$, with $u,v \, \in \mathbb{N}_0$. Solutions with the same $u$ share the same frequency (the greater the $u$, the smaller is the frequency), whilst the $v$ index is related to the the amount of hair within the solution (for the same $u$, hairier solutions have the smallest $v$). In this notation, PSs are  identified by a symbol of the form $u.0$.

We remark that in order for the KBHsPH we have just described to be in the astrophysical BHs mass range, then the boson mass $\mu$ should be ultralight, $i.e.$ $\mu \lesssim 10^{-10}$ eV. Such ultralight bosons are dark matter candidates under the fuzzy dark matter paradigm and could emerge in beyond the standard model scenarios -- see the discussions in~\cite{Arvanitaki:2009fg,Freitas:2021cfi}.

\section{Setup}

\begin{figure}[H] 
	\centering
	\includegraphics[width=0.5\textwidth]{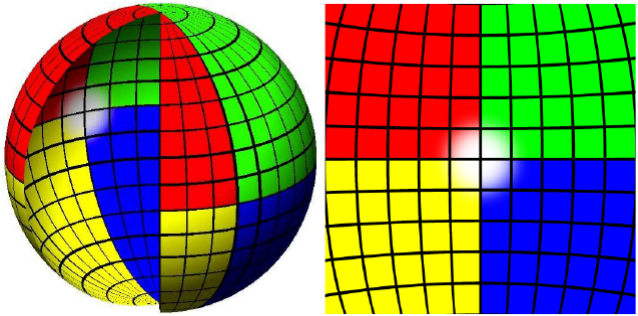}
	\caption{\small \label{fig2} (Left) The full celestial sphere. (Right) Non-distorted field of view seen by the observer (flat spacetime). [Image adapted from~\cite{Cunha_2015}].}
\end{figure}

The generation of synthetic lensing images of PSs and KBHsPH can be obtained by numerically evolving null-geodesics, which describes light motion in the high frequency limit. These synthetic images correspond to what an observer ($\mathcal{O}$), located at an off-centered position inside a celestial sphere, would see as a consequence of the gravitational lensing caused  by the presence of a PS or BH placed at the center of that same sphere (Fig.~\ref{fig2}, left panel).

Synthetic lensing images were generated by evolving numerically $1024 \times 1024$ light ray trajectories from $\mathcal{O}$'s location, using a \textit{backwards ray-tracing} method~\cite{Cunha2016,Bohn2015}. The observation angles seen by $\mathcal{O}$ (in its frame) determine the initial conditions for the light ray propagation. The observer's field of view always spans $35^{\circ}$ across both horizontal and vertical directions in all the lensing images presented in this paper. The image information is then presented in image coordinates $\l x, y \r$, obtained by multiplying the observation angles by the circumferential radius of the observer (defined below). This scaling procedure removes the typical fall-off behaviour of the angular size of objects at very large distances - see~\cite{Cunha2016} for more details.

To get a better grasp of the distortion introduced by either PSs or KBHsPH it is convenient to paint each of the quadrants of the celestial sphere by a different color, a setup introduced in~\cite{Bohn2015} and popularized in~\cite{Cunha_2015}.\footnote{This setup has been subsequently widely used $e.g.$~\cite{Chaotic,Cunha2017,Cunha2017-2,Wang:2017qhh,Wang:2018eui,Cunha:2018gql,Grover:2018tbq,Cunha:2018cof,Cunha:2018uzc,Wang:2019tjc,Chen:2020qyp,Lima:2021las,Bacchini:2021fig,Junior:2021dyw,Wang:2021ara,Zhong:2021mty,Wang:2021art,Wang:2022kvg}.} Following \cite{Cunha_2015}, the point on the colored celestial sphere immediately in front of $\mathcal{O}$ is marked by a white spot and dubbed {\bf F}. This white spot is apparent in the non-distorted field of view of such an observer, depicted in Fig.~\ref{fig2} (right panel) for reference. 

Unless otherwise stated, the lensing images that will be presented in the next sections are obtained assuming an observer on the equatorial plane ($\theta=\pi/2$) at a circumferential radius of $\tilde{r}=15\,  \mathrm{M_{ADM}}$, with the celestial sphere placed at twice this value. The {\it circumferential radius} $\tilde{r}$ of some point {\bf A} (located on the equatorial plane) is a distance defined by:
\begin{equation}
	\tilde{r} \equiv \frac{\mathcal{P}}{2 \pi} = \sqrt{g_{\phi \phi}} \, .
\end{equation}

 $\mathcal{P}$ is the circumference of the circle that includes point {\bf A}, with that circle determined by the orbits of the azimuthal Killing vector field $\partial_\phi$ on the equatorial plane. Hence, the value of $\mathcal{P}$ can be obtained by the following integral:
\begin{equation}
	\mathcal{P} = \int_{0}^{2 \pi} \sqrt{g_{\phi \phi}}\, d \phi = 2 \pi \sqrt{g_{\phi \phi}} \, .
\end{equation} 

Having concluded the discussion of the observation setup, we shall now move in the next section to PS lensing.

\section{Lensing by Proca Stars}

\begin{figure}[H]
	\begin{tabular}{ccc}
		\subfloat{\includegraphics[width=0.3\textwidth]{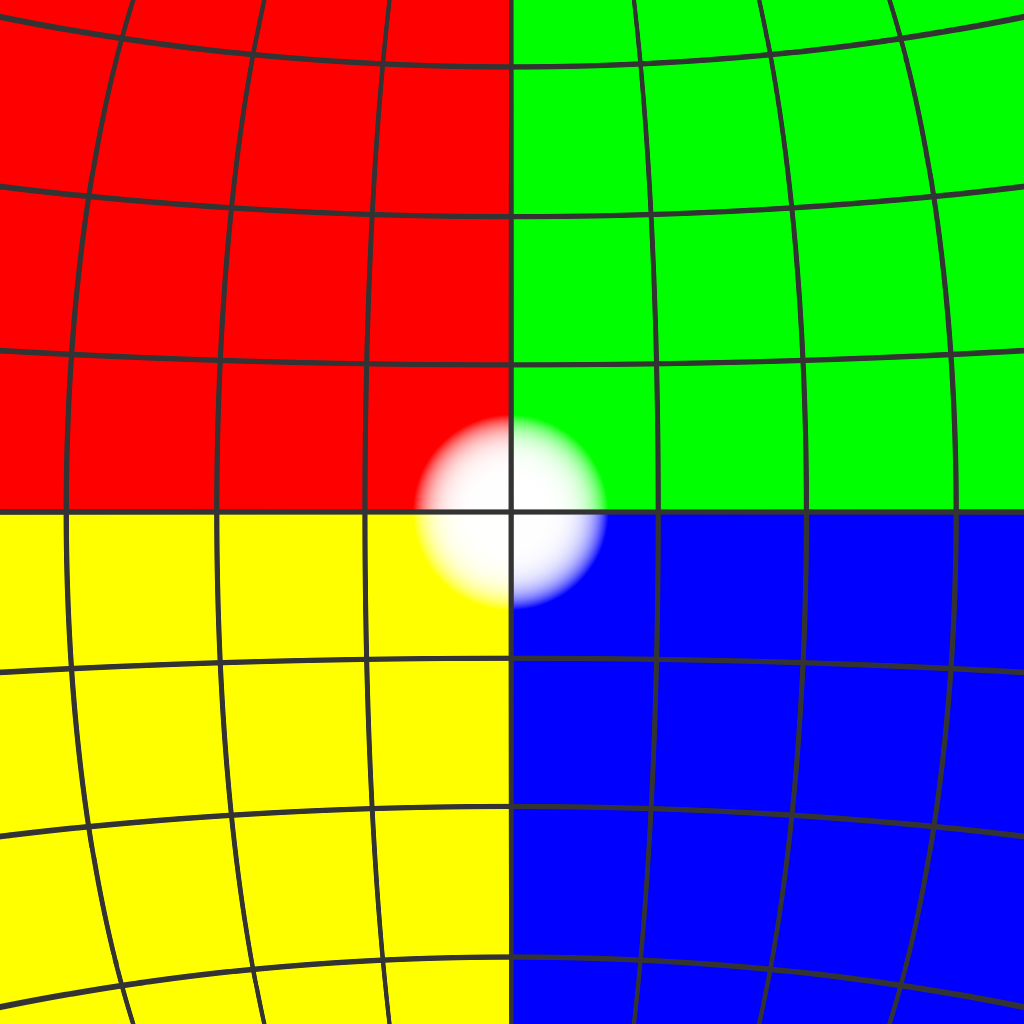}} &
		\subfloat{\includegraphics[width=0.3\textwidth]{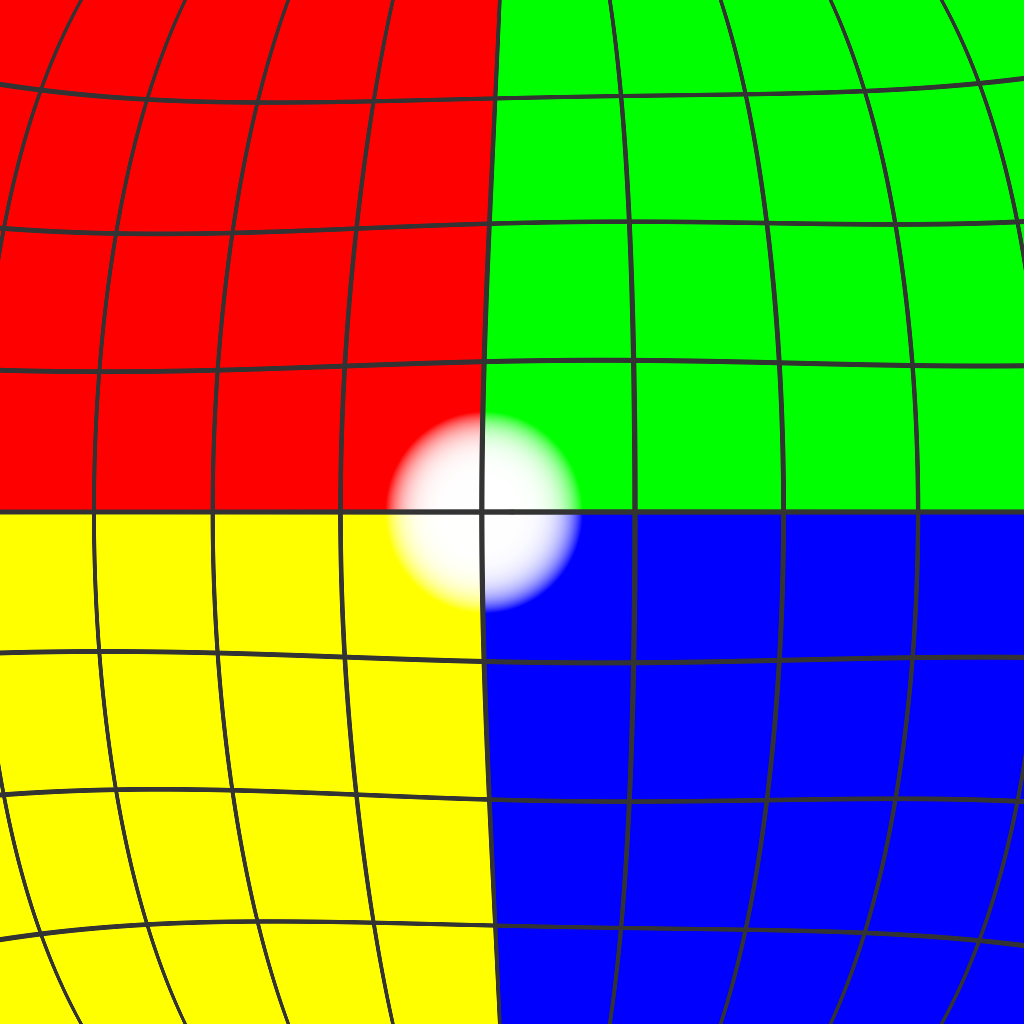}} &
		\subfloat{\includegraphics[width=0.3\textwidth]{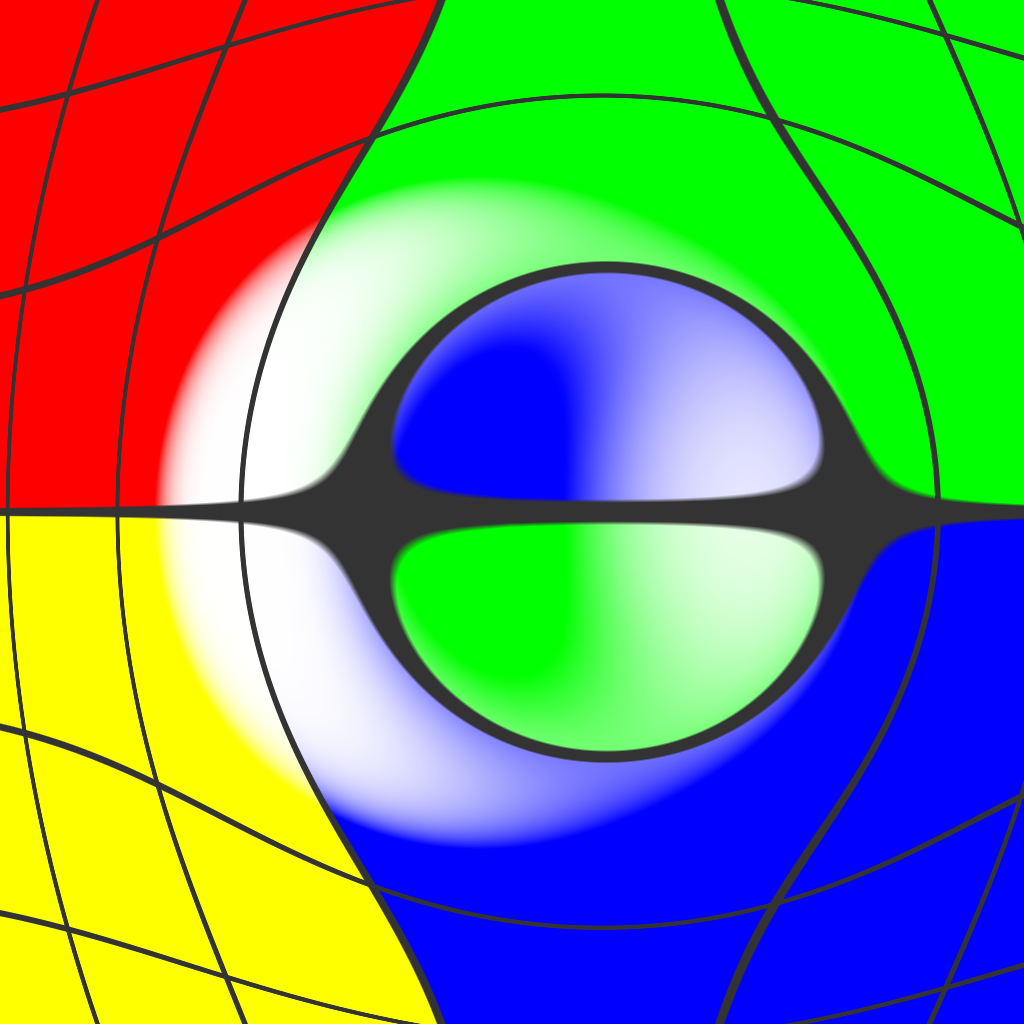}}  \\
		\subfloat{\includegraphics[width=0.3\textwidth]{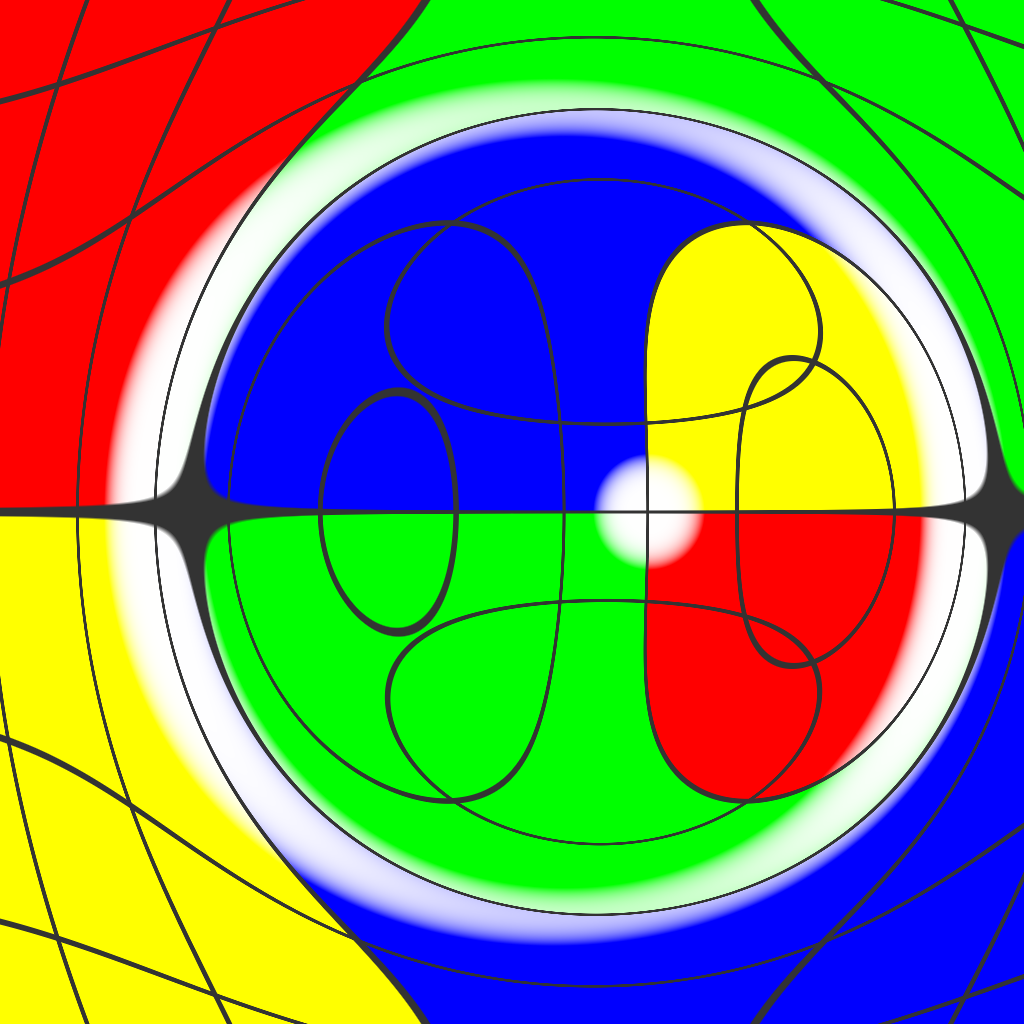}}&
		\subfloat{\includegraphics[width=0.3\textwidth]{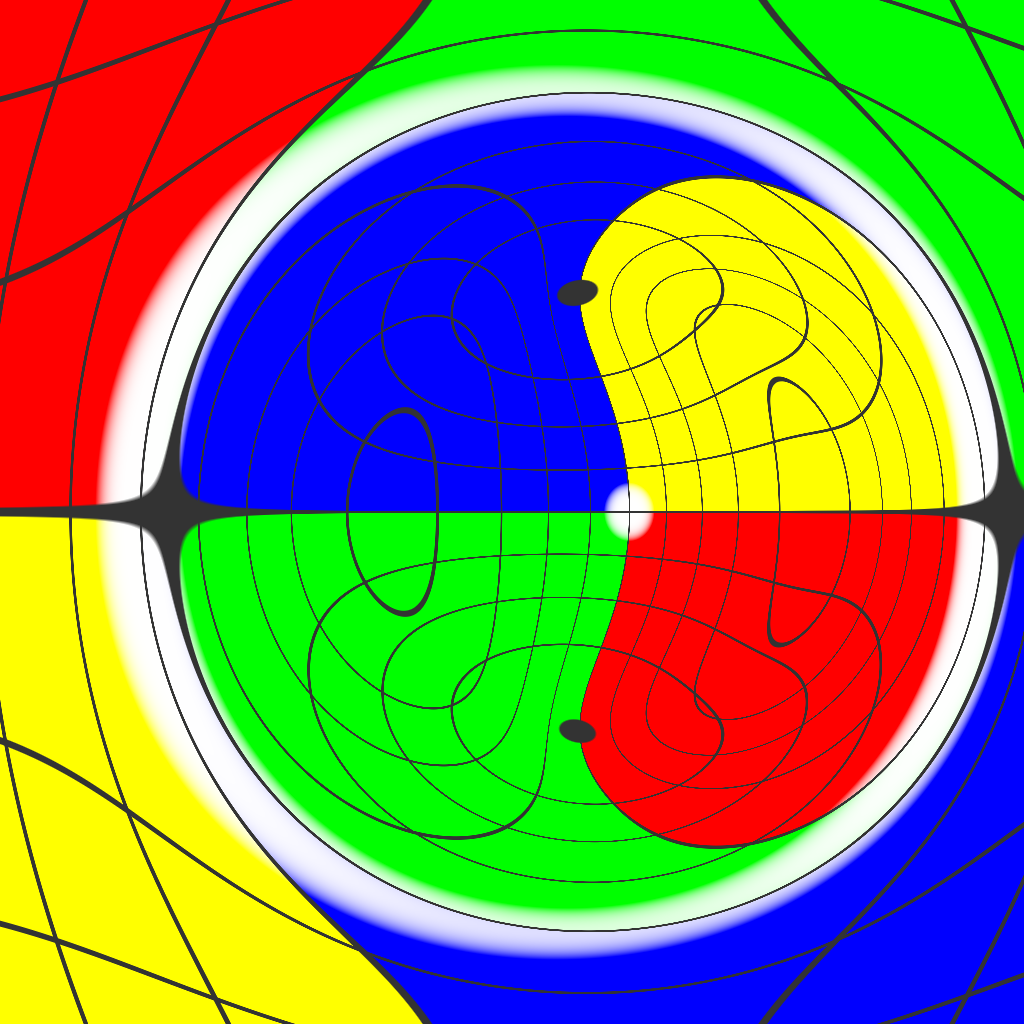}} &
		\subfloat{\includegraphics[width=0.3\textwidth]{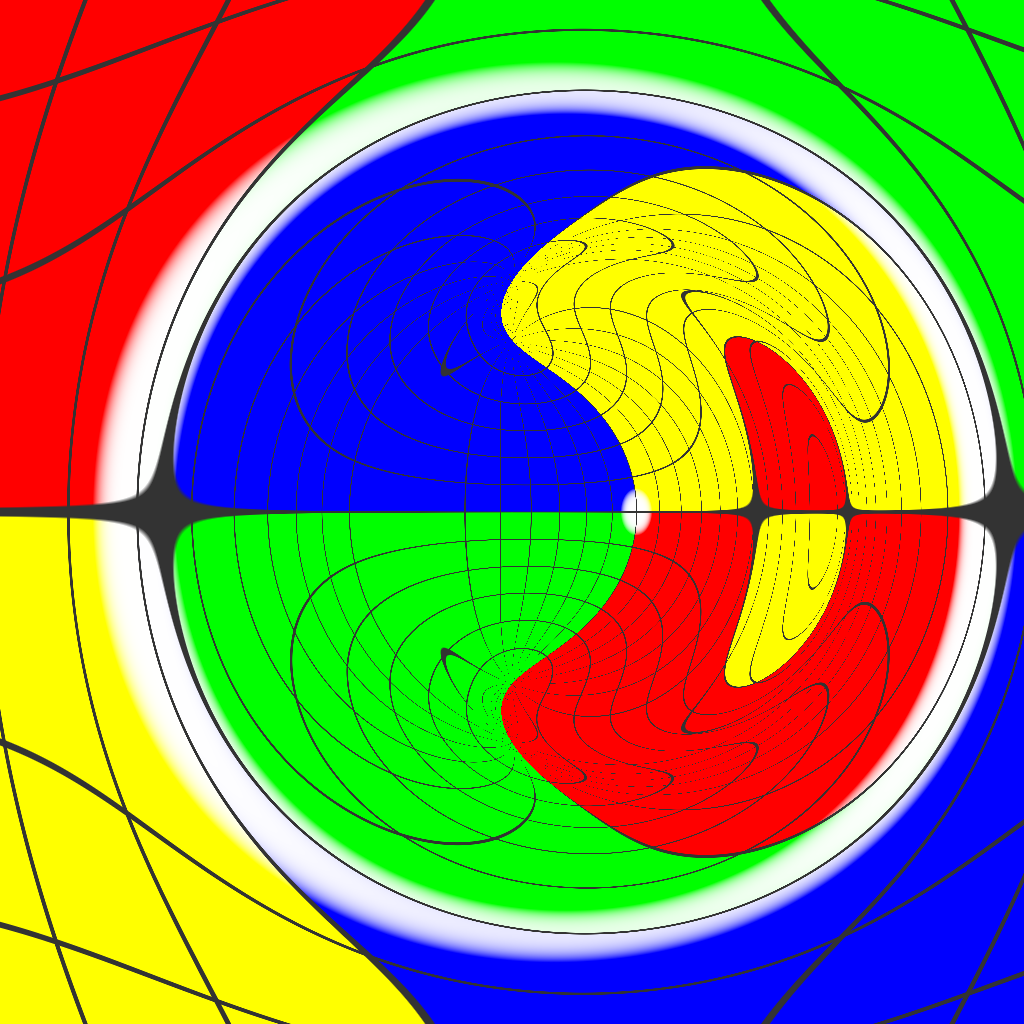}} 
	\end{tabular}
	\caption{\small \label{fig4} Lensing by PSs. From left to right: (top) $\omega^{1.0,3.0,4.0} =0.99; 0.95;0.9$; (bottom)  $\omega^{5.0,6.0,7.0} =0.85; 0.80;0.75$. }
\end{figure}

\begin{figure}[H]
	\begin{tabular}{ccc}
		\subfloat{\includegraphics[width=0.3\textwidth]{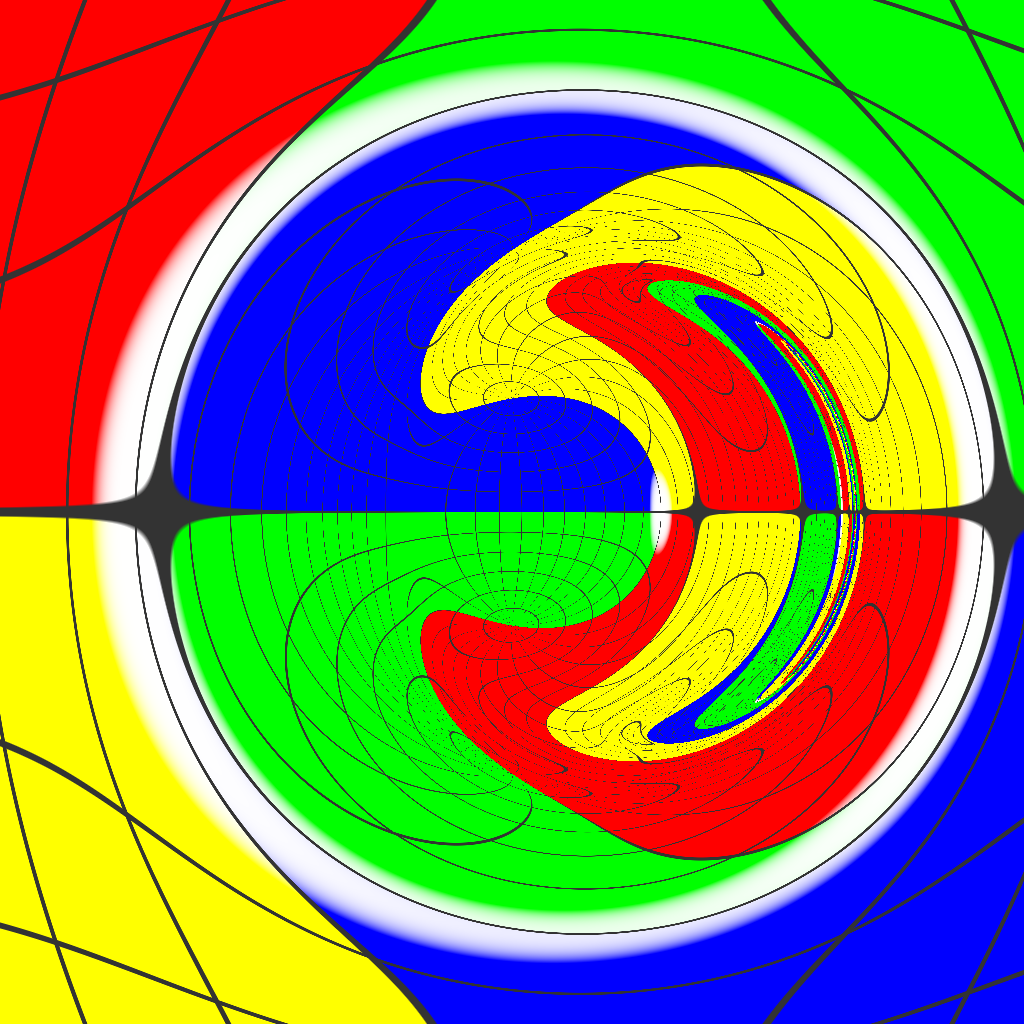}} &
		\subfloat{\includegraphics[width=0.3\textwidth]{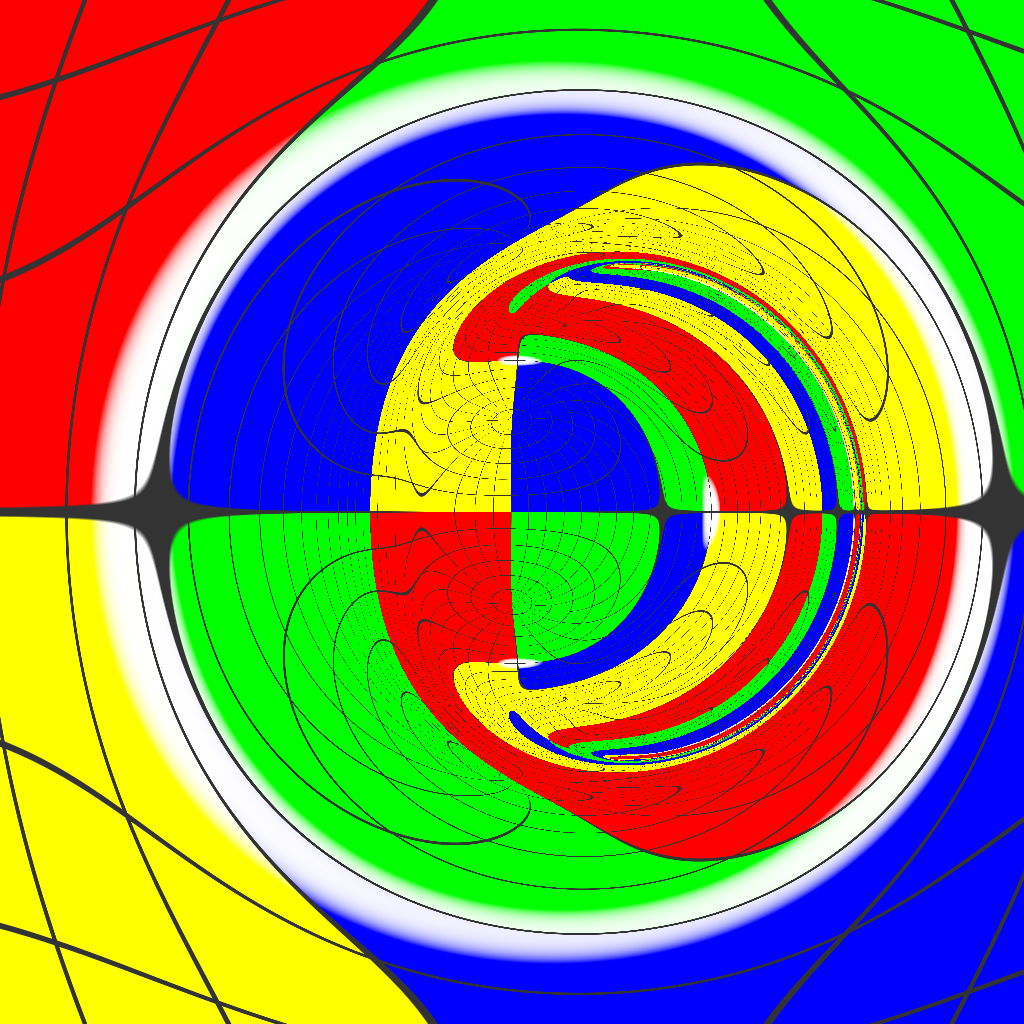}}  &
		\subfloat{\includegraphics[width=0.3\textwidth]{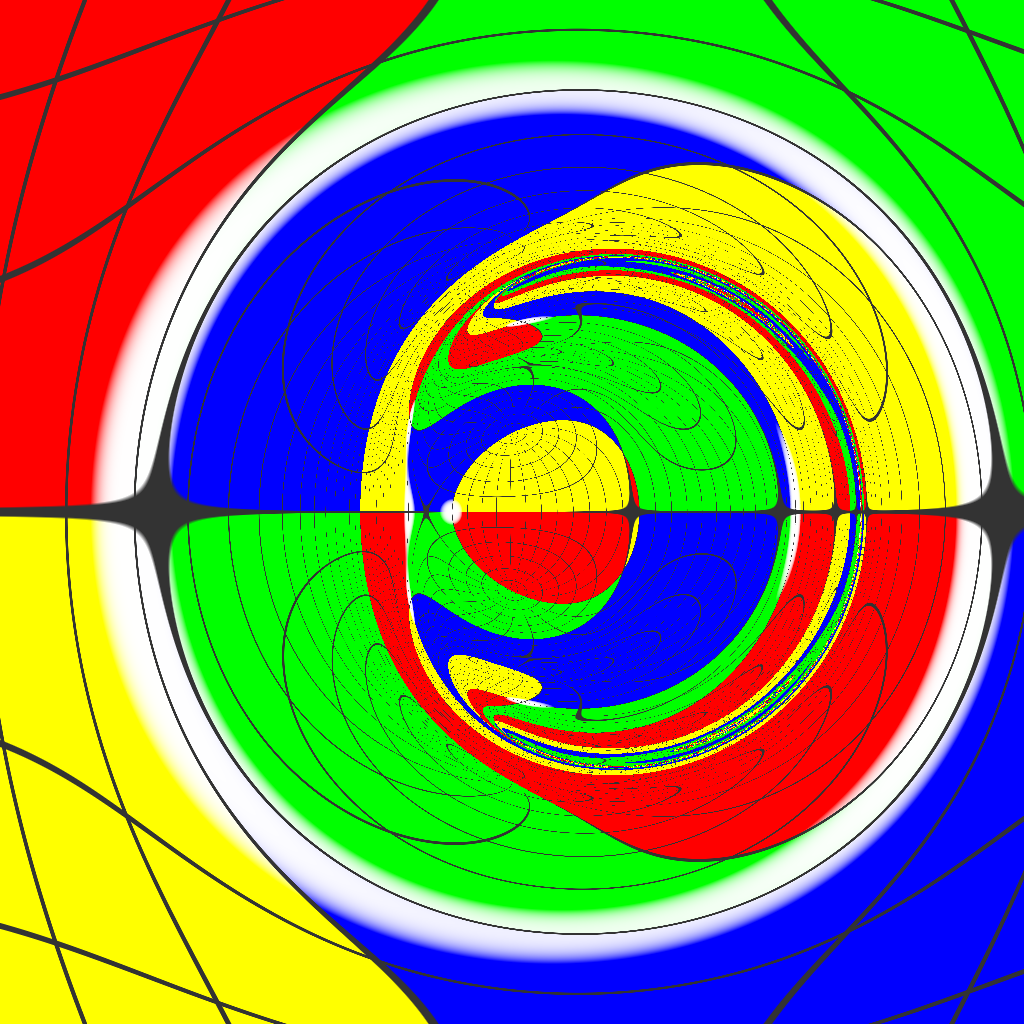}} \\
		\subfloat{\includegraphics[width=0.3\textwidth]{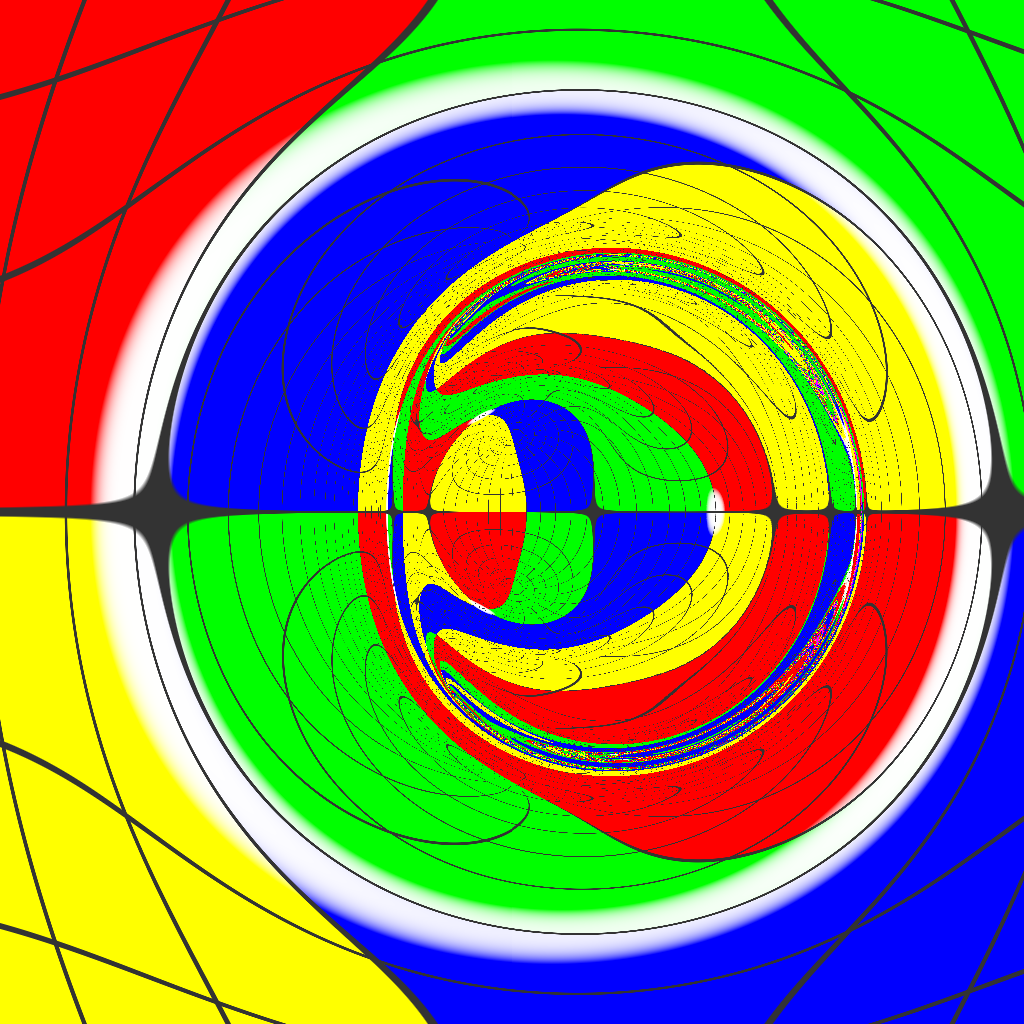}} &
		\subfloat{\includegraphics[width=0.3\textwidth]{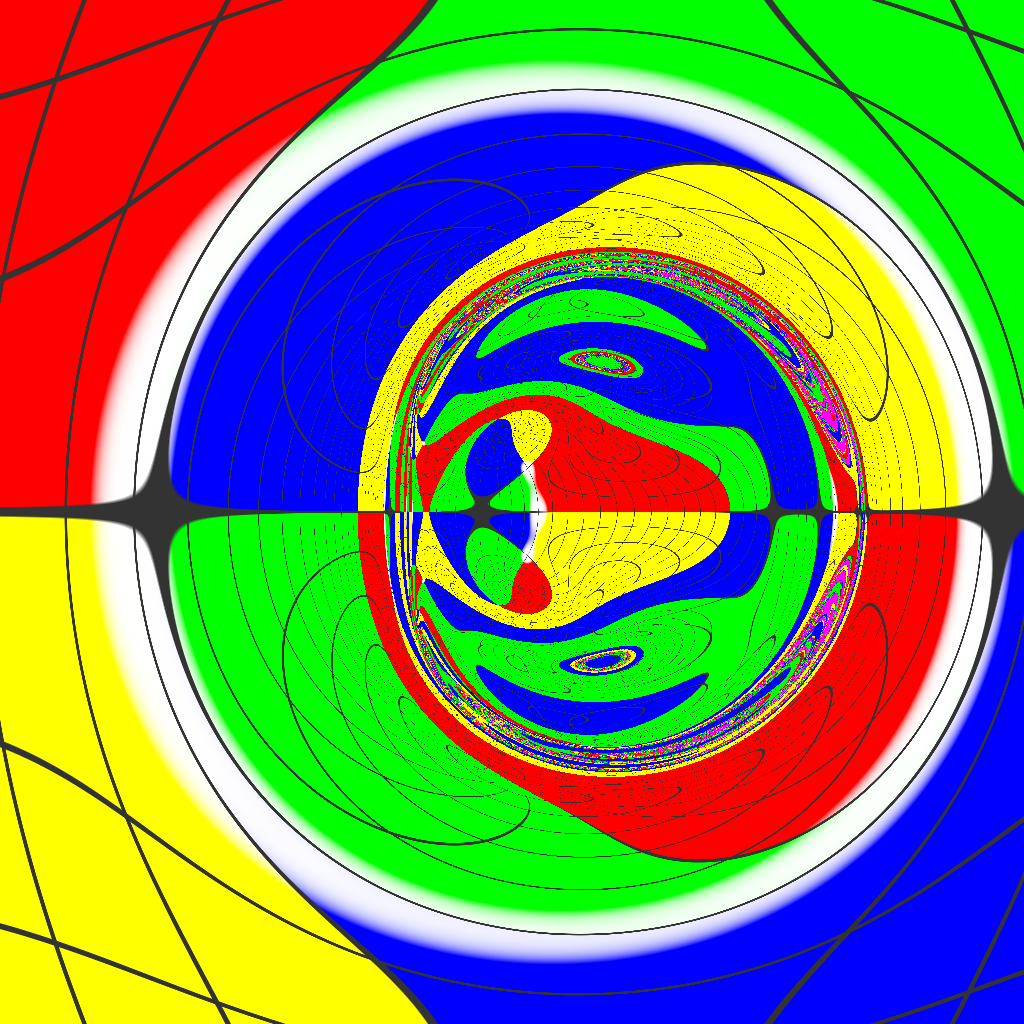}}& 
		\subfloat{\includegraphics[width=0.3\textwidth]{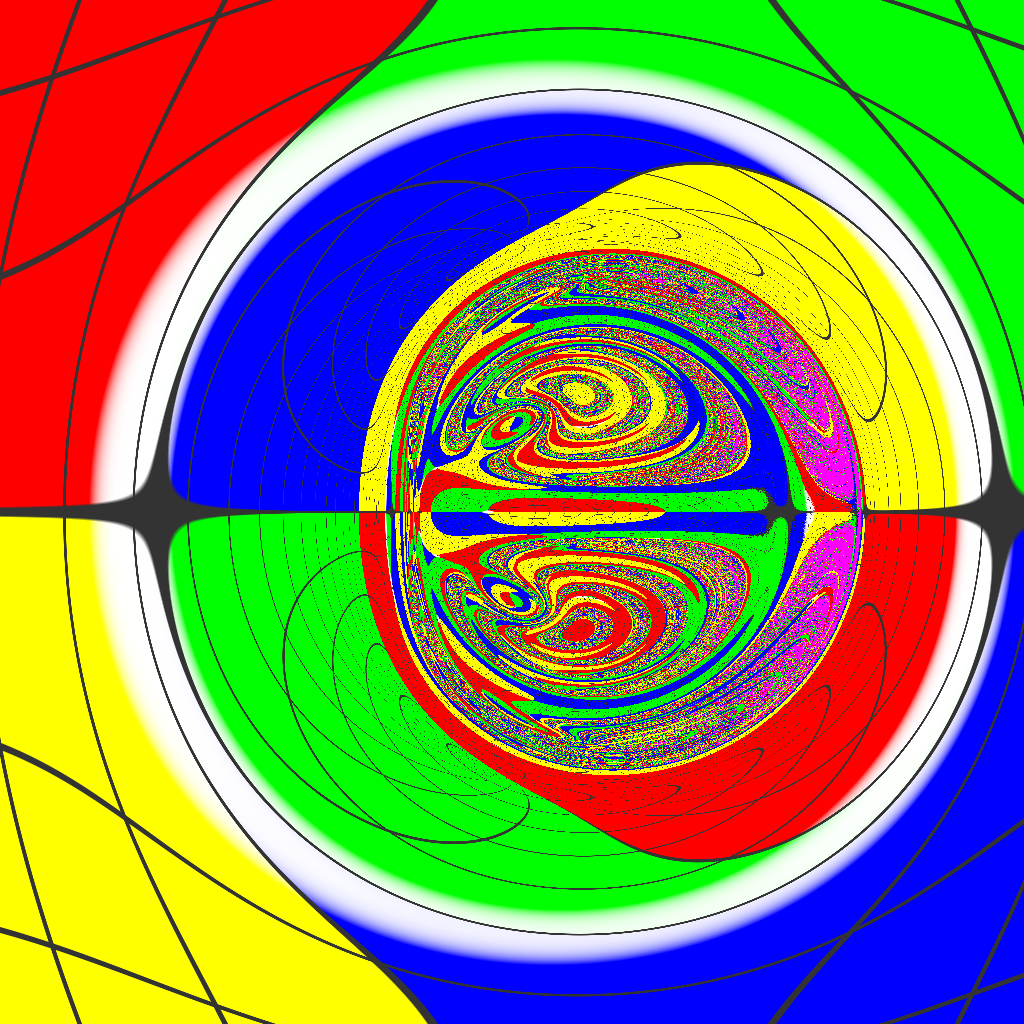}} 
	\end{tabular}
	\caption{\small \label{fig5} Lensing by ultra-compact PSs. From left to right: (top) $\omega^{8.0,9.0,10.0} =0.70; 0.65; 0.60$; (bottom)  $\omega^{11.0,12.0,13.0} = 0.55; 0.50; 0.47$. Note that the lensing images of solutions with $\omega^{12.0,13.0}$ show a few pink colored pixels. These pixels correspond to light rays whose integration time becomes so large that the endpoint cannot be resolved within the numerically allocated integration time, becoming trapped into \textit{pockets} of the effective potential \cite{Chaotic}. }
\end{figure}

\begin{figure}[H]
	\begin{tabular}{cc}
		\subfloat{\includegraphics[width=0.5\textwidth]{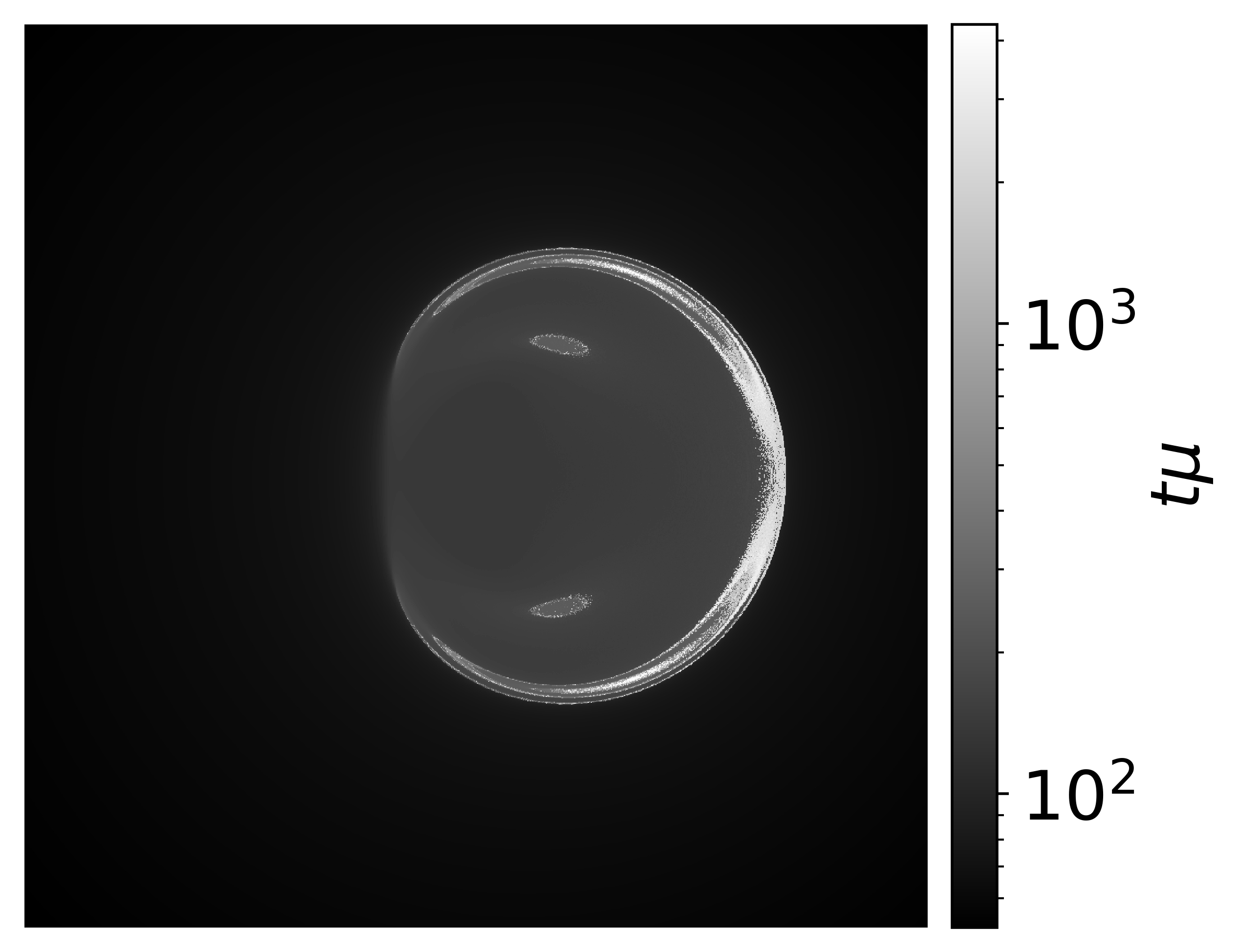}} &
		\subfloat{\includegraphics[width=0.5\textwidth]{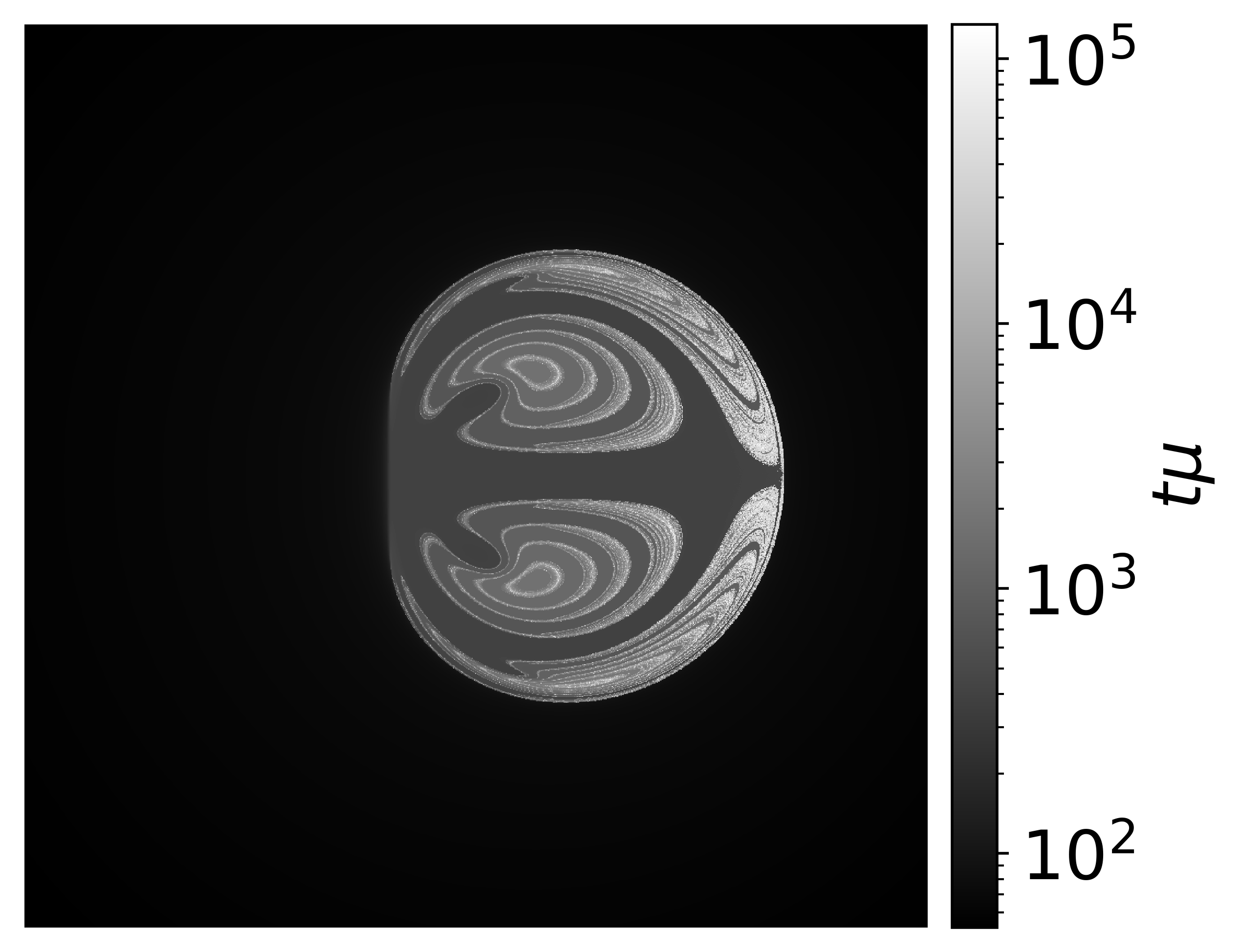}} 
	\end{tabular}
	\caption{\small \label{fig6} Time delay map associated to scattering orbits in the PS solutions 12.0 (left)  and 13.0 (right). Comparison with the bottom middle and right panels of Fig.~\ref{fig5} illustrates how the more pixelated (chaotic) image regions correspond to light ray trajectories that require much larger integration times to complete.}
\end{figure}

It is helpful to start with a discussion on the lensing due to PSs (see Fig. \ref{fig4}-\ref{fig6}).
To gain insight on the gravitational effects of the PSs, it is pedagogical to follow a sequence of solutions with increasingly stronger gravitational effects. Thus, we start from the PS solution 1.0,  with $\omega^{1.0}=0.99$, which is the closest one to vacuum (for which $\omega=1$)~\footnote{For ease of notation we quote the frequencies of the solutions in units of $\mu$; $i.e.$ $\omega/\mu\rightarrow \omega$.}, and we then move along red spiral in the space of solutions, see Fig.~\ref{fig1}.

For solution 1.0 we find no noticeable distortion on the background - Fig. \ref{fig4} (top left panel) - when compared with the flat-spacetime reference image in Fig.~\ref{fig2}. By moving to solutions 2.0 and 3.0 the (still very weak) lensing distortion only increases mildly  - see Fig. \ref{fig4} (top middle panel). Then, a new qualitative feature arises for solution 4.0 with $\omega^{4.0}_{ER1}=0.90$: the appearance of the first Einstein ring, as the white spot at {\bf F} opens up and encloses two inverted copies of regions of  the sky belonging to two of the quadrants of the celestial sphere - see Fig. \ref{fig4} (top right panel).

As we move further along the spiral to solution 5.0, the Einstein ring takes a more elliptical shape and copies of all four quadrants can now be seen inside the ring - Fig. \ref{fig4} (bottom left panel). Similarly to rotating scalar BSs~\cite{Cunha_2015}, the side rotating away from $\mathcal{O}$ appears more amplified, and {\bf F} is progressively shifted to one side. As we move even further along the PS spiral to solution 7.0 and $\omega^{7.0}_{ER2}=0.75$, then we find the emergence of a second Einstein ring - see Fig. \ref{fig4} (bottom right panel). Likewise to rotating BSs, these new rings also have a squashed "D-shape".

The first appearance of a light ring marks the transition from the compact to the ultra-compact regime~\cite{Cardoso_2014}. This light ring appearance is observed (in our selection of solutions) for the PS solution 8.0, with $\omega^{8.0}_{LR1}=0.70$, cf. Fig.~\ref{fig5} (top left panel); then an infinite number of copies -- and a self similar structure -- is expected to arise~\cite{Bohn2015}. 

Even more compact PSs solutions start to exhibit image features that were shown to be associated with chaotic scattering in the scalar case~\cite{Chaotic}. Solutions 12.0 and 13.0, with respectively $\omega^{12.0}=0.50$ and $\omega^{13.0}=0.47$, are both examples of PSs that display these features - see Fig.~\ref{fig5} (bottom middle and right panels). 

In addition, the authors in~\cite{Chaotic} discussed how (in the scalar case) there is a correlation between the chaotic pattern in the lensed image and a large integration time $t$, computed along null geodesics of those regions. Following a similar analysis, we present in Fig.~\ref{fig6} the time-delay map associated to PS solutions 12.0 and 13.0 and, also here, a correlation between the integrated time and chaotic behaviour is apparent.

Finally, in Table \ref{tab1} we summarize some key features of the PSs data set by indicating the frequency (in units of $\mu$) of the solution at which they first appear.\footnote{Some of this information was recently reported in~\cite{Cunha:2022gde}.} For comparison, we also present the same information but for the rotating BSs case \cite{Cunha_2015}. We observe that the emergence of the first light ring precedes that of an ergo-region, which is first spotted at $\omega \simeq 0.602$. This is generic feature: it was recently proved that any stationary, axisymmetric and asymptotically flat spacetime in $1+3$ dimensions with an ergo-region must have at least one light ring outside the ergo-region \cite{ghosh2021light}. Thus, some of the PSs solutions that we present here provide another example where the converse statement of the above mentioned theorem does not hold. 

The fact that ultracompact spinning PSs do not feature an ergo-region is useful in isolating imprints of the light ring instability from those of ergo-region sourced instabilities - see $e.g.$, \cite{Maggio_2019}. The light ring instability has been suggested to be associated with the presence of stable light rings in the spacetime \cite{Keir:2014oka,Cardoso_2014, Cunha_2017_3}, and ultracompact PSs necessarily have both a stable and an unstable LR~\cite{Cunha_2017_3}. For the case of PSs this instability was recently shown to lead to a migration of the ultracompact PSs to non-ultracompact ones~\cite{Cunha:2022gde}.  

\begin{table}[H]
	\centering
	\begin{tabular}{l|c|c|c|c}
		\hline
		\hline
		Family & $\omega$ first Einstein ring & $\omega$ for Multiple rings & $\omega$ of Light rings  & $\omega$ for ergo-region \\ \hline
		Proca  & 0.90              & 0.75         & 0.711    & 0.602    \\ \hline
		Scalar & 0.92             & 0.75         & 0.747   & 0.658     \\ \hline
	\end{tabular}
	\caption{\label{tab1} Main features of the PSs and BSs dataset and the corresponding frequencies at which they first emerge.}
\end{table}

As a final comment, we remark that the whole analysis of the lensing features by spinning PSs has a clear qualitative parallelism with the analogous study of their scalar cousins~\cite{Cunha_2015,Chaotic}.

\section{Lensing by KBHsPH}

Let us now consider the lensing images and shadows of KBHsPH. Here, besides the frequency $\omega$, the solutions are also parameterized by the radial coordinate of the event horizon, $r_H$.  In general, the larger the value of $r_H$, the smaller is the deviation from Kerr.

\begin{figure}[H]
	\begin{tabular}{cccc}
		\subfloat{\includegraphics[width=0.225\textwidth]{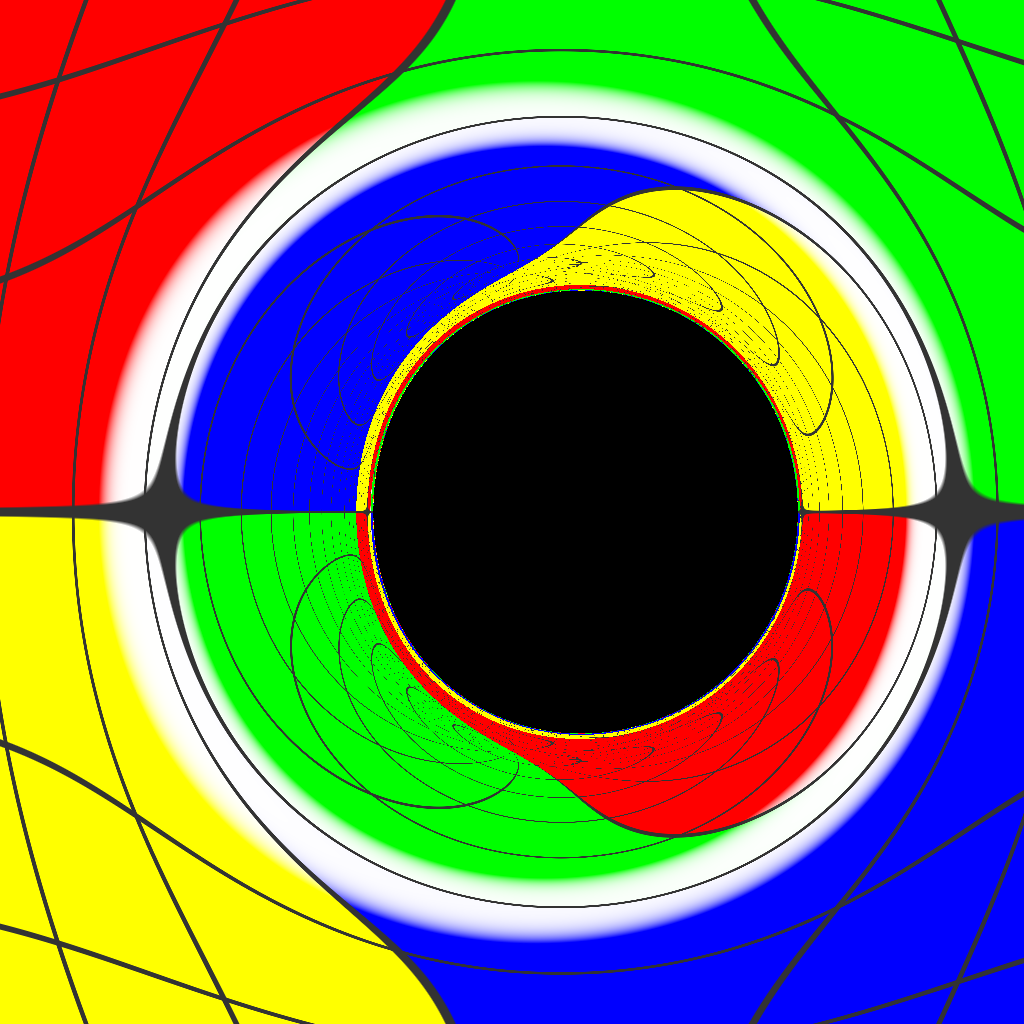}} &
		\subfloat{\includegraphics[width=0.225\textwidth]{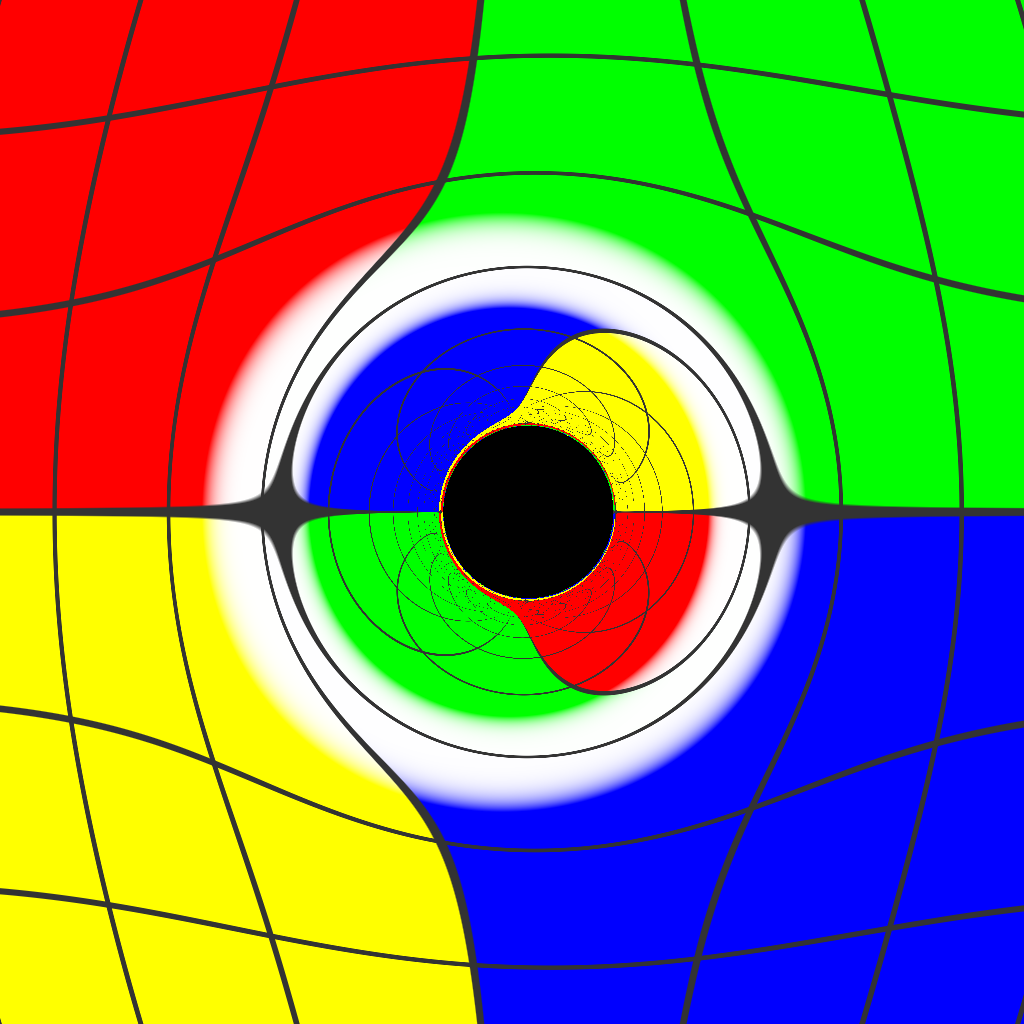}}  &
		\subfloat{\includegraphics[width=0.225\textwidth]{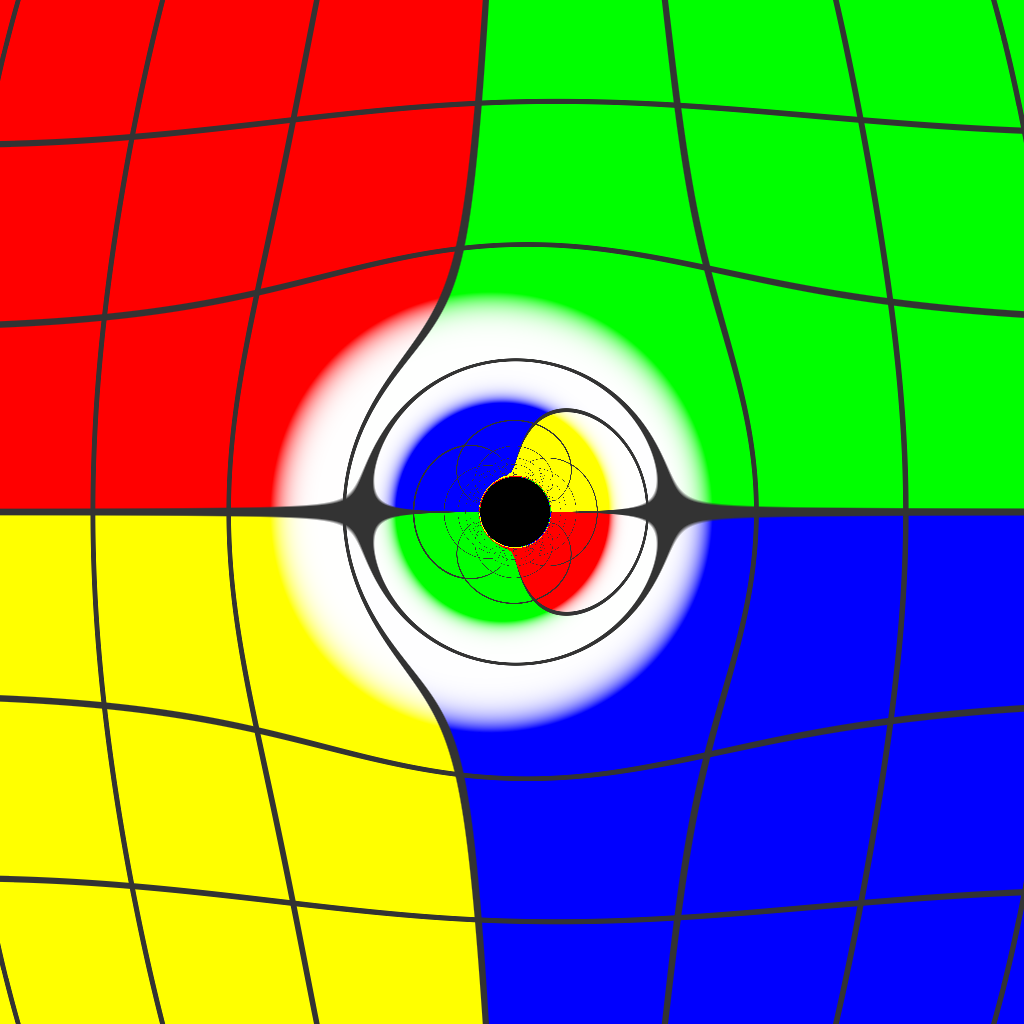}} &
		\subfloat{\includegraphics[width=0.225\textwidth]{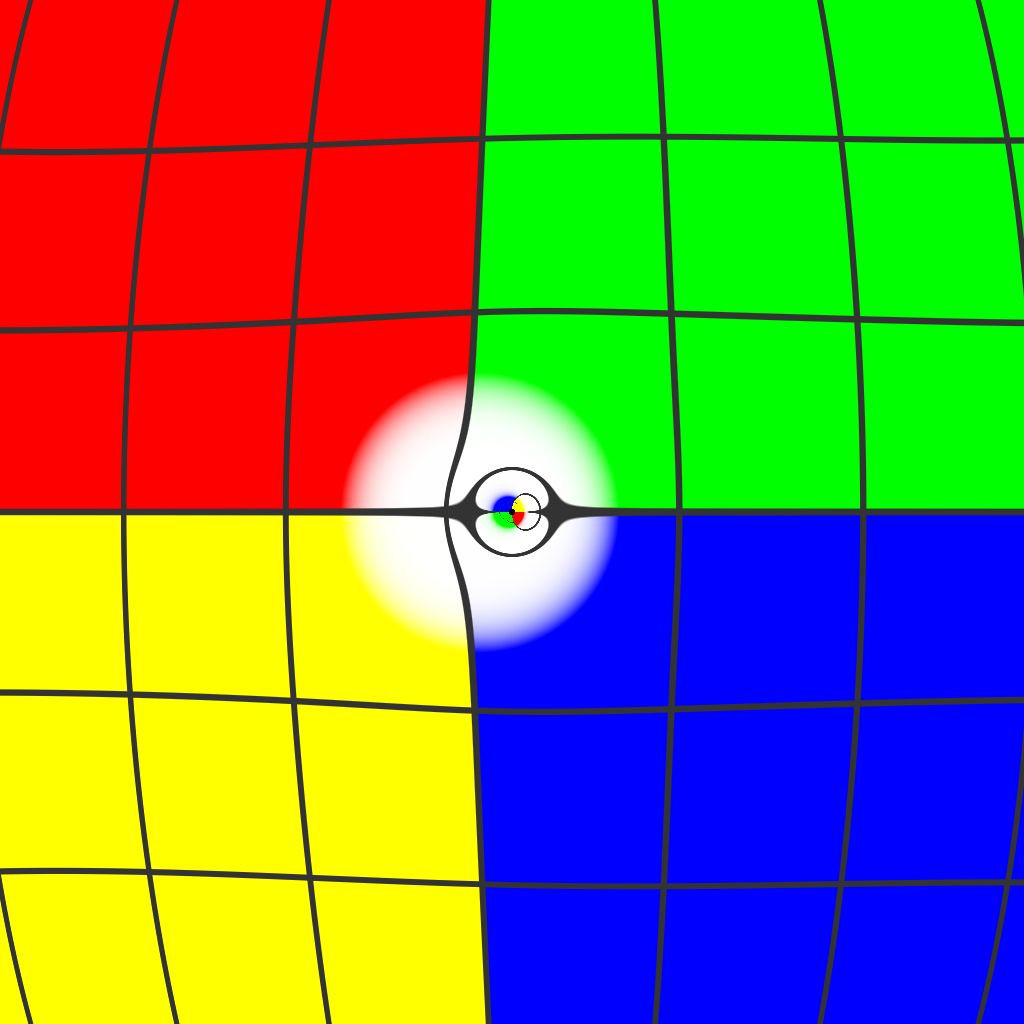}} \\
		\subfloat{\includegraphics[width=0.225\textwidth]{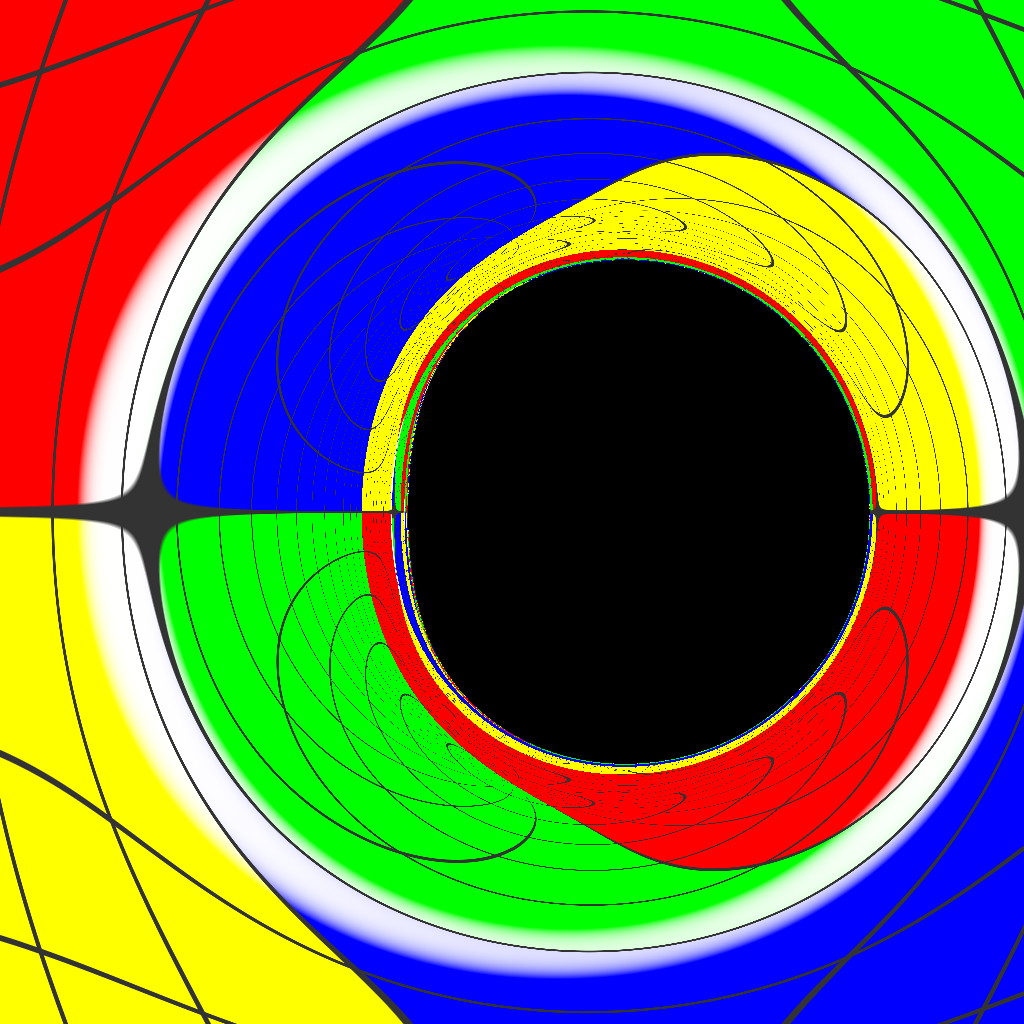}} &
		\subfloat{\includegraphics[width=0.225\textwidth]{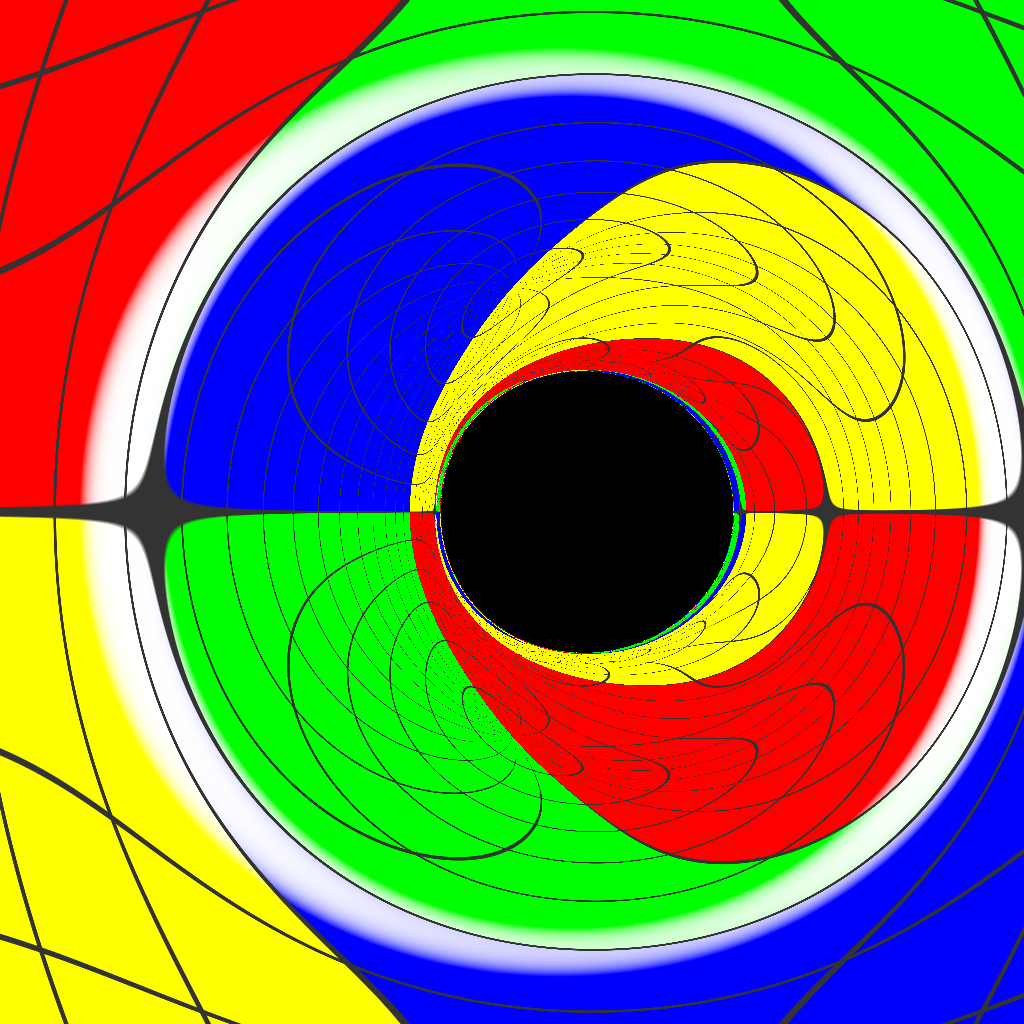}}  &
		\subfloat{\includegraphics[width=0.225\textwidth]{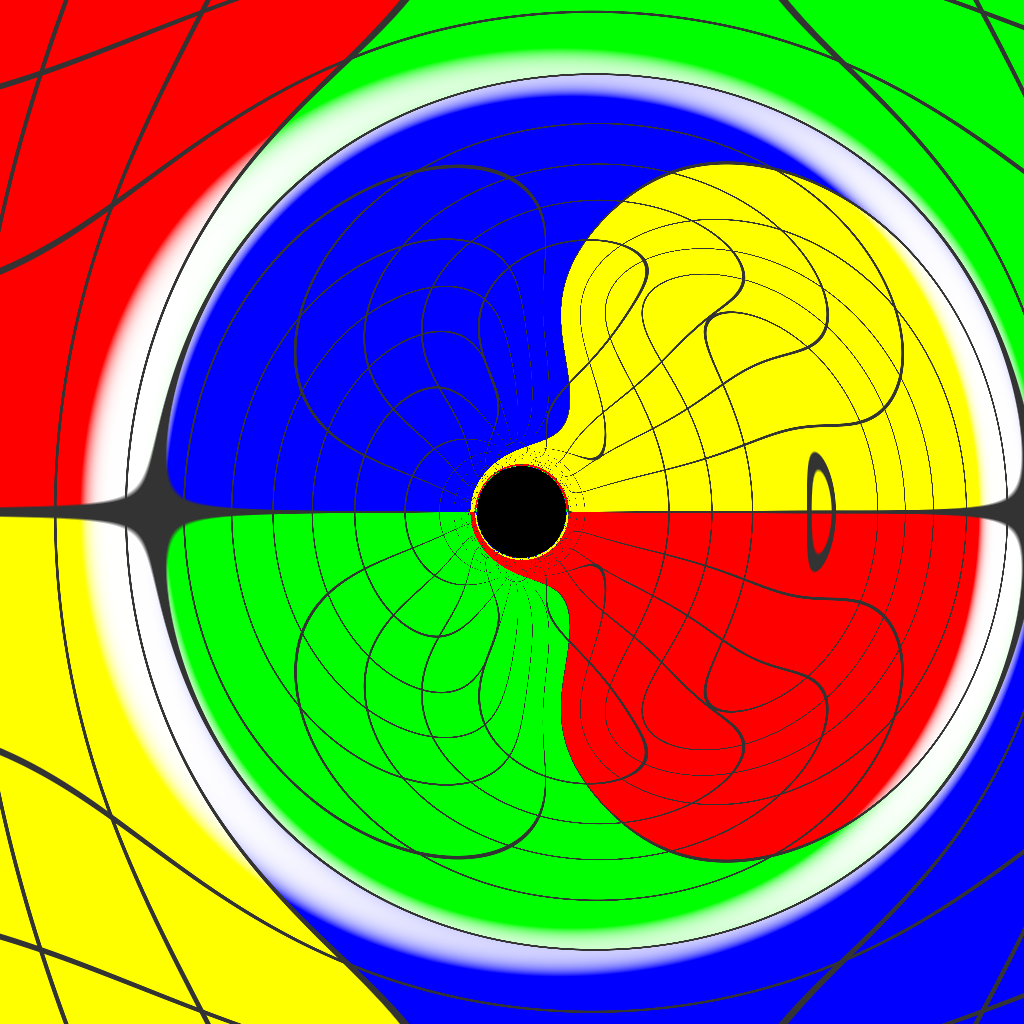}} &
		\subfloat{\includegraphics[width=0.225\textwidth]{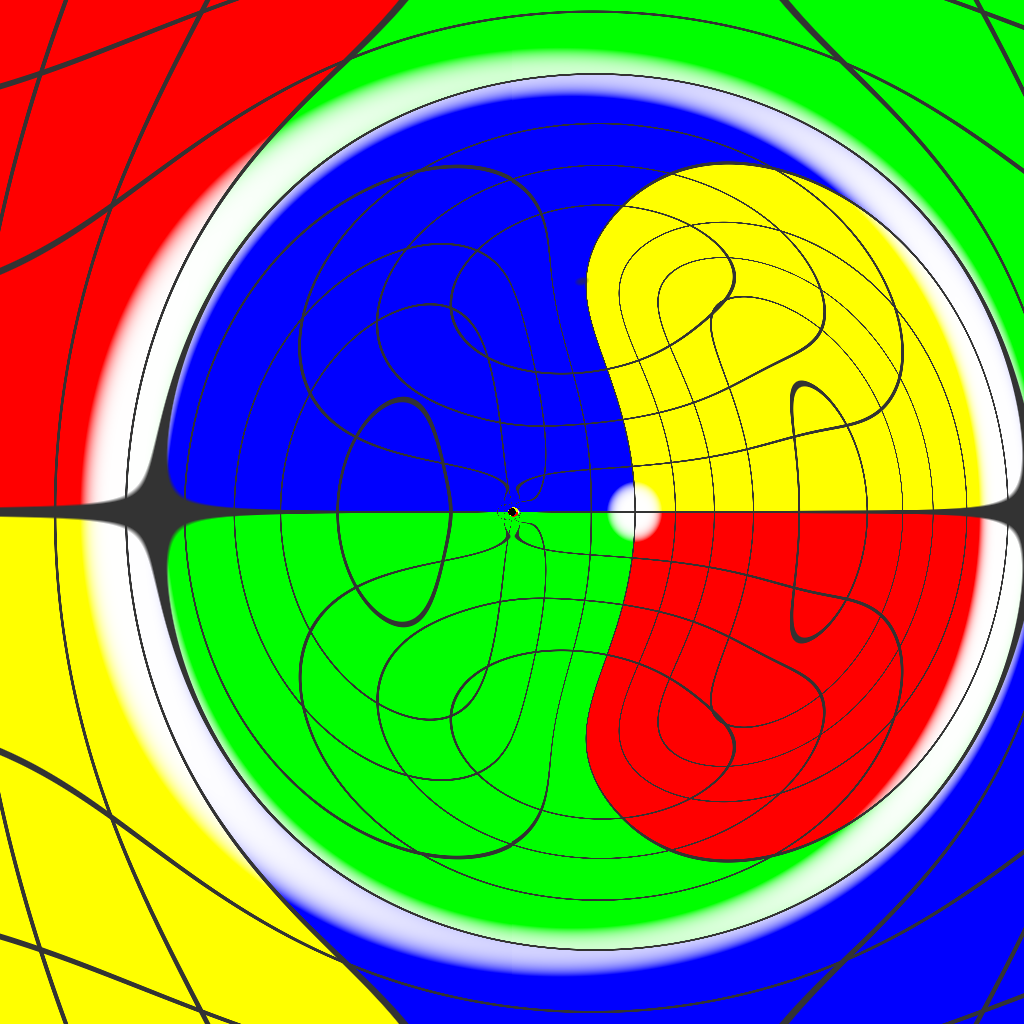}}
	\end{tabular}
	\caption{ \small \label{fig7a} Subsets  $\omega^{3.4,3.3,3.2,3.1}=0.95$  (top) and  $\omega^{6.4,6.3,6.2,6.1}=0.80$ (bottom) of KBHsPH solutions. The value of the horizon mass decreases from left to right.}
\end{figure}

Similarly to the scalar case, KBHsPH have Kerr-like shadows near the existence line (whence they bifurcate from Kerr solutions) for the same mass, angular momentum and comparable observation conditions. This is the case for instance of solution $3.4$ - Fig. \ref{fig7a} (top left panel). Considering then the sequence of solutions $3.v$ (with $v$ $\in \left\{ 1,2,3,4\right\}$), one observes a monotonically decreasing shadow size as $v$ also decreases - see Fig. \ref{fig7a} (top panels, from left to right). The decreasing shadow size seems a natural consequence of the correspondingly smaller fraction of the horizon mass to the ADM mass. We remark that the sequence of KBHsPH solutions $3.v$ bifurcate from the PS solution 3.0, with frequency $\omega^{3.0}=0.95$.

We can consider next the sequence of solutions $6.v$, with $v$ $\in \left\{ 1,2,3,4\right\}$ - see Fig.~\ref{fig7a} (bottom panels), which bifurcates from a Kerr solution at a higher spin. This manifests in the shadow of solution 6.4 having a slightly more $D$-like shape (a distinctive feature of a spinning Kerr observed from the equatorial plane), at least in comparison to solution 3.4 in the previous sequence - see bottom left panel of Fig. \ref{fig7a}.

Another curious feature, apparent along the sequence $6.v$, is the (approximately) composite nature of these hairy BHs, $i.e.$ the lensing effects are akin to what one would expect of an horizon in the center of a PS. In particular, for the solution $6.1$, the bosonic part clearly dominates the geometry, and the lensing image closely resembles that of the PS solution 6.0 analyzed in the previous section - Fig.~\ref{fig4} (bottom middle panel).

\begin{figure}[H]
	\begin{tabular}{ccc}
		\subfloat{\includegraphics[width=0.3\textwidth]{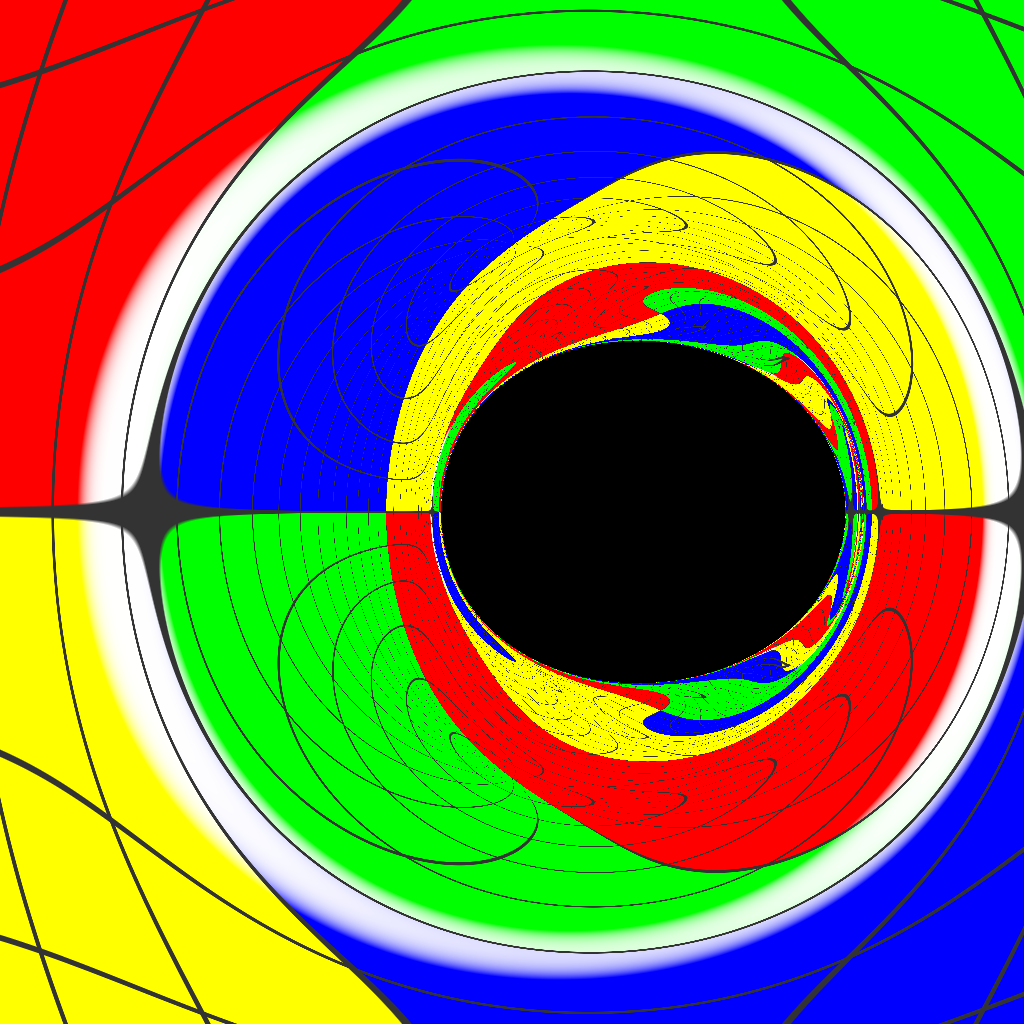}} &
		\subfloat{\includegraphics[width=0.3\textwidth]{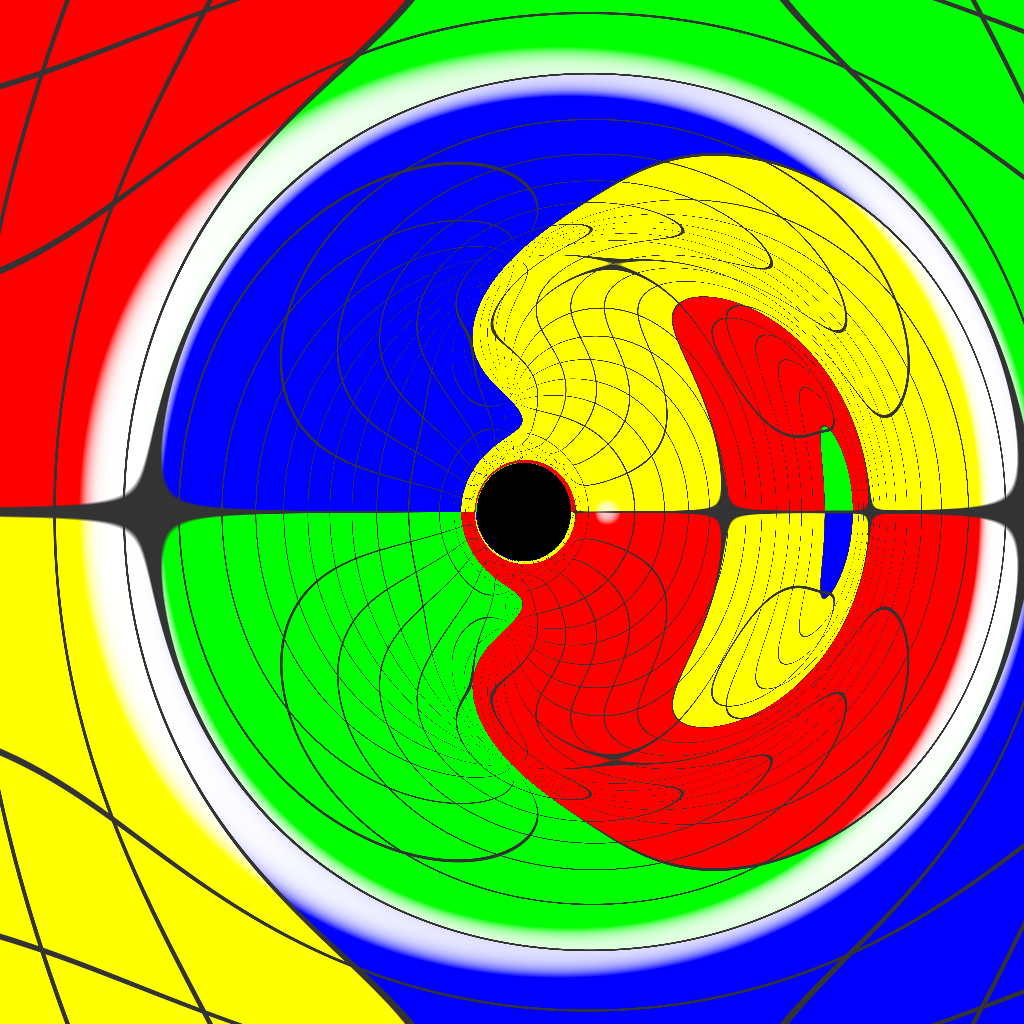}}  &
		\subfloat{\includegraphics[width=0.3\textwidth]{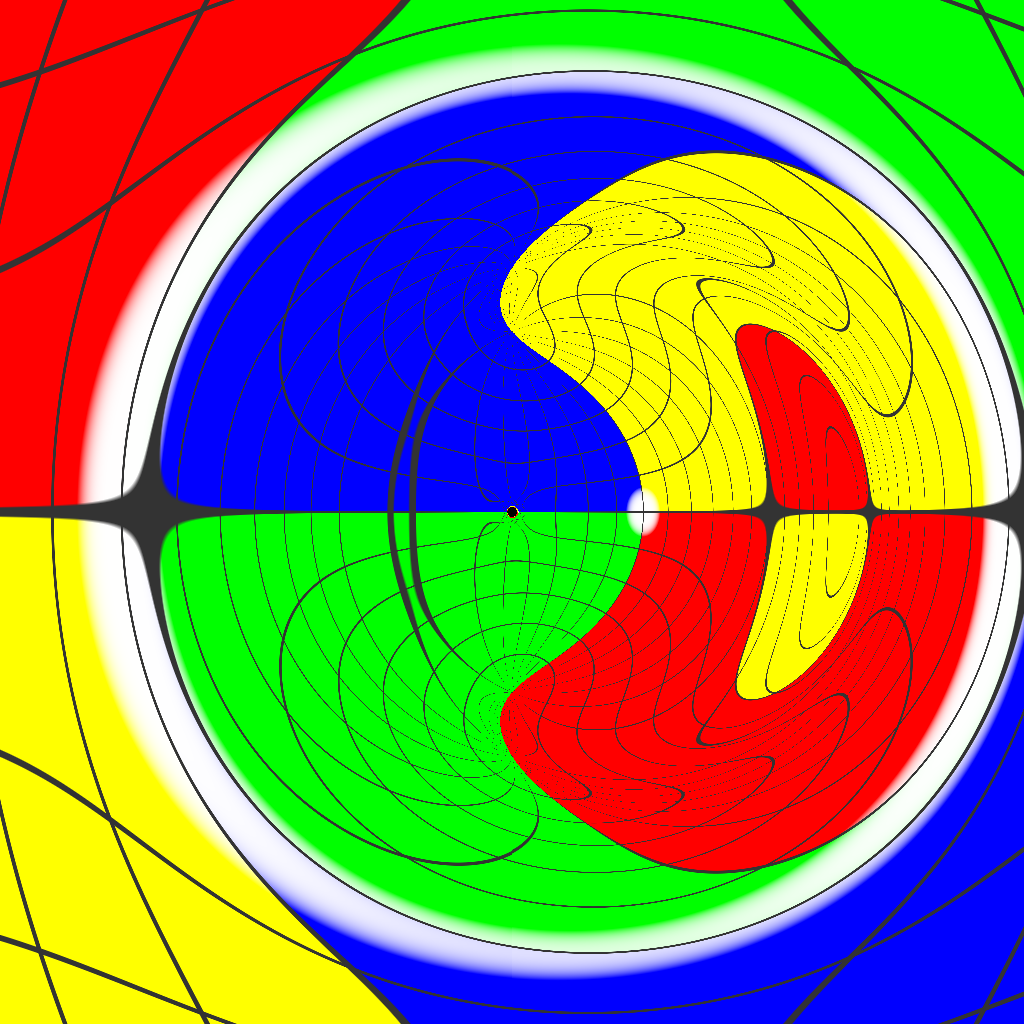}} \\
		\subfloat{\includegraphics[width=0.3\textwidth]{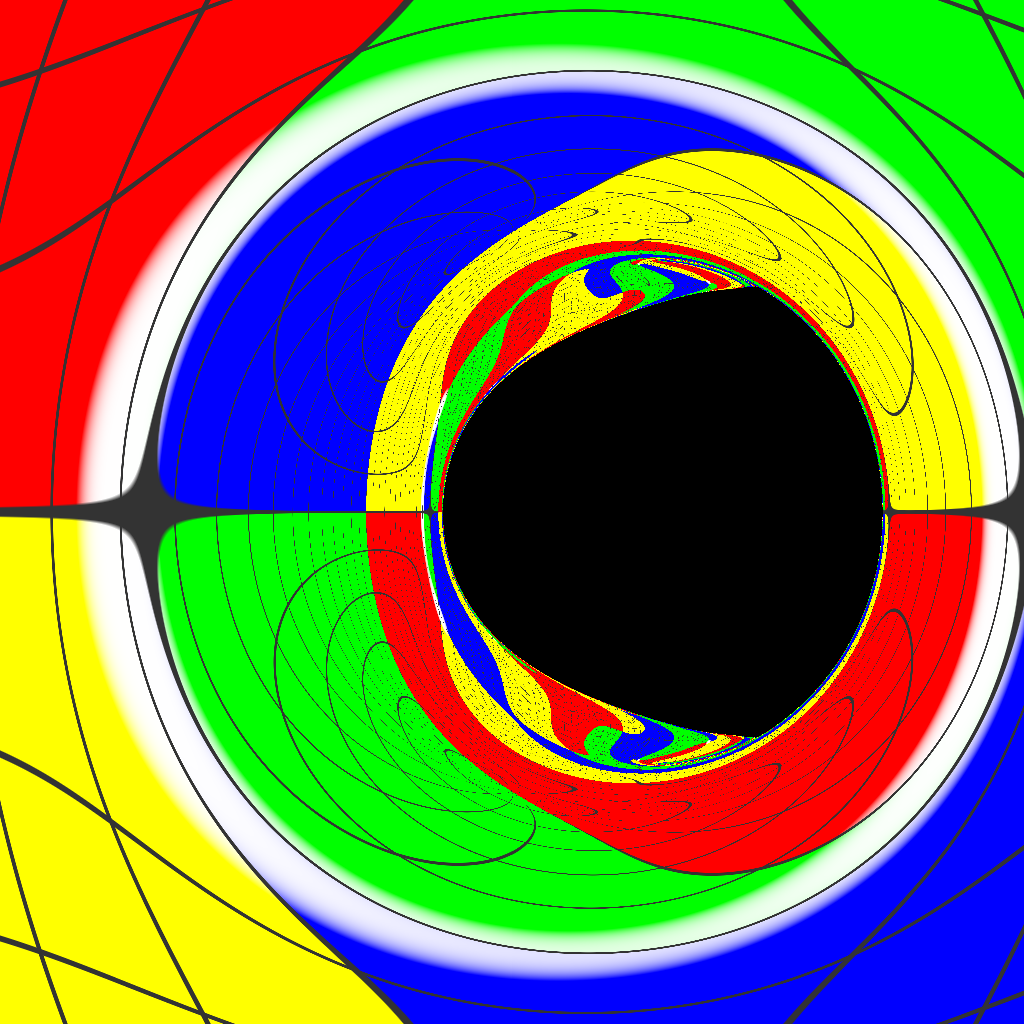}} &
		\subfloat{\includegraphics[width=0.3\textwidth]{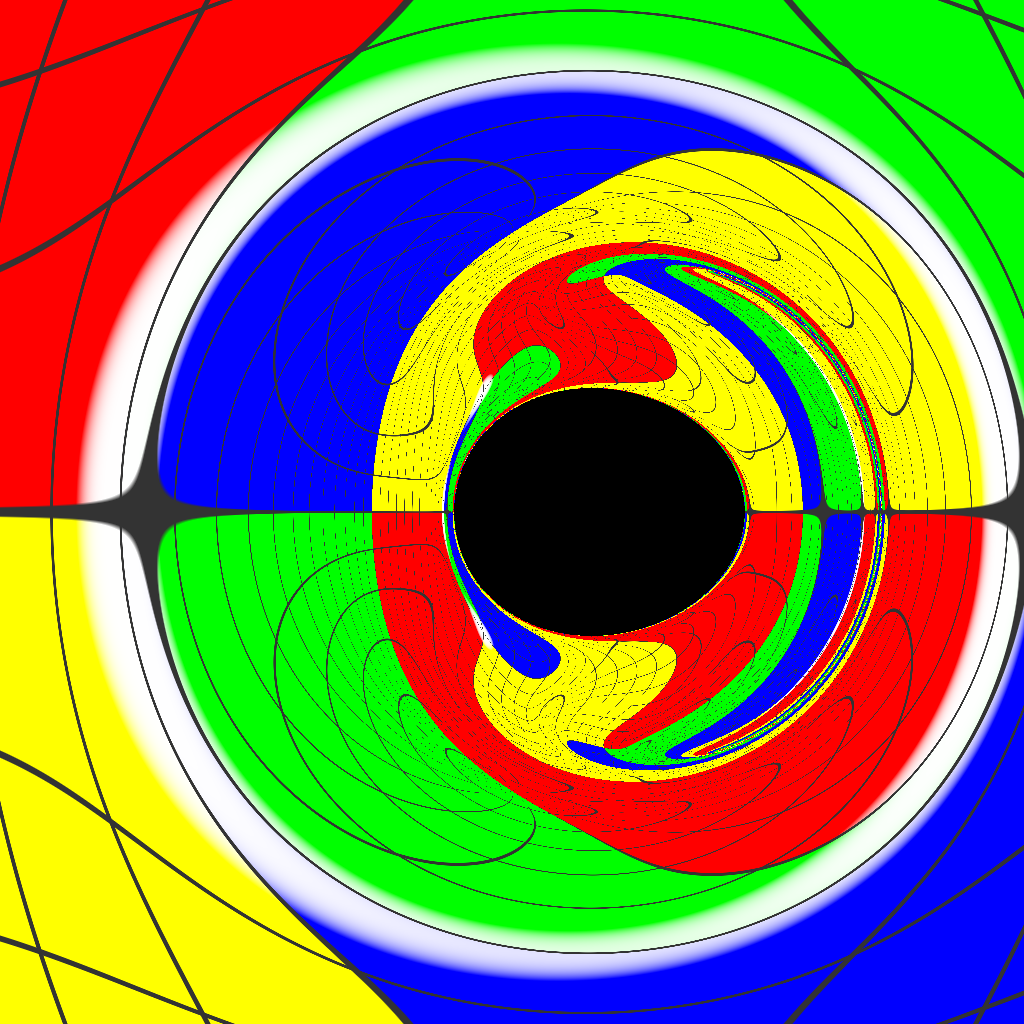}}  &
		\subfloat{\includegraphics[width=0.3\textwidth]{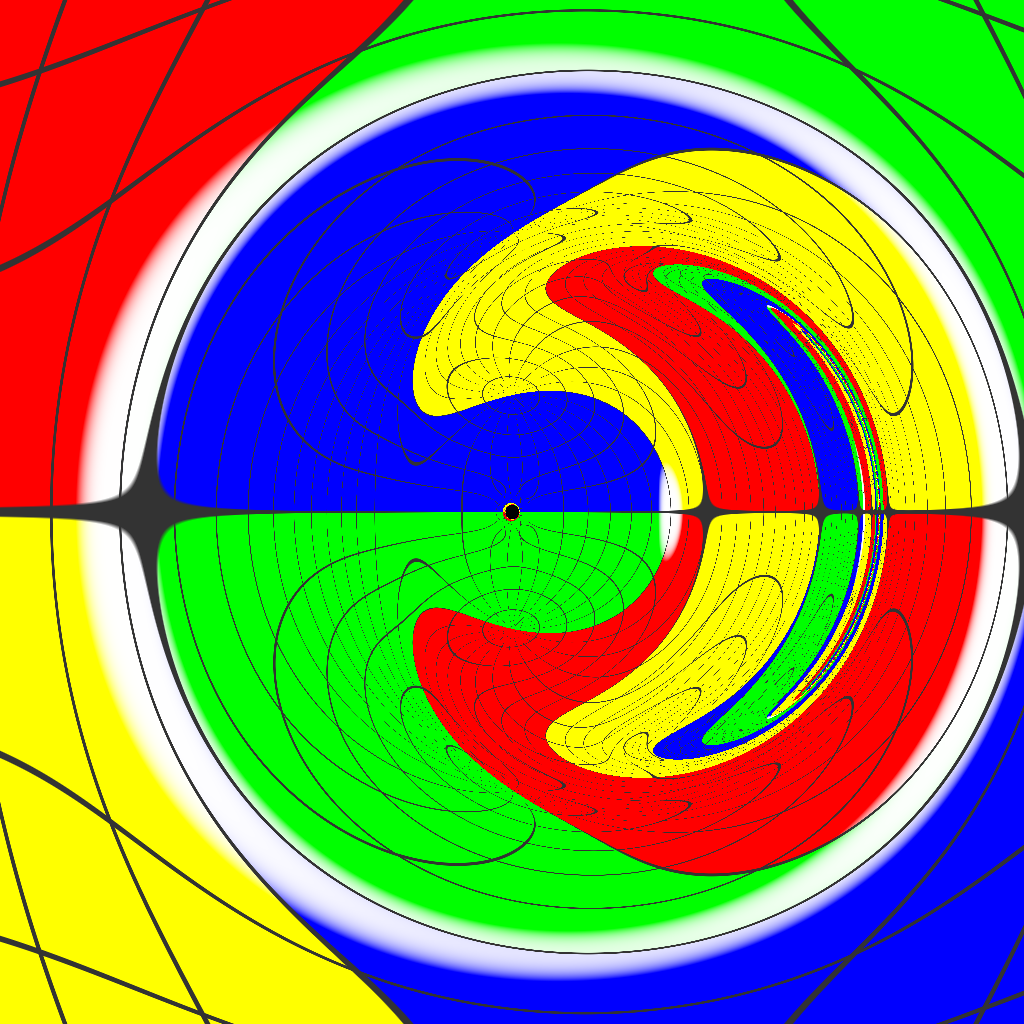}}
	\end{tabular}
	\caption{\small \label{fig7b} Subsets  $\omega^{7.3,7.2,7.11}=0.75$  (top) and  $\omega^{8.3,8.2,8.1}=0.70$ (bottom) of KBHsPH solutions. The value of the horizon mass decreases from left to right.}
\end{figure}

\begin{figure}[H]
	\begin{tabular}{ccc}
		\subfloat{\includegraphics[width=0.3\textwidth]{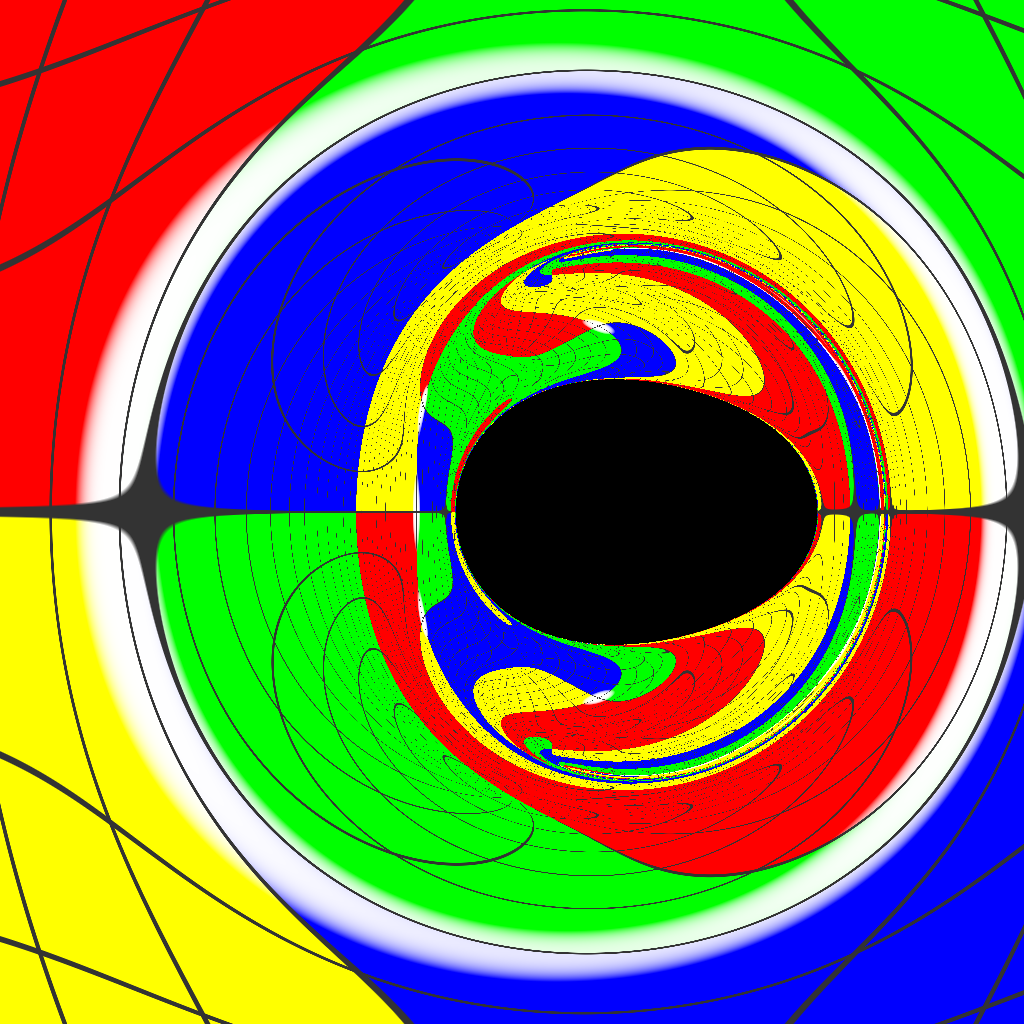}} &
		\subfloat{\includegraphics[width=0.3\textwidth]{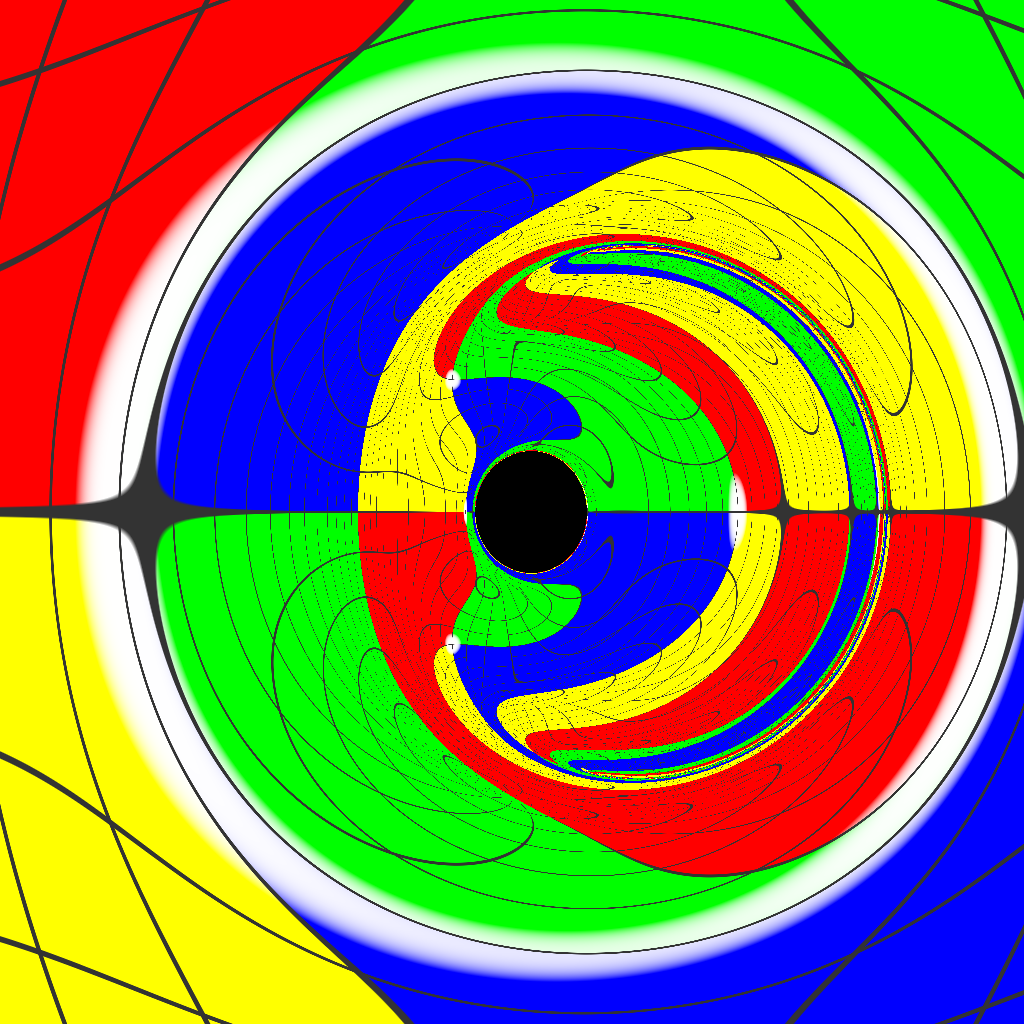}}  &
		\subfloat{\includegraphics[width=0.3\textwidth]{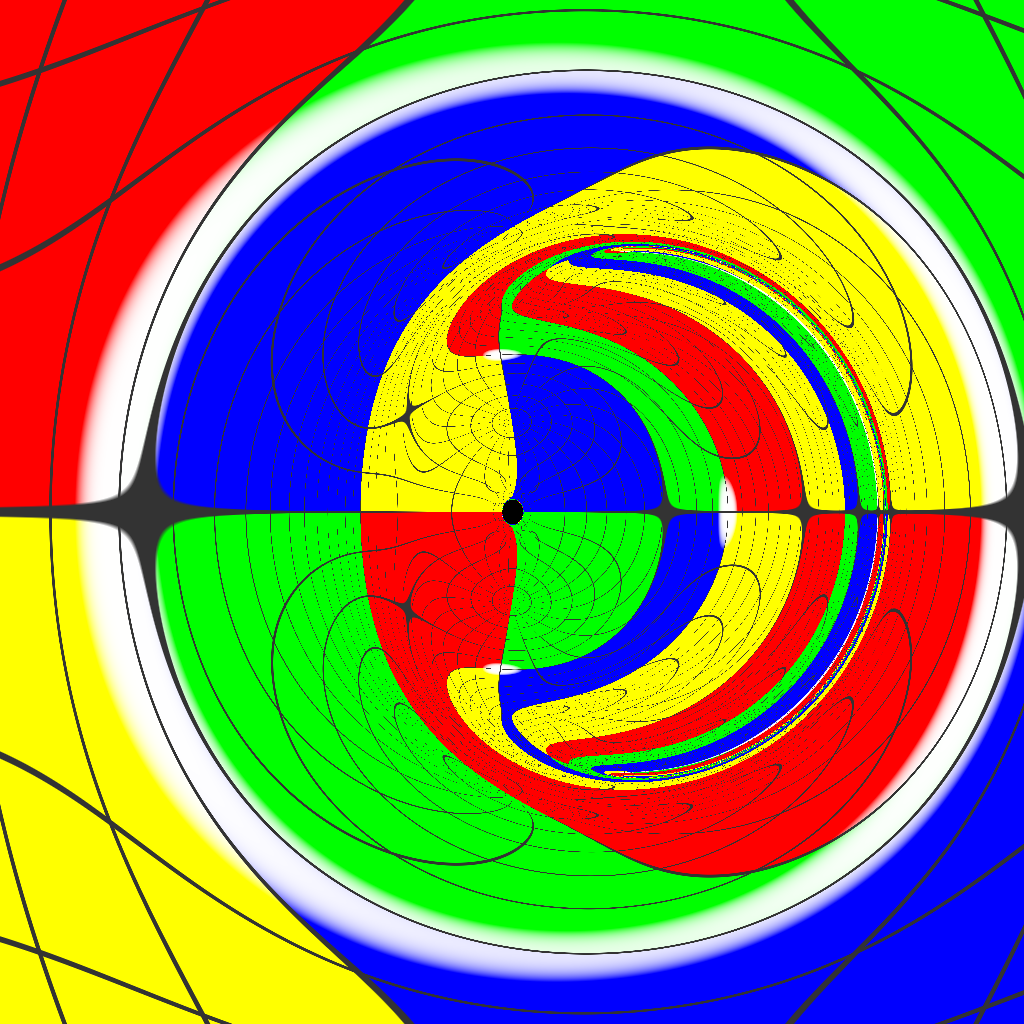}} \\
		\subfloat{\includegraphics[width=0.3\textwidth]{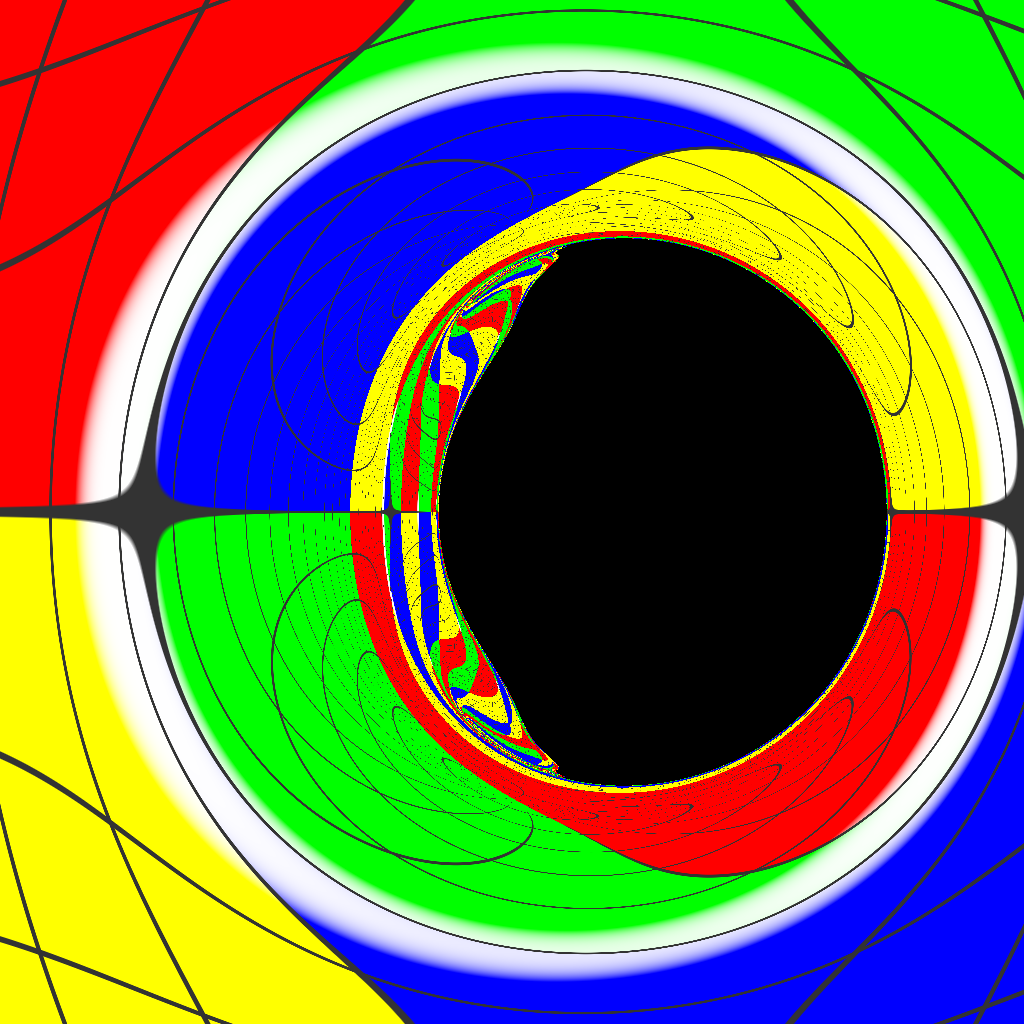}} &
		\subfloat{\includegraphics[width=0.3\textwidth]{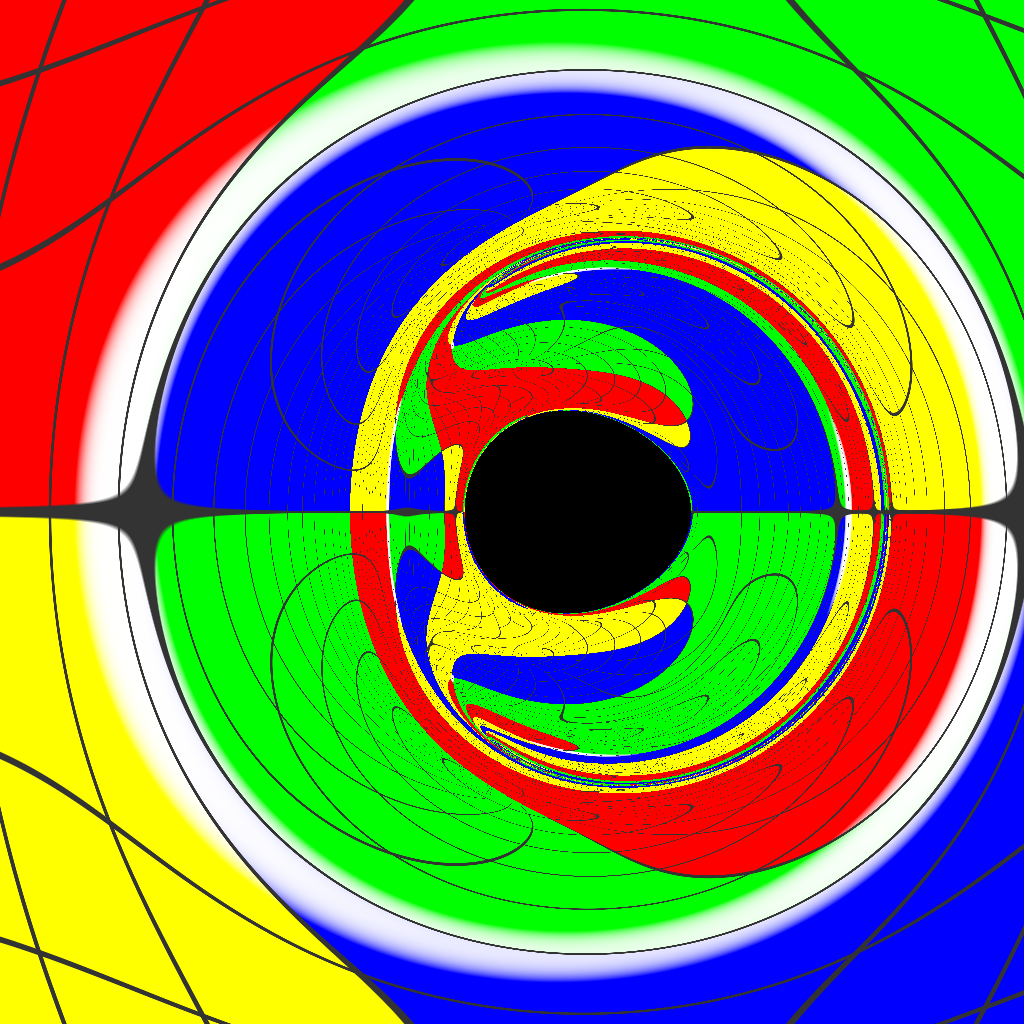}}  &
		\subfloat{\includegraphics[width=0.3\textwidth]{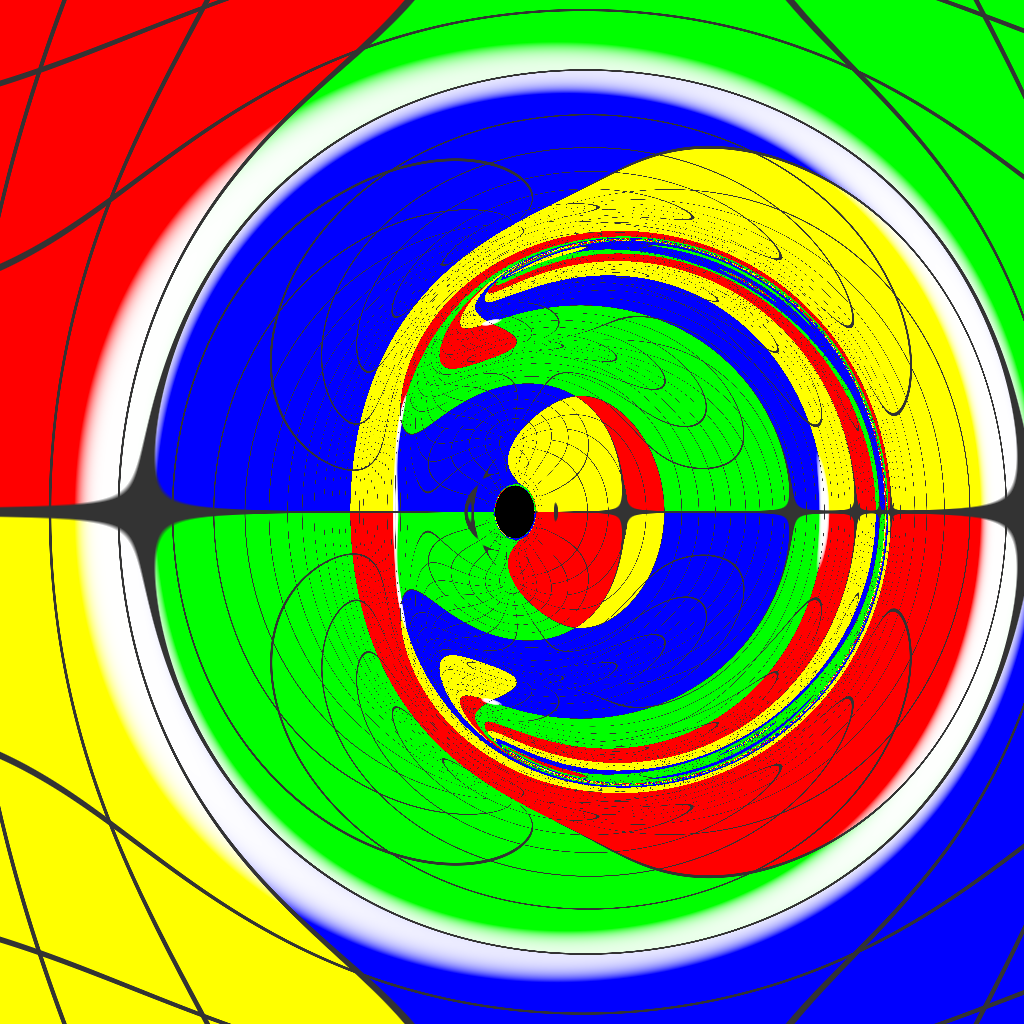}}
	\end{tabular}
	\caption{\small \label{fig8} Subsets  $\omega^{9.3,9.2,9.1}=0.65$  (top) and  $\omega^{10.3,10.2,10.1}=0.60$ (bottom) of KBHsPH solutions. The value of the horizon mass decreases from left to right.}
\end{figure}

Clear non-Kerr features emerge in the sequences with smaller frequency $\omega$,  for instance in the subsets $7.v$ and $8.v$ (cf. Fig.~\ref{fig7b}), as well as $9.v$ and $10.v$ (cf. Fig.~\ref{fig8}), where $v$ $\in \left\{ 1,2,3\right\}$. An example of such non Kerr-like features are peculiar ``egg-like" shadow shapes, $e.g.$ $7.3$ and $8.2$ - respectively top-left and bottom-middle panel of Fig.~\ref{fig7b}. Another notable image characteristic is for example the existence of a ``cuspy" shadow edge~\footnote{Another study of BHs with cuspy shadows was presented in \cite{Qian:2021qow}.}, $e.g.$  $8.3$ and $10.3$  - bottom left panel of respectively Fig.~\ref{fig7b} and Fig.~\ref{fig8}.

Some of these image features can be connected to a complex FPO structure.
For instance, solution 8.3 has two additional light ring orbits with respect to Kerr, one of them being stable. This is consistent with the PS limit solution 8.0, that has two light rings (one stable and another unstable).
Solution 8.3 contrasts with some of the other KBHsPH solutions already discussed, $e.g.$ the sequence $6.v$, which (like Kerr) only possessed two light rings, both of them unstable. 

In the next section we shall attempt to explain some of these features by an in-depth analysis of the FPO structure of these solutions.

\section{Fundamental photon orbits}

\subsection{Overview}

In the Kerr spacetime, the edge of the BH shadow is closely related to the notion of SPOs~\cite{Teo2003,Bardeen:1973tla}. More specifically, points along the shadow edge define the necessary initial condition for null geodesics to asymptotically approach SPOs, which are unstable bound orbits of constant (Boyer-Lindquist) radial coordinate. Light rings in particular are a subset of these SPOs - the planar ones - and define the two shadow points on the equator, in the case of observations at $\theta=\pi/2$. 

In more general spacetimes, such as the ones discussed in this paper, the concept of SPOs can be generalized. However, the condition $r = const.$ is not geometrically invariant, since it is not preserved by coordinate mixing of $r$ and $\theta$. In addition, typically there is no key property that singles out a particular coordinate chart, such as the geodesic separability that exist in the Kerr case for Boyer-Lindquist coordinates~\cite{Cunha2017}. 

Nevertheless, the concept of FPOs has been introduced in \cite{Cunha2017}, which generalizes SPOs and include the latter as a subset. Although its definition is more general, in the specific metric ansazt~\eqref{Eqansatz} a FPO is simply a null geodesic that describes a periodic trajectory when projected in $\l r, \theta  \r$ space. 

The dynamics of the null geodesic flow can be derived from the Hamiltonian $\mathcal{H}=\frac{1}{2}g^{\mu \nu} p_{\mu}p_{\nu}=0$, where $p_{\mu}$ is the photon's 4-momentum. In terms of the first integrals $p_t \equiv - E$ and $\Phi \equiv p_{\phi}$, we can define an effective potential $V \l r, \theta \r$ and a kinetic term $T \geq 0 $ such that $2 \mathcal{H} =T+V$. In particular we have 
\begin{equation}
	T= g^{rr} p_r^2 + g^{\theta \theta}p_{\theta}^2\geq 0 \, ,
\end{equation}

\noindent and 
\begin{equation}
	V=g^{tt}E^2 - 2 g^{t \phi}E \Phi + g^{\phi \phi} \Phi^2  \leq 0 \, .  
\end{equation}

Given that the kinetic term $T$ is never negative, from the equation above we conclude the boundary for the allowed region in $\l r, \theta \r$-space is given by the condition $V=0$. Each FPO is confined to its own allowed region, which depends on the value selected for the impact parameter $\eta \equiv \Phi/E$. 

We will follow the classification of FPOs proposed in \cite{Cunha2017}, as follows: FPOs are identified by a symbol $X^{n_r \pm}_{n_s}$, where $X=\{O,C\}$, and $n_r$, $n_s$ $\in \mathbb{N}_0$, such that:
\begin{enumerate}
	\item $X=O$ ($C$) if the orbit is open (closed);
	\item A plus (minus) sign is used if the orbit  is even (odd) under the $\mathbb{Z}_2$ reflection symmetry around the equatorial plane;
	\item $n_r$ is the number of times that the orbit crosses the equatorial plane;
	\item $n_s$ is the number o self-intersection points.
\end{enumerate}
In addition, FPOs can be classified according to their stability under small trajectory perturbations. In Kerr spacetime all FPOs outside the horizon are unstable but, as discussed in \cite{Cunha2017}, this not always true for other spacetimes.

\begin{figure}[H]
	\centering
	\begin{subfigure}[b]{0.32\textwidth}
		\centering
		\includegraphics[width=\textwidth]{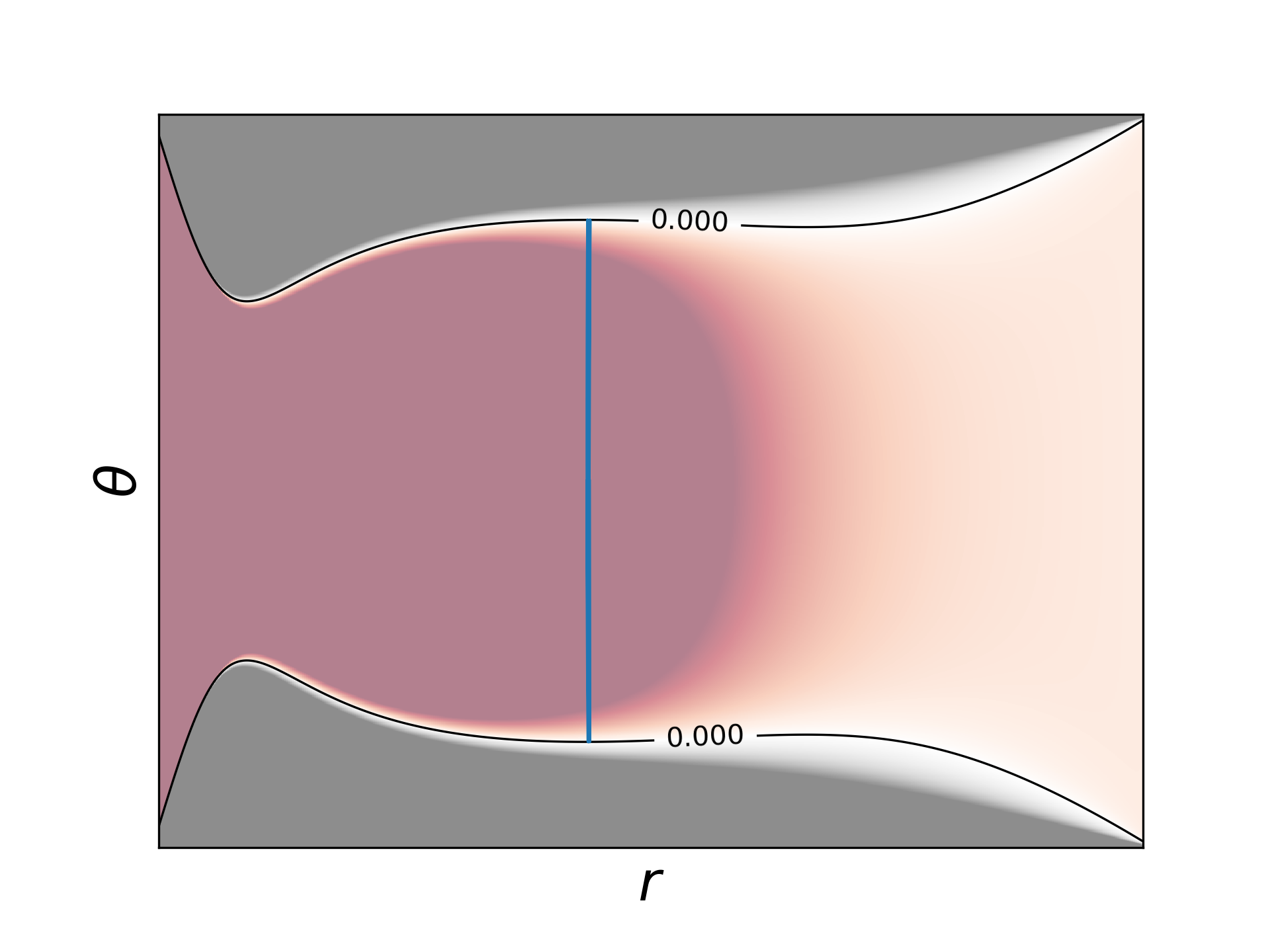}
		\includegraphics[width=\textwidth]{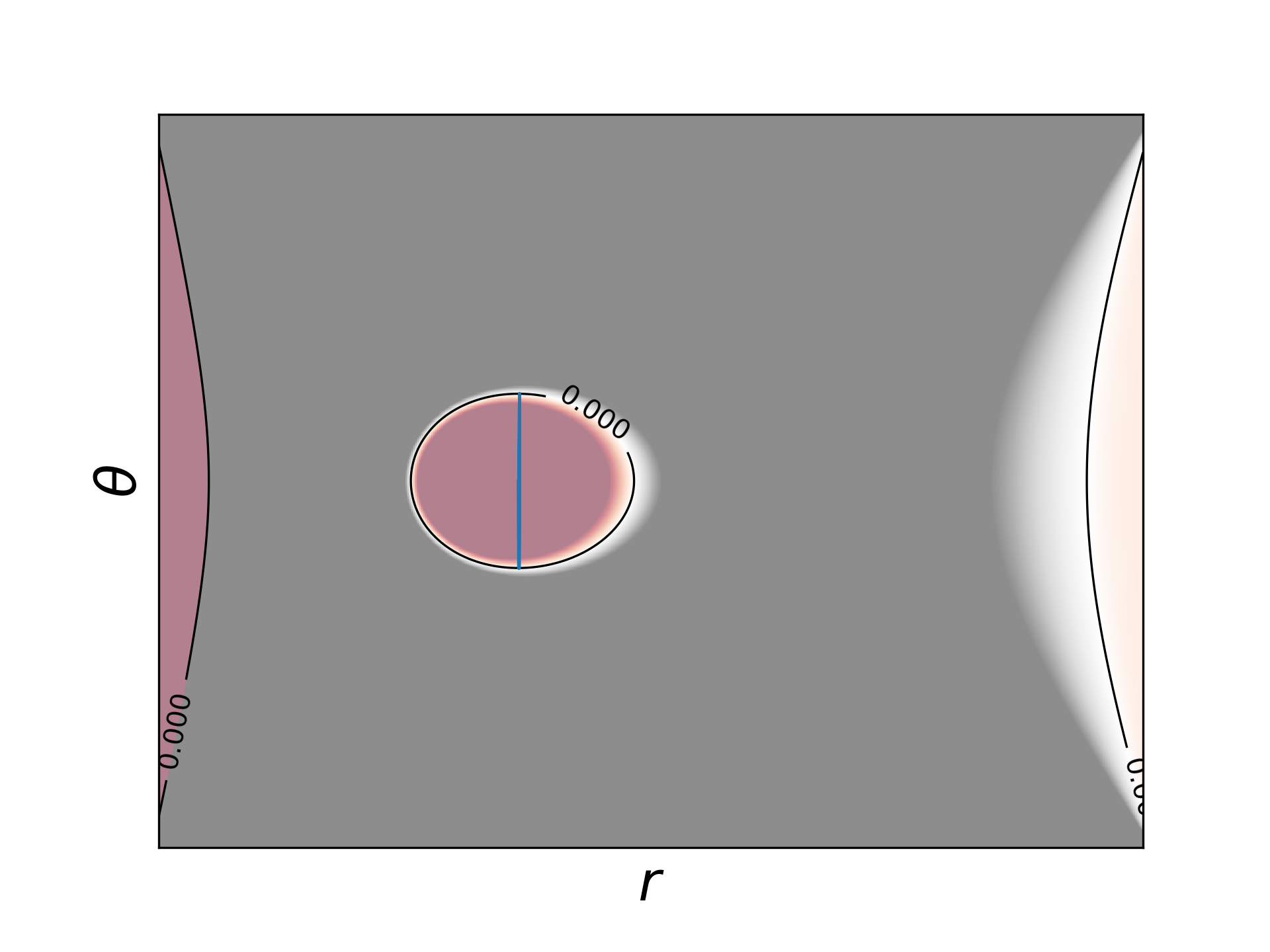}
		\caption{Class $\mathcal{O}^{1+}_{0}$}
		\label{type1}
	\end{subfigure}
	\hfill
	\begin{subfigure}[b]{0.32\textwidth}
		\centering
		\includegraphics[width=\textwidth]{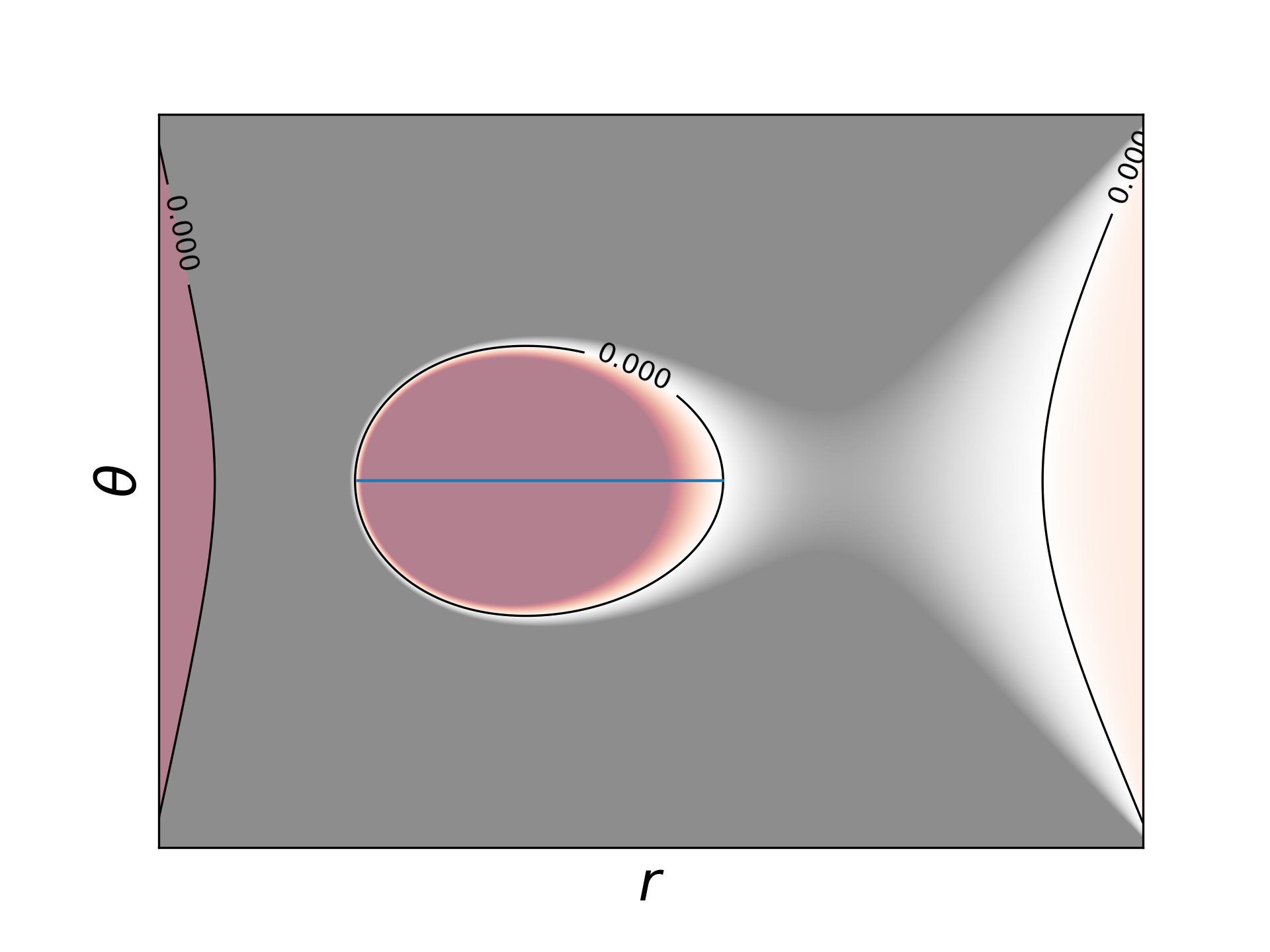}
		\includegraphics[width=\textwidth]{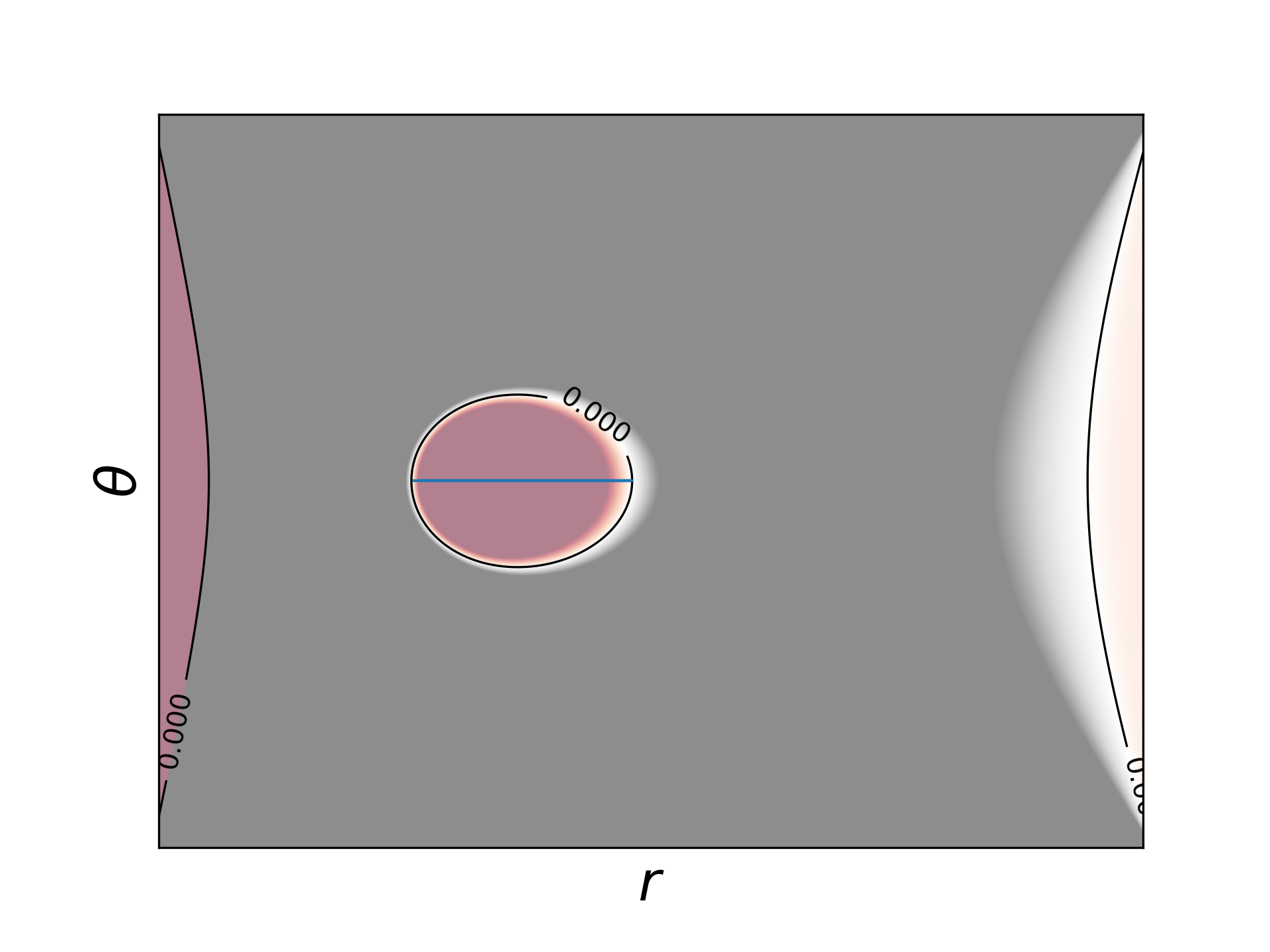}
		\caption{Class $\mathcal{O}^{0+}_{0}$}
		\label{type2}
	\end{subfigure}
	\hfill
	\begin{subfigure}[b]{0.32\textwidth}
		\centering
		\includegraphics[width=\textwidth]{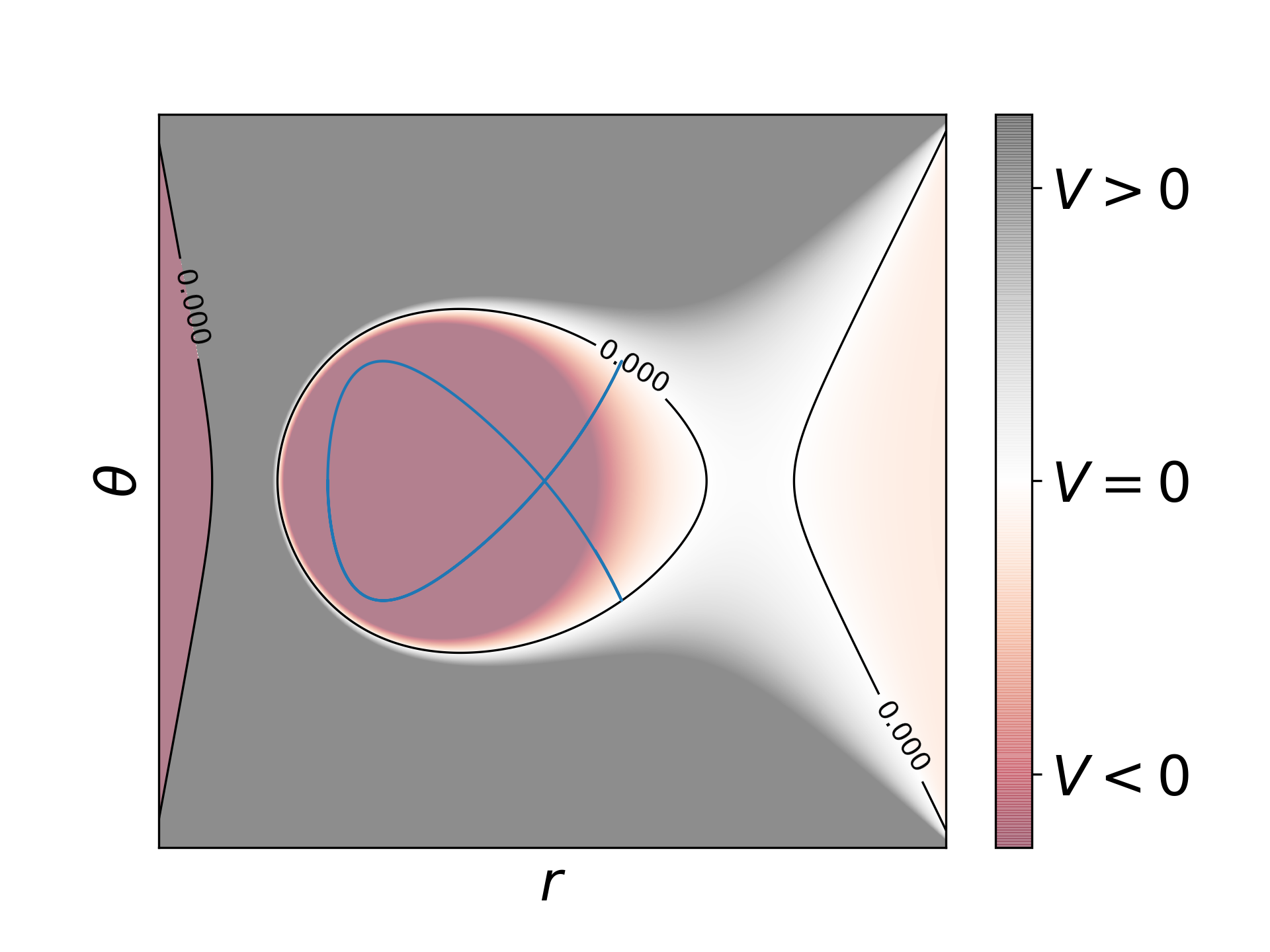}
		\includegraphics[width=\textwidth]{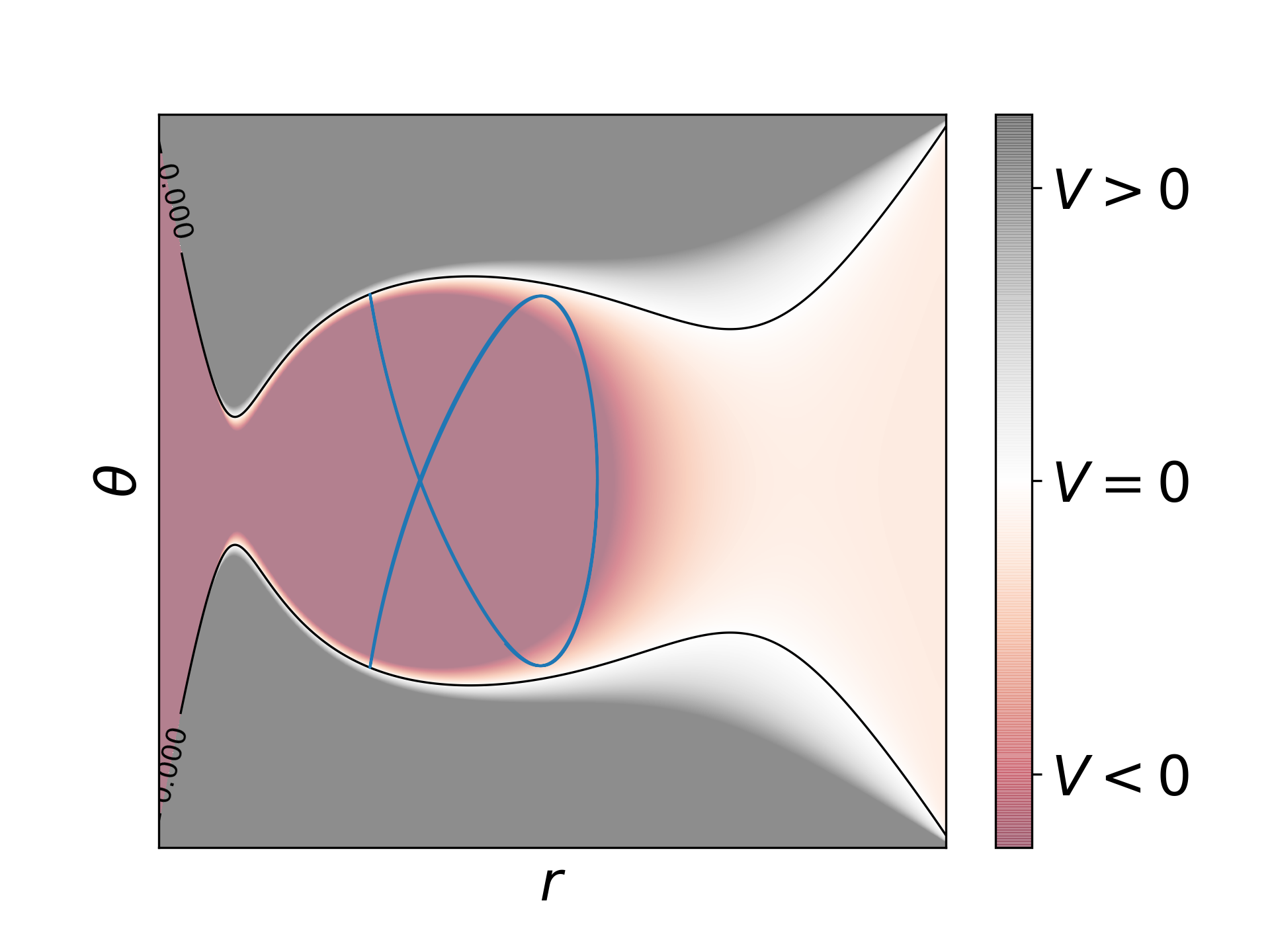}
		\caption{Class  $\mathcal{O}^{2+}_{1}$}
		\label{type3}
	\end{subfigure}
	\caption{\small \label{fig_fpos} Representation, in the $\l r, \theta \r$ plane, of the FPOs classes featured in the paper. The heat map represents to the values taken by the effective potential $V\l r, \theta \r$.}
\end{figure}

Although the SPOs of the Kerr spacetime allow for very intricate motion around the BH (as thoroughly discussed in \cite{Teo2003}), all of them can be classified  as either $\mathcal{O}^{0+}_{0}$ (for light rings) or $\mathcal{O}^{1+}_{0}$. As we will see in the next section, this scenario changes drastically when we consider KBHsPH. In Fig. \ref{fig_fpos} we provide an example of all the classes of FPOs that we discuss in this paper (with the exception of light rings). The existence of some of these \textit{exotic} FPOs have aleady been reported. For instance, FPOs of type $\mathcal{O}^{0+}_{0}$ that are not light rings have been found in Kerr BHs with scalar hair \cite{Chaotic}, whilst some FPOs with more than one self-intersection where known to be possible for some PS models~\cite{Cunha2017}.

In the following sections we address, in a more systematic way, how these different FPOs classes emerge within the space of solutions, while also arguing for a close connection with the features seen in the lensing images discussed previously.

\subsection{Analyzing lensing images using FPOs}

Kerr BHs possess two light rings with opposite rotation senses, both of them unstable: one for a negative impact parameter, $\eta^{LR}_{-}$, and another for a positive one, $\eta^{LR}_{+}$. A continuum family of SPOs exists between the two light rings, with $\eta^{LR}_{-} \leqslant \eta \leqslant \eta^{LR}_{+}$, all of them related to one or more points of the shadow edge (see \cite{Cunha2017}).

\begin{figure}[H] 
	\centering
	\includegraphics[width=0.5\textwidth]{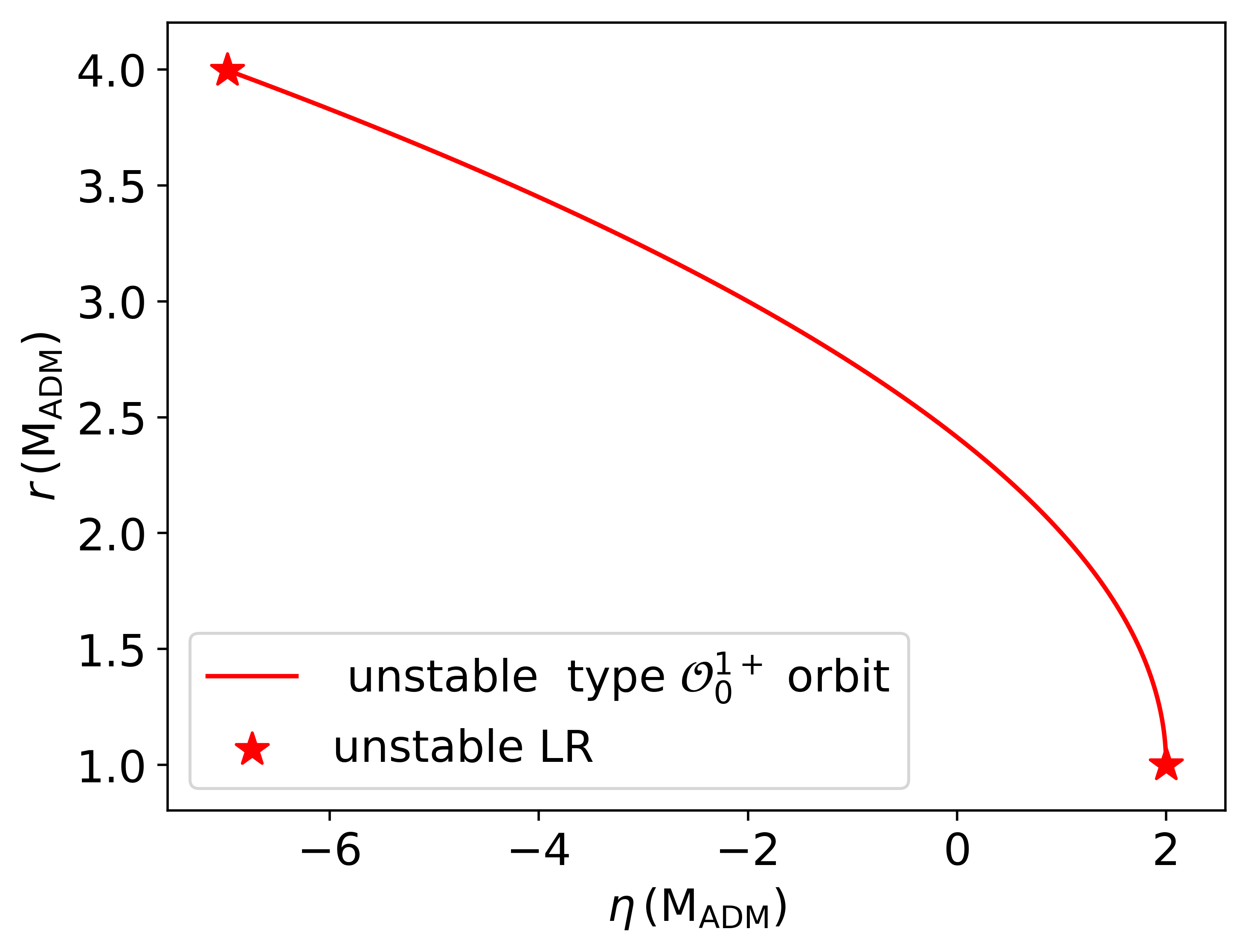}
	\caption{\small \label{FPO1} Diagram of the FPO family of an extremal Kerr BH.}
\end{figure}

We represent in Fig.~\ref{FPO1} the SPO family for an extremal Kerr BH in terms of the impact parameter $\eta$ and the corresponding Boyer-Lindquist radius coordinate $r$. 
Critically, each individual SPO (which is also an FPO) is uniquely labelled by its radial coordinate at the crossing point with the equatorial plane.

A similar method can be used to label FPOs around KBHsPH, provided that the FPOs intersect the equator. Such a labelling criteria is necessary because FPOs around KBHsPH will no longer be described (generically) by a single radial coordinate along its trajectory. The radial coordinate of the FPO at the equatorial plane crossing point is a geometrically well-defined choice, since that point is invariant under the $\mathbb{Z}_2$ reflection symmetry.

It is not unexpected that there exists a subset of hairy BH solutions whose FPO family structure does not differ very much from the Kerr case. For the specific examples we discussed above, in terms of the lensing, this is the case for solutions with frequencies down to $\omega^{7.v}=0.75$. Below this frequency we find lensing images with distinctively non-Kerr shadow features that can be connected to distinctive FPO structures, as we now discuss.

\subsubsection*{Cuspy shadow}

As we progress to smaller frequencies, solution 8.3 (with $\omega^{8.3}=0.70$) is one of the first whose shadow displays some striking deviations from Kerr  - Fig. \ref{fig7b} (bottom left panel). In \cite{Cunha2017}, the authors also analysed a hairy BH solution with a cusp on the shadow edge, and then related this feature to the existence of two (discontinuous) different FPO branches related to the shadow edge. Although the solution studied in \cite{Cunha2017} corresponds to a different family of hairy BHs solutions\footnote{The solution analyzed in \cite{Cunha2017} is an example of an excited KBHPH, rather than fundamental as the ones discussed here.}, we verified that the same interpretation of the shadow cusp also applies here, as we shall now discuss, reinforcing the idea that FPOs are a useful non-trivial tool when it comes to understanding non-Kerr features present in shadows and lensing images.

\begin{figure}[H] 
	\centering
	\includegraphics[width=0.9\textwidth]{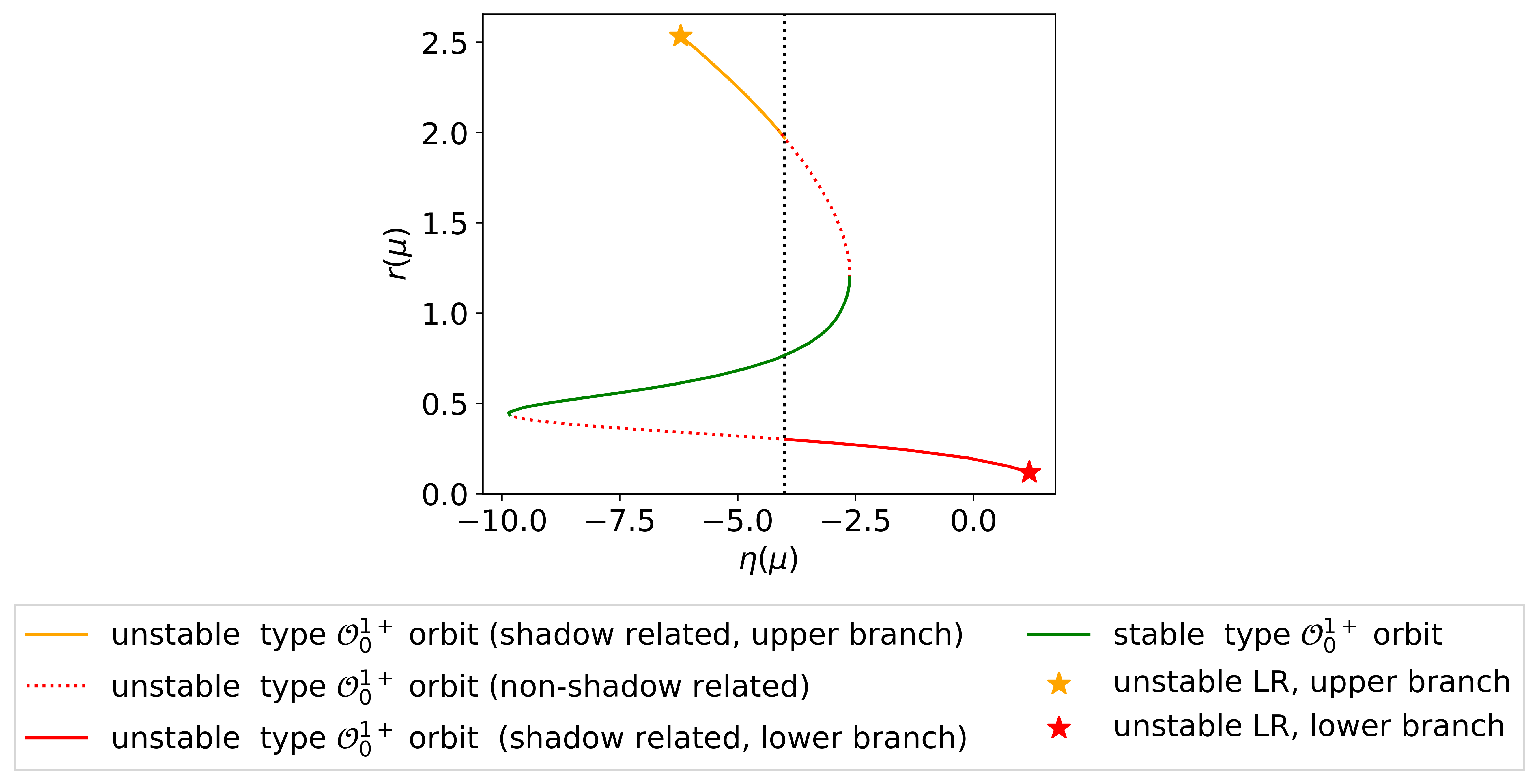}
	\caption{\small \label{FPO2} FPO family diagram for solution 8.3. The black dotted line marks the separation between the two FPO branches related to the shadow edge.}
\end{figure}

\begin{figure}[H] 
	\centering
	\includegraphics[width=0.6\textwidth]{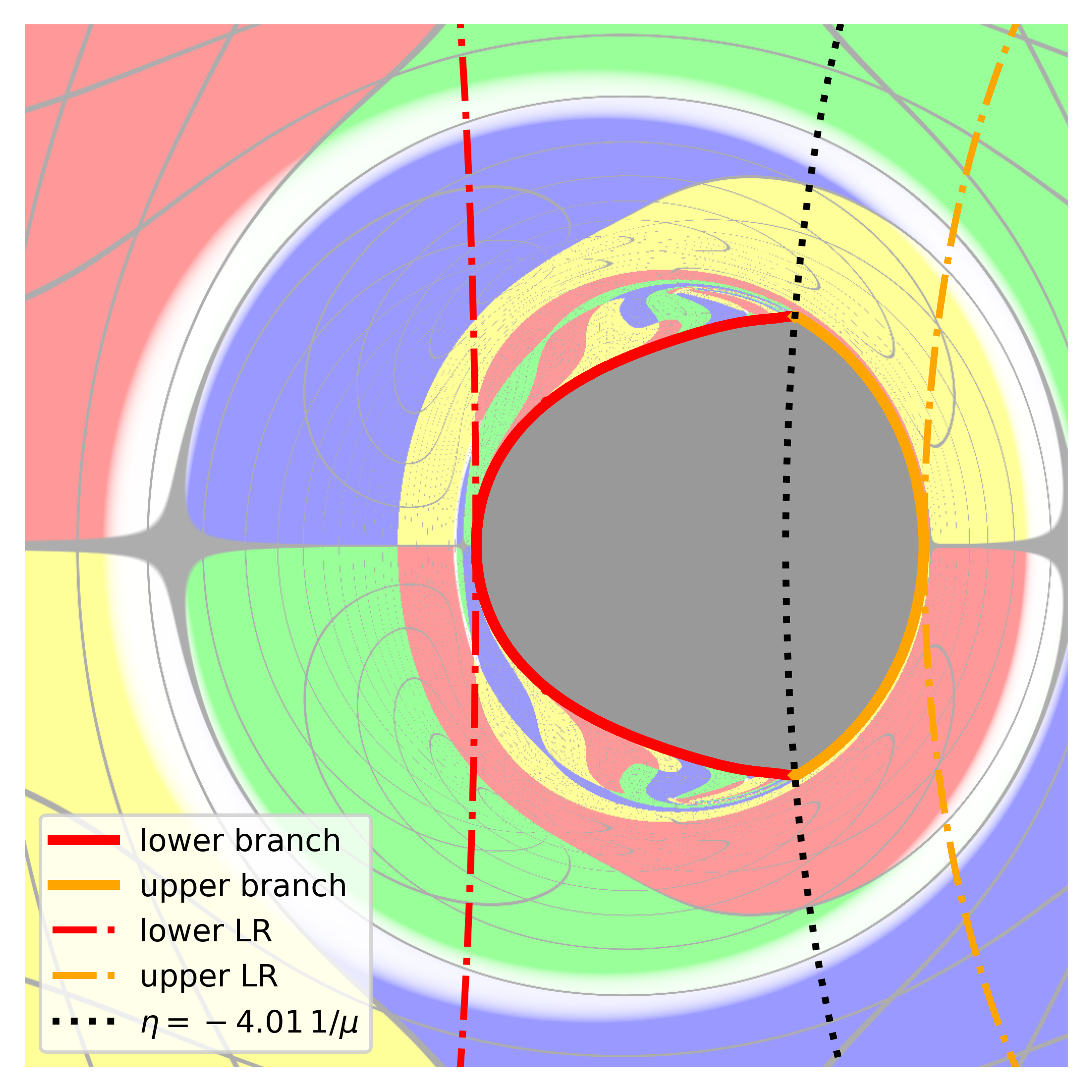}
	\caption{\small Different sections of the shadow contour of solution 8.3, connected to distinct FPO branches, are represented by solid lines. In addition, points in the image with constant impact parameter $\eta$ are displayed by either dotted or dotted-dashed lines. The impact parameter $\eta=-4.01/\mu$ marks the separation between the two FPO branches, and corresponds to the vertical dotted line shown in Figure~\ref{FPO2}. Dotted-dashed lines with the impact parameter of both light rings (LRs) are displayed for reference, intersecting the shadow edge at the points determined by the light ring orbits.\label{FPO3}}
\end{figure}

In Fig.~\ref{FPO2} we present the FPO family diagram for solution 8.3, where it is clear that there are two distinct and discontinuous branches of unstable FPOs that are shadow related - the red and yellow solid lines. The red line is dubbed the \textit{lower} branch, occurring for larger (albeit some still negative) impact parameters, whereas the yellow line is dubbed the \textit{upper} branch, occurring for smaller (and all negative) impact parameters. 
These two branches actually connect to each other via some other FPOs that are not shadow related, some being stable and others unstable. Each branch of the shadow related FPOs terminates at a light ring, on one end, and at a particular FPO, on the other end. The impact parameter of the latter is the same for both branches, which allows the shadow edge to be continuous, but not smooth (as a curve). This explains the cusp seen in the shadow edge of solution 8.3, 
which is displayed in Fig.~\ref{FPO3} together with an illustration of how each shadow point is mapped to a corresponding FPO branch.

Curiously, solution $8.3$ is also fairly close in the domain of existence of KBHsPH to solution 8.2, which has two additional light rings (one stable and the other unstable) with respect to 8.3 - see  Fig.~\ref{fig1}. This realization, together with an examination on how extra light rings might emerge by deforming the FPO family diagram (see also the discussion of the next section), leads to the conjecture that the presence of a cuspy shadow could signal the near-emergence of a second pair of light rings in the space of solutions.  It would be interesting to further explore this idea on general grounds.

\subsubsection*{Ghost shadows}

\begin{figure}[H] 
	\centering
	\includegraphics[width=1\textwidth]{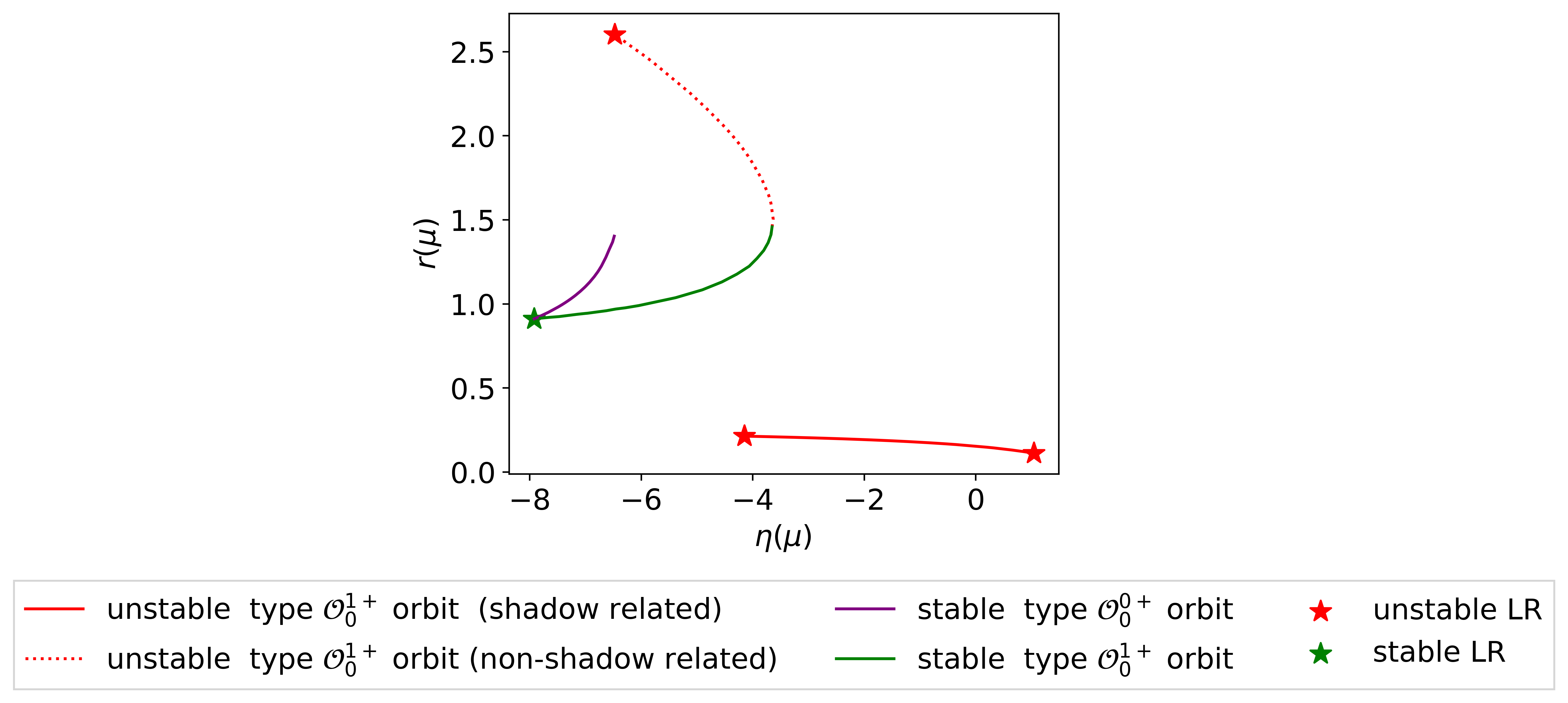}
	\caption{\small \label{FPO4}  FPO branch diagram for solution $8.2$. Since orbits of type $\mathcal{O}^{0+}_{0}$ do not exit the equatorial plane,  the radius used to represent these orbits corresponds to the radial mean value.}
\end{figure}

In Fig.~\ref{FPO4} we present the FPO family diagram for solution $8.2$, which contains four light rings in total. The lensing image of this solution can be found in Fig. \ref{fig7b} (bottom middle panel). With the emergence of the second light ring pair, we now have two fully disconnected branches: the shadow related branch (the lower branch, which completely defines the shadow edge, therefore being smooth with no cusp) and a non-shadow related branch (the upper one).

This upper branch can be divided into three different sub-branches: a stable\footnote{Note that stable branches never contribute to the shadow edge.} and unstable $\mathcal{O}^{1+}_{0}$ type orbits and a stable  $\mathcal{O}^{0+}_{0}$ type one.

Orbits of type $\mathcal{O}^{0+}_{0}$ discussed here are restricted to the equatorial plane and, thus, there are several available options when it comes to labelling these orbits with a single coordinate $r$ value.  In this paper, the radial value used  to represent type $\mathcal{O}^{0+}_{0}$ orbits corresponds to the mean radial value, defined as the average of the maximum and minimum $r$ values reached along the geodesic motion. By labelling these FPOs in such a manner, the $\mathcal{O}^{0+}_{0}$ branch upper end point in Fig.~\ref{FPO4}, $i.e.$ with largest radius, coincides precisely with the radius at which the the stable $\mathcal{O}^{1+}_{0}$ branch connects to the unstable one.

In \cite{Cunha2017} the authors discussed how unstable FPOs non-related to the shadow could be associated to a set of lensing patterns attached to the shadow edge, dubbed "eyelashes". A similar feature can be seen in the lensing of $8.3$ (cf. Fig. \ref{fig7b}, lower left panel). After the emergence of a second pair of light rings in solution 8.2, the eyelashes appear to become disconnected from the shadow, forming a pixelated banana-shaped strip that can be seen in the lensing image - see bottom middle panel of Fig. \ref{fig7b}. This feature has been dubbed ``ghost shadow'' in the literature~\cite{Cunha_2018}, and it seems plausible that it is a consequence of the existence of the unstable FPO branch that is non-related to the shadow. This question merits a rigorous formulation and analysis which is, however, outside the scope of this paper.

\begin{figure}[H] 
	\centering
	\includegraphics[width=1\textwidth]{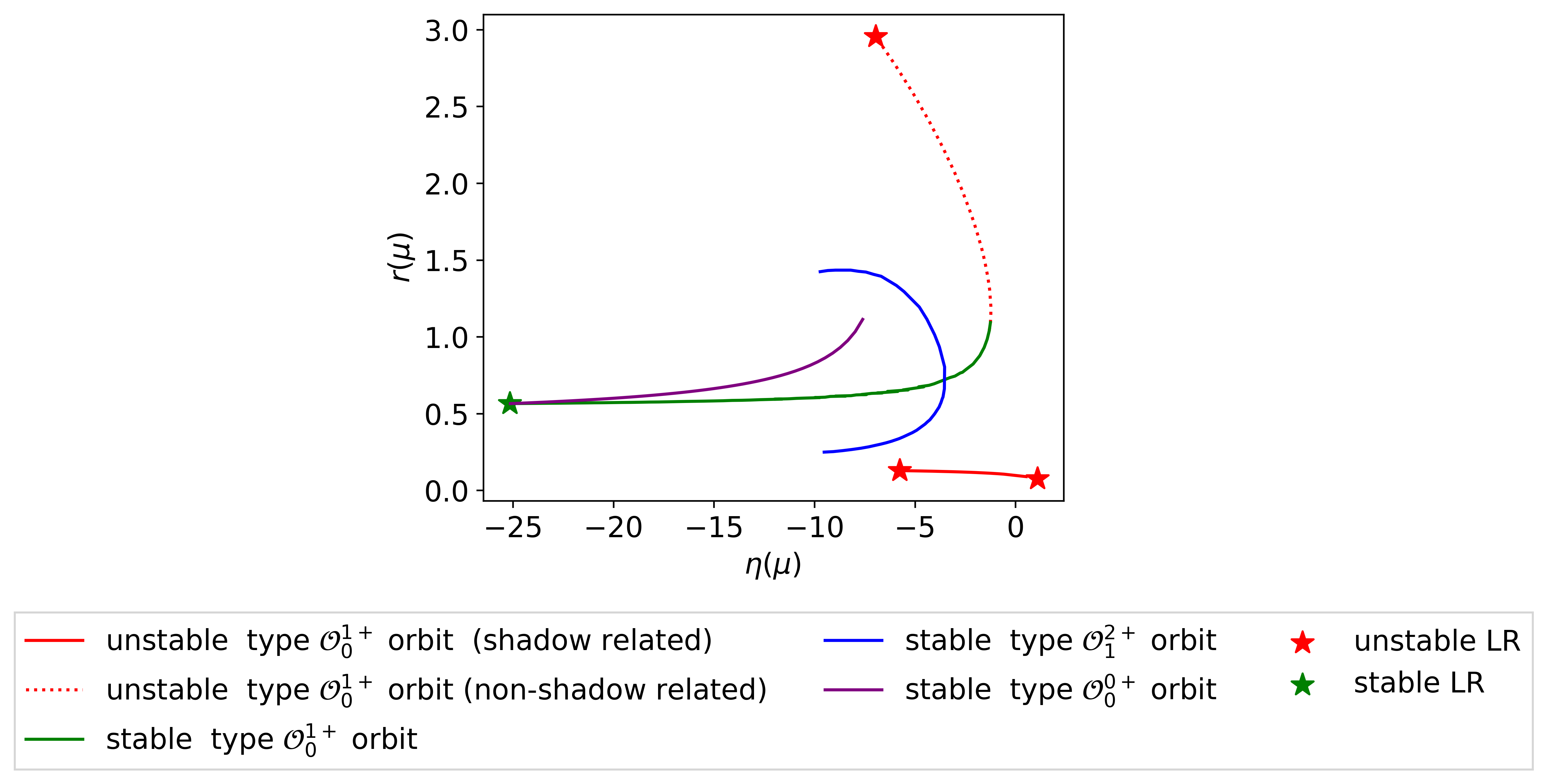}
	\caption{\label{FPO9}  FPO branch diagram for solution $9.3$.}
\end{figure}

Shifting our attention to solution $9.3$ (Fig. \ref{fig8}, top left panel), FPOs of type $\mathcal{O}^{2+}_{1}$ can also be found (cf. Fig. \ref{FPO9}). These orbits were already reported to exist in PSs \cite{Cunha2017}, but it was still unknown whether they could be present in KBHsPH, and how they were related to the other FPOs branches. In Fig. \ref{FPO9} we see that $\mathcal{O}^{2+}_{1}$ FPOs (blue line)  bifurcate from the already familiar $\mathcal{O}^{1+}_{0}$ stable branch (the green line). We verified that the $\mathcal{O}^{1+}_{0}$ branch separates the new $\mathcal{O}^{2+}_{1}$ into two sub-classes: the two different orientations shown in Figure \ref{type2}. Since these are stable FPOs, it is not possible to establish a direct link between them and the shadow image. Nonetheless, it might be interesting to study these FPOs on a mathematical level: in particular, it is not evident what is special about the bifurcation point at which these FPOs emerge in the family diagram. It remains an open problem whether the emergence of these different FPOs can be associated to conserved topological quantities, in a similar spirit to the topological charge associated to the existence of light rings~\cite{Cunha_2017_3,Cunha_2020}, 

Additional FPOs classes, other than the ones discussed in this paper, might still be present in the solutions that we discuss here. The numerical analysis implemented in this work focused on FPOs that crossed the equatorial plane perpendicularly at least once. If other FPOs that do not verify this condition were to be present, then they would have been overlooked by the analysis herein.

\section{Analyzing the astrophysically viable space of solutions}

In the previous sections we have been addressing theoretical aspects of the lensing and shadows of KBHsPH and their solitonic limit, PSs. We have focused, in particular, on distinctively non-Kerr features, to clearly exhibit the exoticness that can occur in this model. We would now like to consider the potential observability of this model via current or near future observations. As such, hereafter we shall focus on the solutions that are astrophysically viable, in the sense that a plausible formation mechanism exists. As discussed in \cite{Cunha_2019}, these are solutions that one can expect to be formed from a Kerr BH via superradiance (see also~\cite{Herdeiro_2022} for hairiness limit of such solutions), and whose superradiant instabilities are only relevant on very long, possibly cosmological, time scales~\cite{Degollado_2018}. 

\begin{figure}[H]
	\centering
	\includegraphics[width=0.65\textwidth]{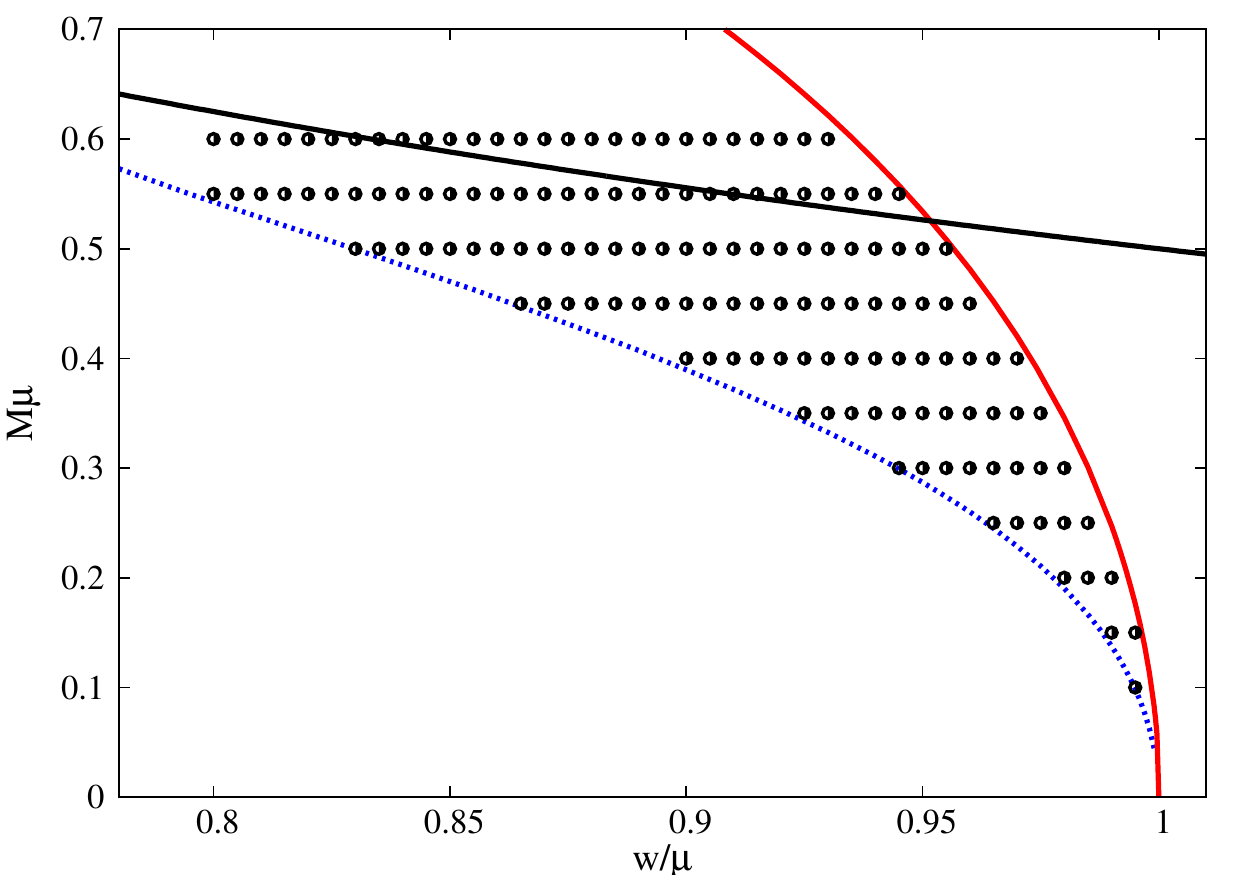}
	\caption{\label{domain}  Domain of existence of the astrophysically viable solutions. This is a sub-region of Fig.\ref{fig1}. As in that figure, the red solid line is the PS limit and the blue dotted line is the Kerr limit. The black solid line is the extremal Kerr line. The black dots highlight the solutions used in our analysis of this and the following sections.}	
\end{figure}

In Fig. \ref{domain} we display the part of the domain of existence containing the solutions that we have considered for the observability study. It comprises solutions for which $0.1 \leq M \mu \leq 0.6$. Only a subset of these solutions might actually be formed through superradiance, depending on their fraction of hair $p$. The latter is defined as the ratio between the (Komar) mass of the BH horizon $M_{BH}$ to the total mass of the spacetime $M$: 
\begin{equation}
p\equiv 1-\frac{M_{BH}}{M} \ .
\end{equation} 
Fully dynamical numerical evolutions of the superradiant instability around a Kerr BH, driven by a complex vector field, were shown to form KBHsPH with $p\lesssim 0.1$ (see~\cite{East:2017ovw,Herdeiro_2017,Herdeiro2016} for details). As discussed in \cite{Herdeiro_2022}, the bound $p\lesssim 0.1$ can be expected to hold in general regardless of the spin of the bosonic field, as long as the development of the superradiant instability from Kerr is assumed to be approximately conservative. Numerical evolutions have indeed suggested the latter to be the case, at least in the cases studied thus far~\cite{East:2017ovw,Herdeiro_2022}. Nevertheless, even if such an assumption were not to hold, there is an overall thermodynamic bound on the maximum rotational energy that can be extracted from Kerr, setting a conservative upper limit of $p\lesssim 0.29$.

For a wider analysis, the region represented in Fig. \ref{domain} also includes solutions that do not verify the condition $p< 0.1$ (or even $p< 0.29$). Such solutions are also valid equilibrium solutions of the Einstein-Proca system of equations, and one cannot dismiss the possibility of formation by alternative channels other than the growth of the superradiant instability - see $e.g.$~\cite{Sanchis-Gual:2020mzb}. Moreover, the inclusion of solutions with $p>0.1$ allow us to have a better interpolation sample for the shadow size throughout the solution domain.

In order to make contact with the observational data we shall work with a simple measure of the BH shadow size: the areal radius $S$, defined as

\begin{align} 
S \equiv \sqrt{\frac{\mathcal{A}}{\pi}} \, ,
\end{align}

\noindent where $\mathcal{A}$ is the area of the shadow, which in general depends on the distance and inclination of the observer with respect to the BH. The areal radius can then be related to the angular size $\vartheta$ of the BH shadow, which is measured by the EHT observations. For a BH with mass $M$, located at a perimetral distance $\mathcal{R}$ from the observer, the relation between $\vartheta$ and $S$ is provided via the expression

\begin{align} \label{eq_angular}
\vartheta = \l S/M\r \frac{M}{\mathcal{R}} \, . 
\end{align}

Both EHT observations of M87* and Sgr A* are, so far, consistent with the Kerr metric within the current uncertainty level of $\simeq 10\%$ of the shadow angular size~\cite{EHT1, EHT2022_1}. However, having such a (fairly) large uncertainty in the shadow size measurement leaves the door open to alternative Kerr models that also might be consistent with observations, namely KBHsPH. This possibility raises the interesting question of how much Proca hair might exist outside the horizon, provided that its shadow is still consistent with the EHT observations and their uncertainties. In the remainder of the paper we shall focus on this issue by analysing how much the areal radius of KBHsPH shadows deviate from Kerr as function of the amount of hair, $p$, and the parameter $M\mu$. For this deviation analysis we shall compare the shadow of each hairy BH solution with the one from a comparable Kerr BH of equal mass $M\mu$, lying along the Kerr existence line in the KBHsPH solution space. Thus, it is convenient to express the Kerr dimensionless spin parameter, $a$, as function of $M \mu$.

Along the Kerr existence line, the spin parameter $a$ is related to the angular velocity at the horizon, $\Omega_H$ via:

\begin{align} \label{eq_a}
a=	\frac{M^2 \Omega_H}{M^2 \Omega_H^2 +1/4} \, . 
\end{align} 
Then, the data in Fig. \ref{domain} allow us to write an approximation for the function $M \Omega_H (M \mu)$  :

\begin{align} \label{eq_MO}
M \Omega_H \approx b_1 +b_2 M \mu +b_3 M^2 \mu^2 \, , 
\end{align} 

\noindent with $\l b_1, b_2, b_3 \r = \l -0.00938981,1.16616021,-0.63782197\r$.  By combining equations \eqref{eq_a} and \eqref{eq_MO} one can directly describe the spin of the comparable Kerr BH solutions mentioned above.

\section{Constraining the hair using M87* data}

\subsection{Approximation formula for the shadow viewed from $17^o$ inclination}

Although it is possible to obtain numerically the shadow size of KBHsPH up to a fairly high precision, it will become useful in the following sections to work with an analytical approximation instead. Following~\cite{Cunha_2019}, we start by expressing the Kerr areal radius as a function of the spin and of the polar angle of the observer, $\theta_o$. This can be achieved by interpolating between the two cases where the shadow area can be computed exactly: when the observer is on the rotation axis ($i.e.$, $\theta_o =\{0, \pi \}$), and for the case of an extremal Kerr BH viewed from the equatorial plane in the far-away limit. The resulting expression can be written as~\cite{Cunha_2019}:
\begin{align} \label{kerr_approx}
S_{\rm Kerr} \l a, \theta_o \r \approx S_{\rm Kerr} \l a, axis \r +\frac{2 \left| a\right| \theta_o}{\pi M}  \left[S_{\rm Kerr} \l M, \frac{\pi}{2} \r -S_{\rm Kerr} \l M, axis \r  \right] \, ,
\end{align}

\noindent which gives an error $\lesssim 0.8 \%$.

Since KBHsPH shadows are obtained numerically with the observer at a finite (perimetral) distance from the BH, their size cannot be immediately compared with the Kerr analytical approximation \eqref{kerr_approx}, which is constructed for observers at infinity. Following~\cite{Cunha_2019}, this issue is addressed by i) considering how the numerical shadow size $S$ changes among two observers, respectively $\{S_1,S_2\}$, at large (finite) perimetral distances $\{ \mathcal{R}_1, \mathcal{R}_2 \} \gg M$, and then ii) by extrapolating the value of $S$ to the limit of an infinitely far-away observer. This procedure leads to:

\begin{align}
S_{\infty} \approx S_2 -  \l \frac{S_2-S_1}{1- \mathcal{R}_1 / \mathcal{R}_2  }\r  \, .
\end{align}

\noindent In our analysis we considered the values $\mathcal{R}_1=100M$ and $\mathcal{R}_2=200M$.

In order to describe the KBHsPH areal radius at infinity as viewed from an observer at $\theta_o=17^o$ (which corresponds to the angle between M87*  BH spin and the line of sight \cite{CraigWalker:2018vam}), we propose the following approximation formula:

\begin{align} \label{Shairy_1}
S_{\rm hairy} \l p, M\mu, 17^o \r /M =\l 1- p \r \left[ S_{\rm Kerr} \l a \l M \mu \r, 17^o \r/M + \sum_{i=0}^{2}\sum_{j=0}^{4} \alpha_{ij} (M \mu)^j p^{i+1} \right]\, ,
\end{align}

The parameters $\alpha_{ij}$ used for the fit can be found in the matrix Table \ref{Table_approx1} below.

\begin{table}[H]
	\centering
	\begin{tabular}{l}
	\multicolumn{1}{c}{}  \\ 
$\begin{pmatrix}
\alpha_{00} & \alpha_{01} & \cdots\\
\alpha_{10} & \ddots &  \\
\vdots & & \ddots \\
\end{pmatrix} \equiv \begin{pmatrix}
-1.266 & 18.821 & -90.696 & 267.305 & -236.951\\ 
4.741 & -68.491 & 388.415 & -1066.683  & 1063.879\\ 
5.535 & -66.439 & 226.226 & -152.9 & -225.308\\
\end{pmatrix}$
\end{tabular}
	\caption{\label{Table_approx1} Values for the parameters $\alpha_{ij}$ for an observer at $\theta=17^{\circ}$.}	
\end{table}

The approximation \eqref{Shairy_1} has an average error of $1.1 \%$ when all the solution points represented in Fig. \ref{domain} are considered. However, solutions of particular interest will be the ones that might grow from Kerr BHs via superradiance, which form a solution subset with restricted values of $p$. By restricting ourselves to solutions satisfying the more conservative thermodynamic upper limit of $p \lesssim 0.29$, then eq. \eqref{Shairy_1} has an error smaller than $0.8\%$ overall, and around $0.14\%$ in average.

\begin{figure}[H]
	\centering
	\includegraphics[width=0.7\textwidth]{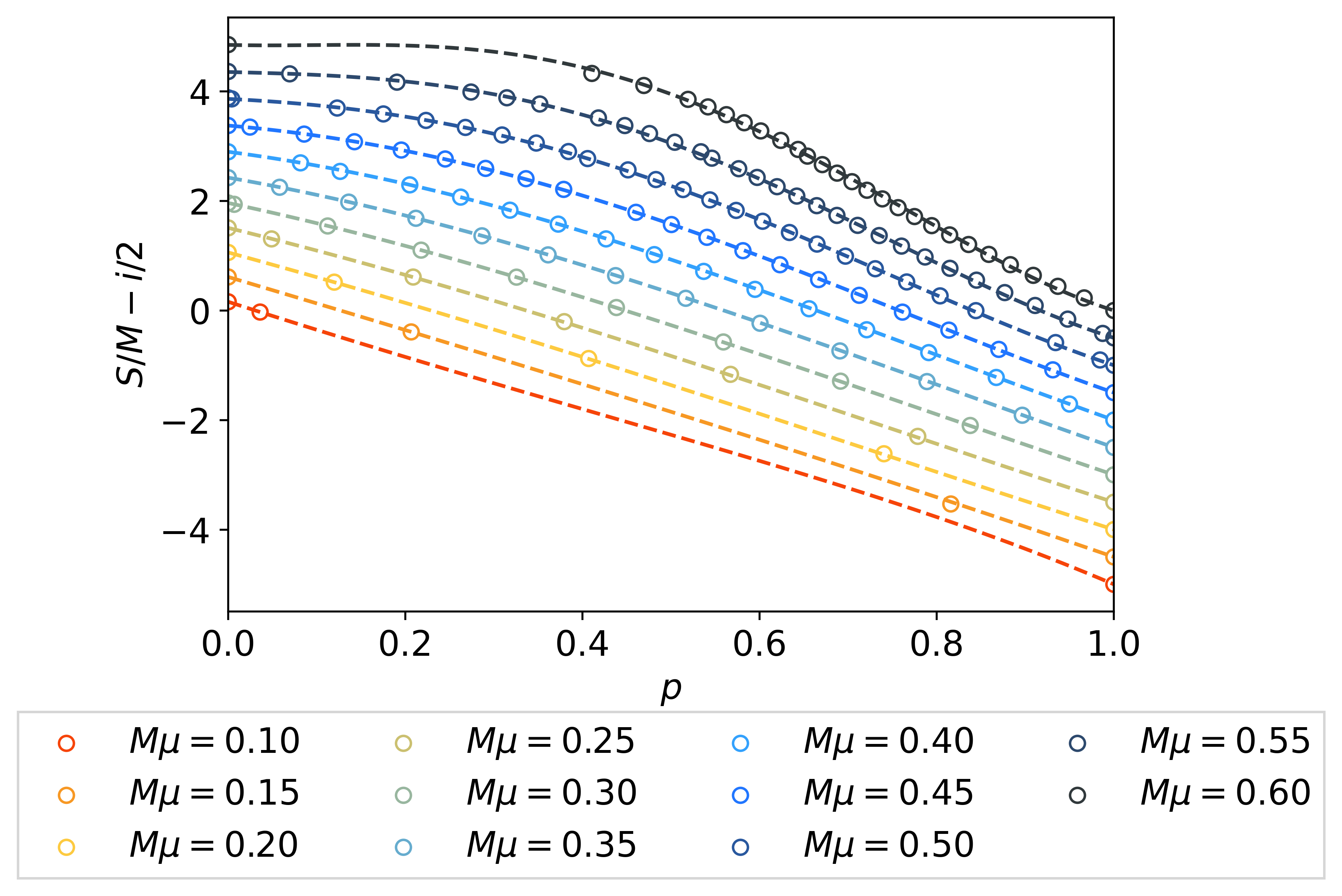}
	\caption{ \label{fig_S_1}\small  Areal radius for hairy BHs (dots) and the analytical approximation (dashed lines) \ref{Shairy_1}. Each line correspond to a constant $M\mu$ value and, to better distinguish them, they are translated by an amount $-i/2$, where $i= \{0,..,10\}$. }	
\end{figure}

The accuracy of this approximation can also be inferred  from Fig. \ref{fig_S_1}, where we show the areal radius for hairy BHs computed via ray-tracing and the function $S_{\rm hairy} \l p, M\mu, 17^o \r /M$.

\subsection{Application to the M87* BH shadow}

Using Eq. \eqref{fig_S_1}, we can now easily inspect how the shadow of KBHsPH deviates from that of the comparable Kerr BH by introducing the deviation $\delta S$:

\begin{align}
\delta S \l p, M \mu \r \equiv 1- \frac{S_{\rm hairy} \l p, M \mu, 17^o \r}{S_{\rm Kerr \l a \l M\mu\r, 17^o \r }} \ . 
\end{align} 

\begin{figure}[H]
	\centering
	\includegraphics[width=0.65\textwidth]{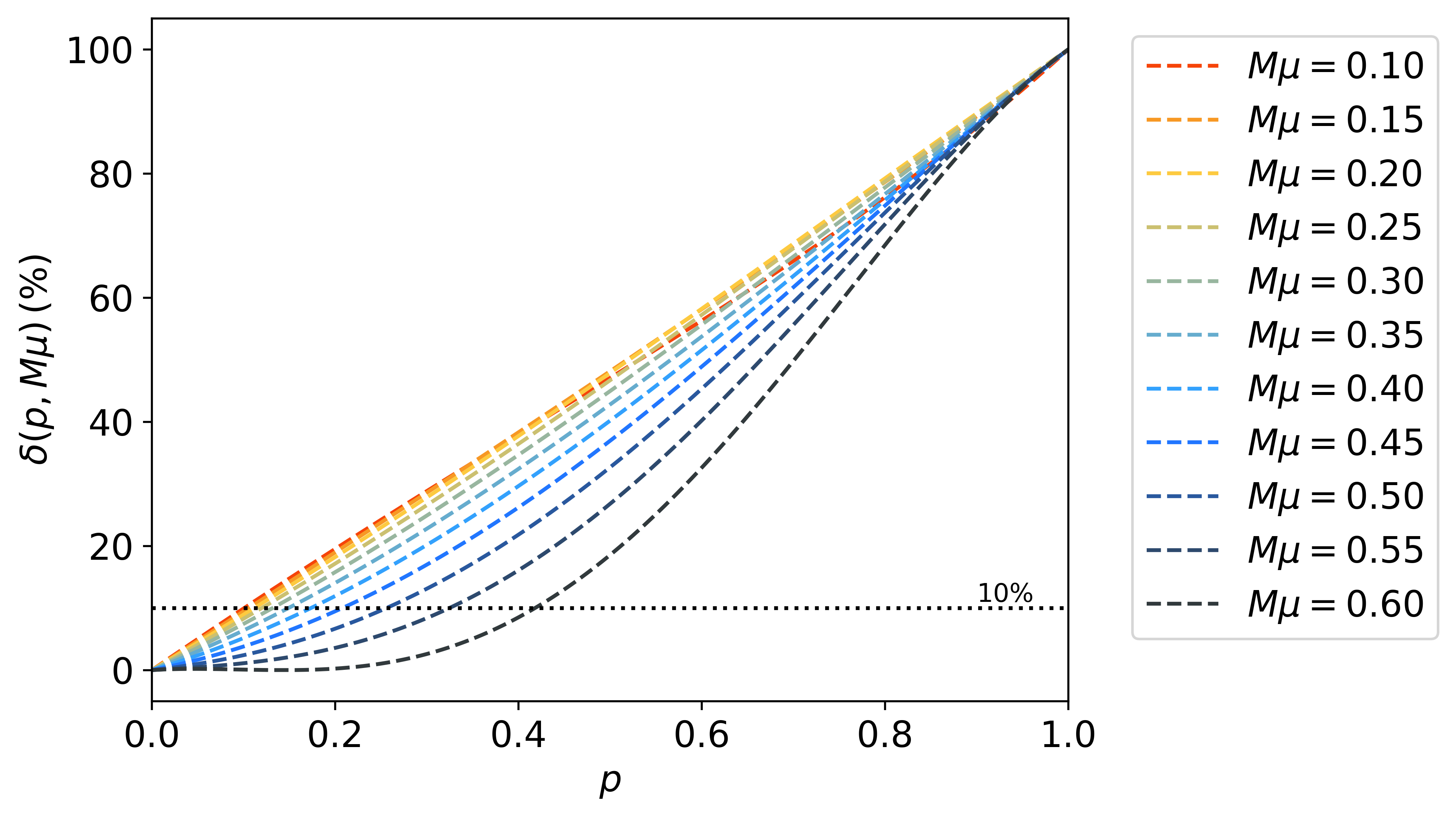}
	\caption{\small \label{fig_dev}Deviation from Kerr as function of hairiness for the different $M \mu$ considered. The horizontal dotted line corresponds to a deviation of $10 \%$, which sets an approximate upper threshold for consistency with the EHT observations of M87*.  }	
\end{figure}

In Fig. \ref{fig_dev} we represent how the deviation $\delta S \l p, M \mu \r$ changes as a function of the amount of hair $p$, for a selection of $M \mu$ values. The profile of $\delta S$ as a function of $p$ depends strongly on the value of $M \mu$ being considered. This comes in sharp contrast with the scalar case discussed in \cite{Cunha_2019}, in which case the deviation was roughly proportional to the fractional amount of hair, regardless of $M\mu$. However, for KBHsPH this is only true for smaller $M \mu$ values. 

Focusing on the solutions with $p \lesssim 0.1$ - the ones that may grow from a Kerr BH via its superradiant instability and are therefore dynamically plausible through this formation channel - then all such solutions are still compatible with the EHT M87* constraint, as in the scalar case. Nonetheless, by considering the possibility that hairier BHs can still be formed via other dynamical channels, such as the merger of bosonic stars \cite{Sanchis_Gual_2020}, one faces the interesting prospect that very hairy BHs with sufficiently large $M\mu$ (up to $p\simeq$ 0.4 in the data considered) could be mistaken by a Kerr BH with the current observational data.

The main implication of this shadow analysis is that the vector (Proca) case introduces a more complex picture than that of the scalar case, described in~\cite{Cunha_2019}. For instance, as one can realise from Fig.~\ref{fig_dev}, KBHsPH can in fact become better Kerr mimickers, since there are solutions whose shadow deviation from the comparable Kerr is below $10 \%$ and that possess up to $40\%$ of the total mass stored on the Proca hair outside the horizon ($i.e.$ $p=0.4$). It is nothing short of surprising that having such a significant deviation on the spacetime geometry can still lead to small deviations on shadow size. This scenario contrasts with the scalar case, wherein a shadow deviation of $10 \%$ disfavoured solutions with $p>0.1$. This is a key difference that we have unveiled between the two models. We will come back to this difference and its possible cause below.

\begin{figure}[H]
	\centering
	\includegraphics[width=0.45\textwidth]{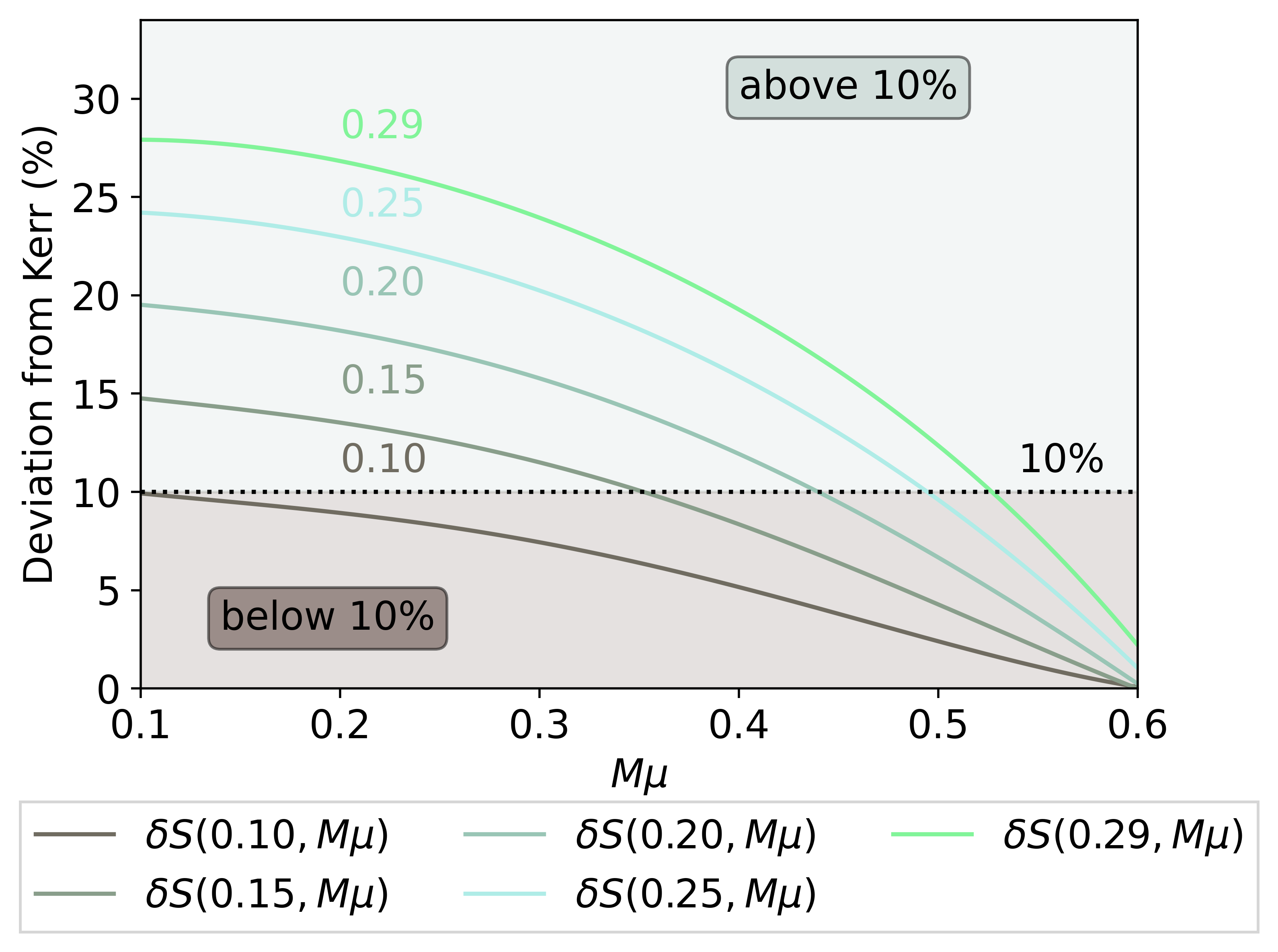}
	\includegraphics[width=0.45\textwidth]{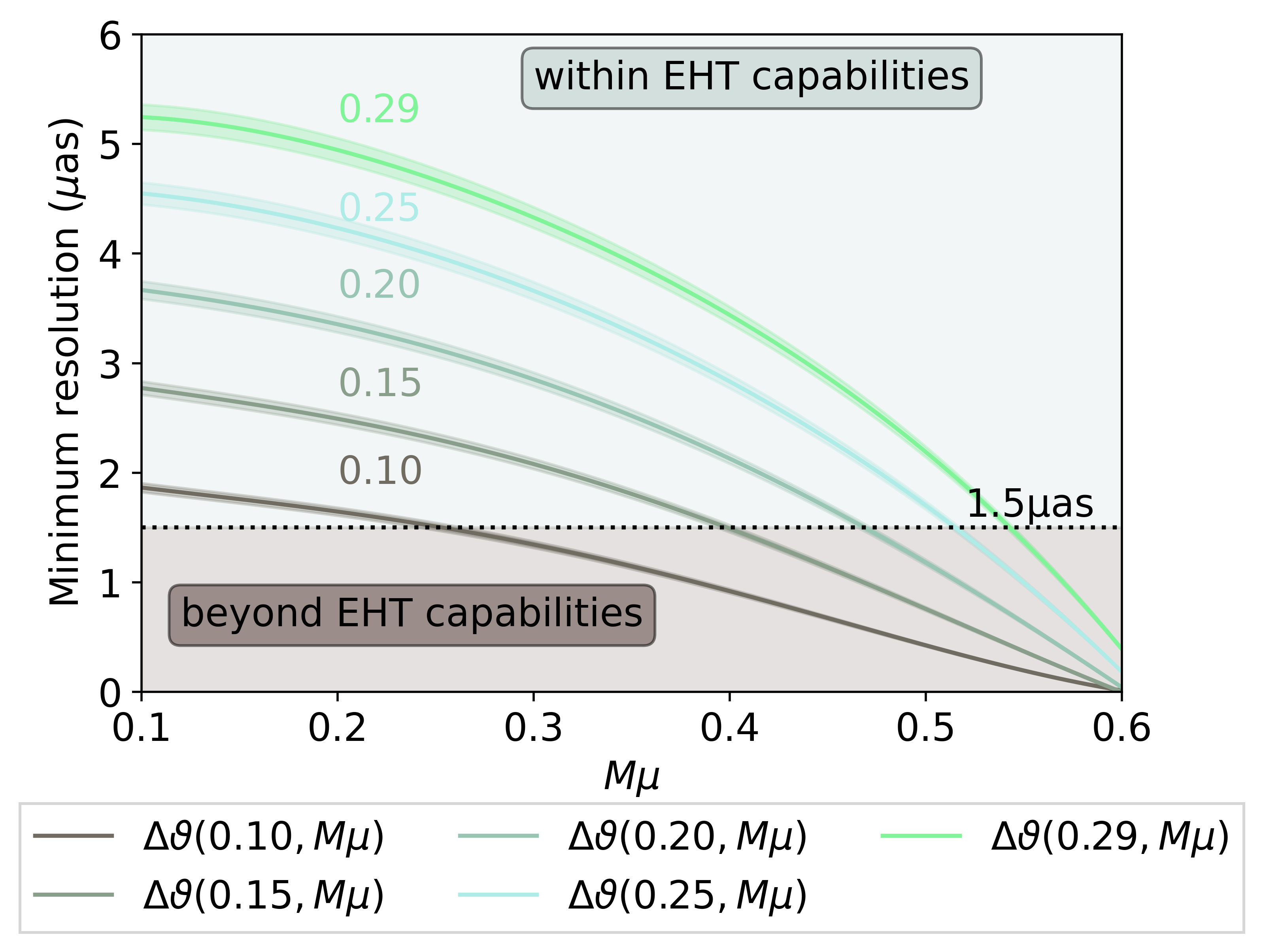}
	\caption{\label{fig_prec} \small On the left panel the shadow deviation from Kerr is represented as function of $M \mu$ for different fixed values of fractional hair $p$. The right panel features a similar diagram, albeit with the minimum M87* angular resolution required to test the model as function of $M\mu$, where the shaded bands account for the uncertainty in the M87* known distance and mass. }	
\end{figure}

According to Fig. \ref{fig_dev}, for a fixed $M \mu$, the greater the value of $p$, the greater is the deviation $\delta S$. Thus, each value of $p$ determines a minimum observation resolution necessary to test that KBHPH model, for a given $M \mu$. This is clear by looking at the left panel of Fig. \ref{fig_prec}, where we represent $\delta S \l p, M\mu \r$ as function of $M \mu$ for fixed values of $p$ from 0.1 to 0.29. From this plot we realise that as we increase $M \mu$, and for a fixed $p$, the more challenging it becomes to distinguish a KBHsPH solution from a Kerr one. 

This degeneracy demands more than our current observational capabilities to rule out the hypothesis that the BHs that we observe are in fact BHs with synchronized Proca hair, at least considering formation from superradiance under the theoretical (thermodynamical) upper limit of $p\simeq 0.29$~\cite{Hawking_colliding,Herdeiro_2022}. The necessary optical resolution to test these models can be obtained from the relation between the shadow areal radius and its angular size via Eq.~\eqref{eq_angular}, which requires knowledge of the mass-distance ratio $\lambda=M/\mathcal{R}$ from M87* to Earth. This ratio $\lambda$ can be inferred from two independent measurements -- one based on star dynamics (hereafter ``star data") \cite{star_data}, and another one based on gas motion (hereafter ``gas data") \cite{gas_data}: 

\begin{align}\label{eq_data}
\begin{cases}
\lambda_{\rm star}= 0.369 \pm 0.022 \, \l \mathrm{\frac{10^9 M_{\odot}}{Mpc}} \r \\
\lambda_{\rm gas}= 0.196^{+0.05}_{-0.04} \, \l \mathrm{\frac{10^9 M_{\odot}}{Mpc}}\r 
\end{cases} \, .
\end{align}

{We shall refer primarily to the star data in our analysis, since gas data is reported to be under tension \textit{even} with the Kerr BH hypothesis \cite{EventHorizonTelescope:2019_6}}.
Using the experimental values for $\lambda$, we can infer how much the angular shadow size of a KBHsPH solution varies from a comparable Kerr solution, as function of $M \mu$:

\begin{align}
\Delta \vartheta (p, M \mu) =\lambda   \frac{S_{\rm hairy}(p, M \mu, 17^o) -S_{\rm Kerr} \l a \l M \mu \r \r}{M}  \, .
\end{align}

We represent this quantity, for $p= \left\{ 0.1,0.15,0.2,0.25,0.29 \right\}$, in the right panel of Fig. \ref{fig_prec}, where we have used the star data. Knowing that the current EHT resolution for the observation of the M87* BH is about $1.5 \ \mathrm{\mu as}$, this figure tells us how much we need to improve the resolution to fully test our model.

 We shall end this section with a more comprehensive and systematic analysis of the parameter region that provides an angular shadow size $\vartheta$ compatible with the EHT observation~\footnote{This value was determined by assuming that the shadow diameter is $10\%$ smaller than the emission ring diameter reported by EHT (in agreement with \cite{EventHorizonTelescope:2019_6}). Although we use the same uncertainty as the one associated with the measurement of the ring, it would be more realistic to expect a slightly larger uncertainty for the shadow diameter (as discussed more thoroughly for the case of Sgr A* \cite{EventHorizonTelescope:2022_6}).} 
 \begin{equation}
\vartheta_{\rm M87*} = \l 18.9 \pm 1.5 \r \mathrm{\mu as} .
 \end{equation} 
 Equation~\eqref{eq_angular} can be first recast in the form $\vartheta(p,\lambda,\,M\mu) = \lambda\, S_{\rm hairy} \l p, M\mu, 17^o \r /M $. Then, for fixed values of $M\mu$, one can scan points in the $\l p, \lambda\r$ plane such that both $\vartheta$ and $\lambda$ fall within the observation error-bars of both $\lambda_{\rm star}$ and $\vartheta_{\rm M87*}$. The error bars can correspond to some multiple of the respective standard deviation $\sigma$ of the measured data, for example $1\sigma$ or $3\sigma$.

\begin{figure}[H] 
	\centering
	\includegraphics[width=0.45\textwidth]{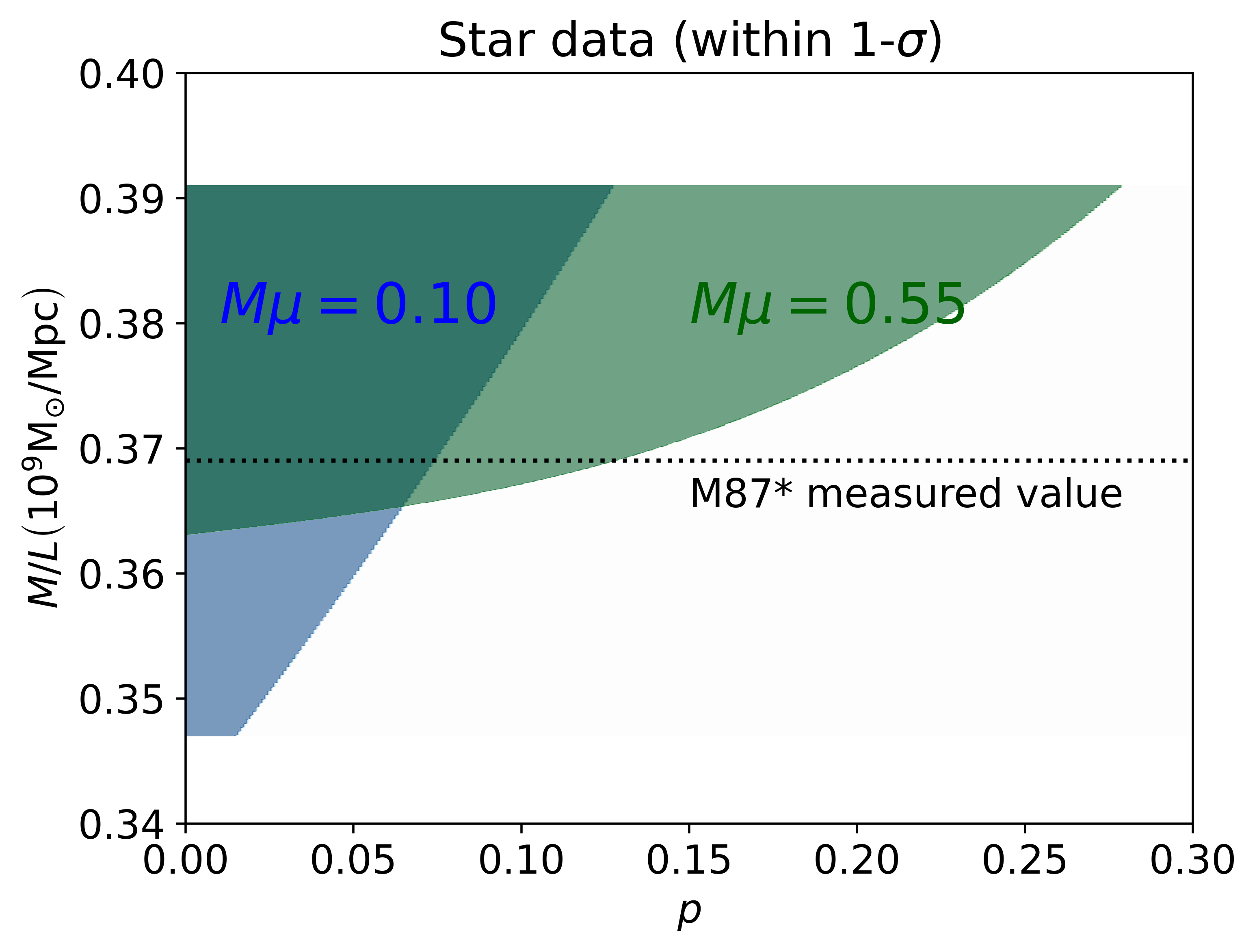}
	\includegraphics[width=0.45\textwidth]{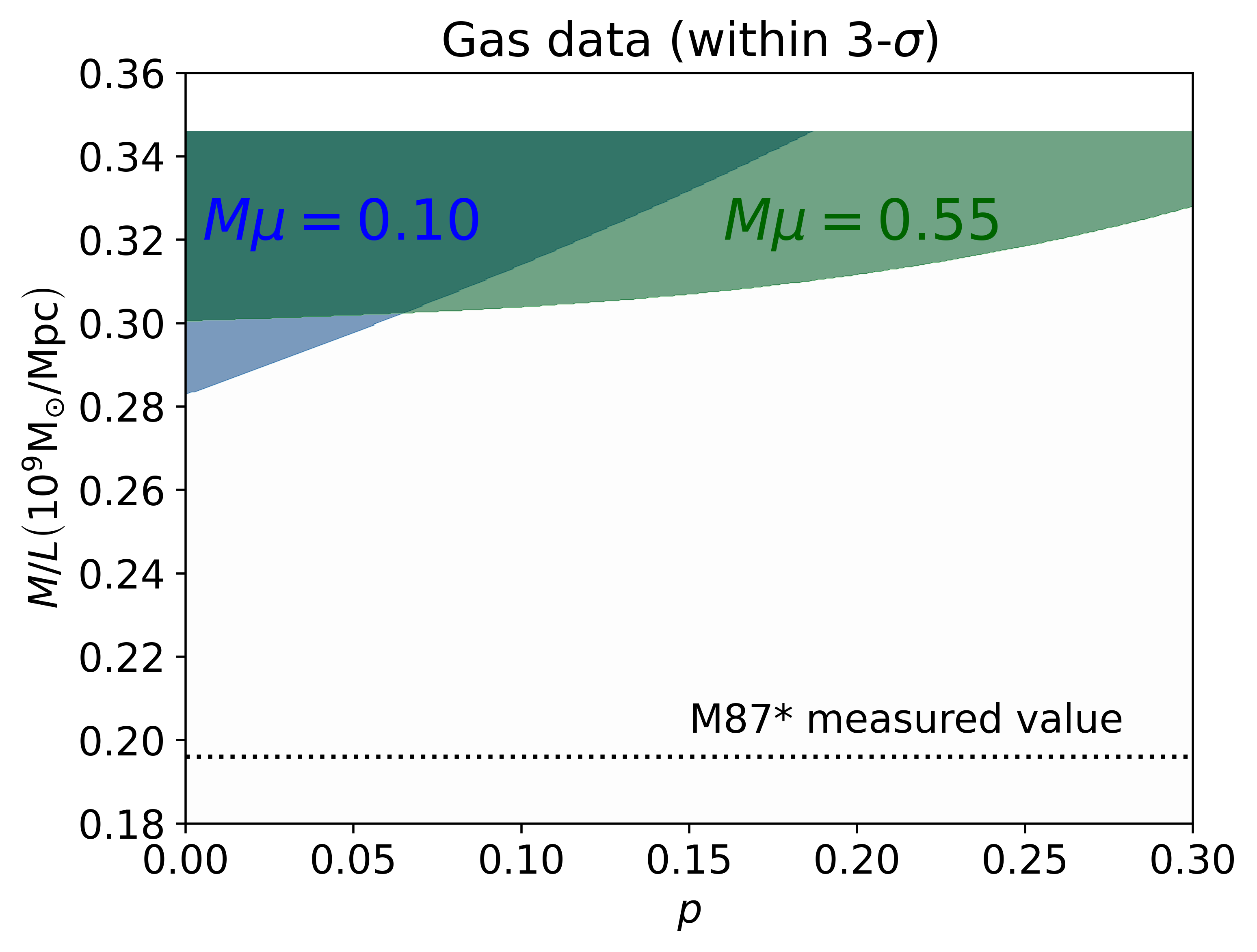}
	\caption{\small \label{fig_sigmas} Shaded domains of $ \l p, \lambda=M/L  \r$ yield a shadow angular size $\vartheta$ consistent with the EHT M87* observation within the uncertainty error-bars of $\{\lambda_{\rm star},\vartheta_{\rm M87*}\}$. Different coloured regions correspond to different values of $M\mu$. Left panel: star data within 1$\sigma$; right panel: gas data within 3$\sigma$.}	
\end{figure}

The result of this analysis is displayed in Fig. \ref{fig_sigmas} for both star data (left panel, and within $1\sigma$) and gas data (right panel, and within $3\sigma$). As one could already expect from Fig. \ref{fig_dev}, and in comparison with the scalar case, there is a much broader range of $p$ values that are still consistent with the EHT data. For instance, if one takes the star data, it is possible to find solutions with $p \gtrsim 0.27$ that are still compatible with the angular size measured with EHT within one standard deviation, for sufficiently large $M\mu$. That is in strike contrast with the scalar case, where one could only get up to $p \approx 0.1.$ 

\section{Constraining the hair using Sgr A* data}

In addition to the M87* analysis of the previous section, we shall perform a similar discussion for the recent EHT observation of Sgr A*, $i.e.$ the BH candidate at the centre of the Milky Way~\cite{EventHorizonTelescope:2022_1}. Unlike M87*, the Sgr A* does not feature a clear jet structure from which one can infer the angle between the line of sight and the angular momentum of the BH ($i.e.$ the observation (inclination) angle, $\theta$). By assuming different inclination angles, the EHT collaboration concluded that the Sgr A* data disfavours $\theta>50^o$. In addition, as reported in \cite{EventHorizonTelescope:2022_1}, the two GRMHD models that better match the imaging data both have $\theta=30^o$. For this reason, we shall consider in our analysis an observation angle of $\theta=30^o$.

Following the same procedure used in the case of M87* , we have similarly found an approximation formula for the shadow size of KBHsPH using the ansatz in Eq. \eqref{Shairy_1}, now with an observation angle of $30^o$ instead of $17^o$. The corresponding parameters $\alpha_{ij}$ can be found in Table \ref{Table_approx2}. 
\begin{table}[H]
	\centering
	\begin{tabular}{l}
	\multicolumn{1}{c}{}  \\ 
$\begin{pmatrix}
\alpha_{00} & \alpha_{01} & \cdots\\
\alpha_{10} & \ddots &  \\
\vdots & & \ddots \\
\end{pmatrix} \equiv \begin{pmatrix}
-1.564 & 22.829 & -112.404 & 321.999 & -290.05\\
4.53 & -65.749 & 382.733 & -1093.222  & 1142.351\\ 
5.939 & -73.042 & 258.157 & -198.798 & -236.928\\
		\end{pmatrix}$
\end{tabular}
	\caption{\label{Table_approx2} Values for the parameters $\alpha_{ij}$ introduced in Eq. \eqref{Shairy_1}, for an observer at $\theta=30^{\circ}$.}	
\end{table}

For values of $p \lesssim 0.3$ the formula gives an error smaller than $0.9 \%$ overall, and around $0.2\%$ in average. 
Using this approximation formula for an observation angle of $\theta=30^o$, we have verified that a plot of the deviation of the shadow size from a comparable Kerr leads to a diagram very similar to Fig.~\ref{fig_dev}, without displaying any qualitative differences (plot not shown for conciseness).

The shadow deviation relatively to Kerr does not change significantly by having an observation at a $30^o$ angle rather than $17^o$. Nevertheless, the EHT observation data of Sgr A* offers us an opportunity to set tighter constraints on the amount of Proca hair that Sgr A* might support, when compared with M87*. Indeed, the mass-to-distance ratio of Sgr A* has been measured much more precisely than for M87*~\cite{EventHorizonTelescope:2022_6}, which allows for a much better estimation of the shadow angular size of KBHsPH.

Following the EHT collaboration, we will base our analysis on two measurements for the mass-to-distance ratio: one by the Very Large Telescope Interferometer (VLTI, \cite{VLTI}), and another by the W. M. Keck Observatory (Keck, \cite{Keck}). Both values have been measured through the observation of the orbital dynamics of the central stellar cluster around Sgr A* (see~\cite{EventHorizonTelescope:2022_6} for a brief summary). The measured values of the mass-to-distance ratio $\lambda$ for Sgr A* are given below:\footnote{This quantity corresponds to $\theta_g$, in the notation followed by the EHT collaboration.}

\begin{align}\label{eq_data2}
\begin{cases}
\lambda_{\mathrm{VLTI}}= 5.125 \pm 0.009 \pm 0.020 \, \l \mathrm{\mu a s} \r \\
\lambda_{	\mathrm{Keck}}= 4.92 \pm 0.003 \pm 0.01 \, \l \mathrm{\mu a s} \r
\end{cases} \, .
\end{align}

In addition, we shall refer to the Sgr A* shadow diameter reported by the EHT~\cite{EventHorizonTelescope:2022_1}: 
\begin{equation}
\vartheta_{\rm Sgr A*}=48.7 \pm 7.0 \,  \mathrm{\mu as}\,.
\end{equation}

\begin{figure}[H] 
	\centering
	\includegraphics[width=0.45\textwidth]{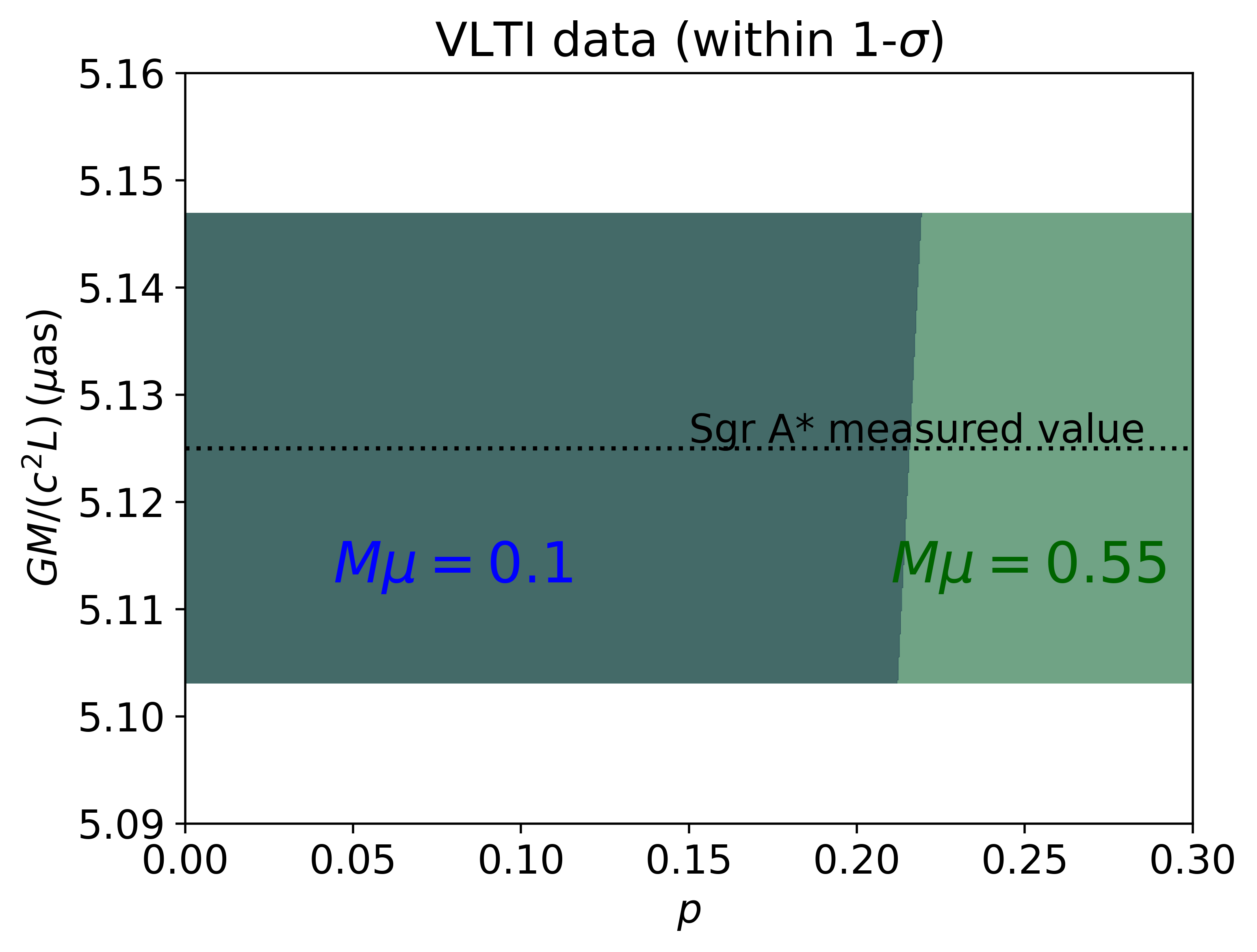}
	\includegraphics[width=0.45\textwidth]{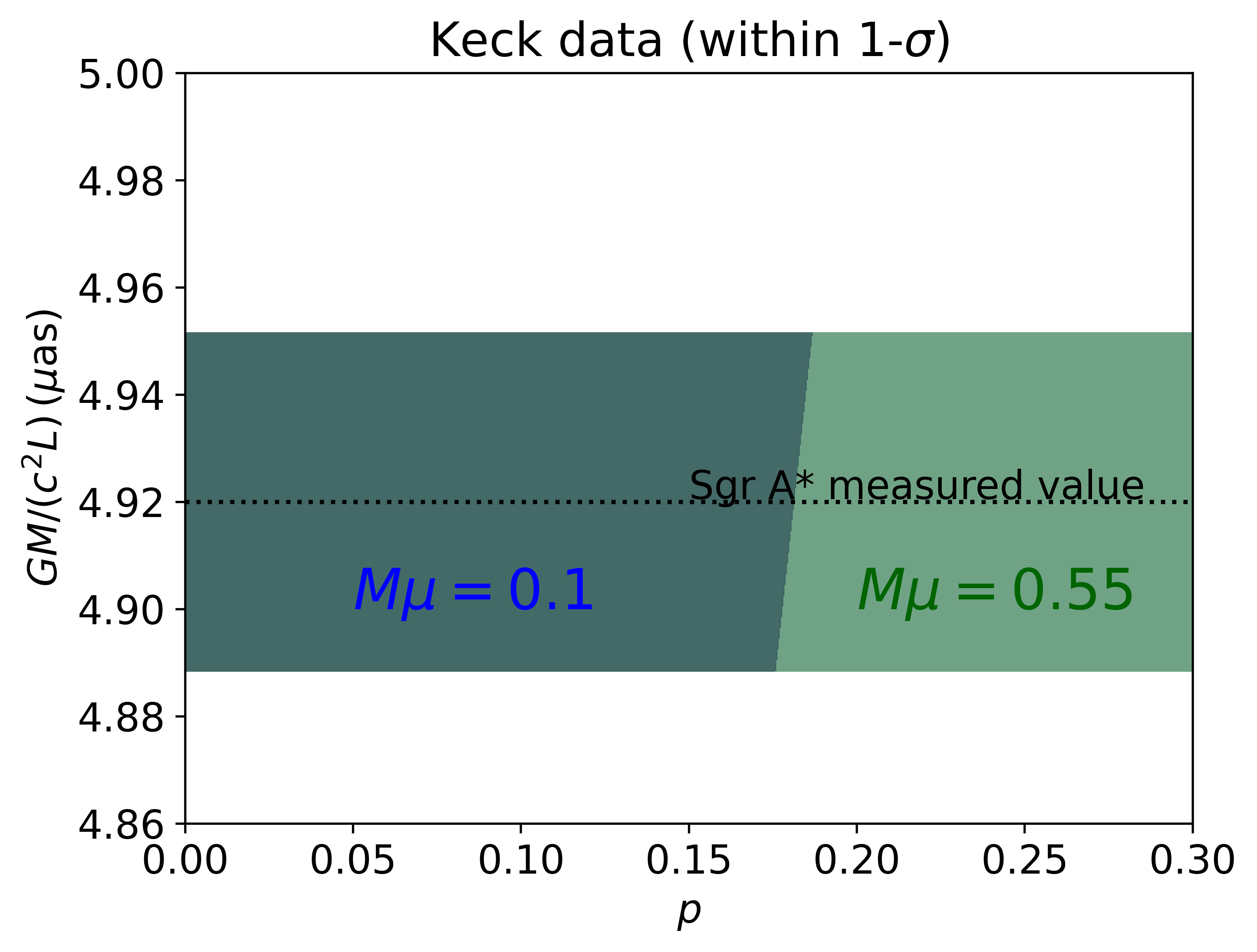}
	\caption{\small \label{fig_sigmas2} Shaded domains of $ \l p, \lambda=M/L  \r$ yield a shadow angular size $\vartheta$ consistent with the EHT Sgr A* observation, within the uncertainty error-bars of $\{\lambda,\vartheta_{\rm SgrA*}\}$. Different coloured regions correspond to different values of $M\mu$. Left panel: VLTI data within 1$\sigma$; right panel: Keck data within 1$\sigma$.  }	
\end{figure}

Following an analysis similar to the one implemented in Fig.~\ref{fig_sigmas}, we have displayed in Fig.~\ref{fig_sigmas2} the region in the plane $\l p, \lambda\r$ compatible with the EHT observation of Sgr A* that is still within the observation error-bar of $\{\lambda,\vartheta_{\rm Sgr A*}\}$. We have considered the VLTI and Keck data separately, one in each of the two plots of Fig.~\ref{fig_sigmas2}. In both cases we are able to find KBHsPH solutions with $30\%$ of the total mass stored in the Proca hair, $i.e.$ with $p=0.3$, and which are still compatible with the EHT shadow observation within 1 standard deviation ($1\sigma$).

\section{Conclusions and Discussion}
Research in BH imaging and the optical appearence of BHs has recently become a thriving field, with many studies on how the EHT data can constrain hairy BHs, $e.g.$~\cite{Dokuchaev:2020wqk,Volkel:2020xlc,Khodadi:2020jij,Glampedakis:2021oie,Islam:2021dyk,Wei:2021lku,Afrin:2021imp,Afrin:2021wlj,Lara:2021zth,Meng:2022kjs,Kuang:2022xjp,Ghosh:2022kit,Saha:2022hcd} as well as many other academic works on the shadows and lensing by ``hairy" spinning (asymptotically flat) BHs - see  $e.g.$ for some recent studies~\cite{Abdujabbarov:2016hnw,Vincent:2016sjq,Cunha_2017,Ayzenberg:2018jip,Hou:2018avu,Wang:2018prk,Shaikh:2019fpu,Contreras:2019nih,Roy:2019esk,Contreras:2019cmf,Chang:2020miq,Wei:2020ght,Xavier:2020egv,Chen:2020aix,Creci:2020mfg,Lee:2021sws,Badia:2021kpk,Wang:2017hjl,Bogush:2022hop} and also the reviews~\cite{Cunha_2018,Perlick:2021aok}.

In this work we have considered the shadows and lensing of a non-Kerr model -- KBHsPH -- that is dynamically robust, in the sense that it appears in a sound physical theory with no known pathologies and it has a plausible formation mechanism, as discussed in the Introduction, at least in some regions of the parameter space. Our study has been both academic, exploring the lensing and shadow features across the full domain of existence, as well as phenomenological, focusing on regions of the parameter space where the solutions are astrophysically more plausible ($e.g.$ free of stable light rings or FPOs, that can source a spacetime instability). 

Our study corroborates the generic expectation: since KBHsPH interpolate between the (vacuum) Kerr solution and the (horizonless) PSs, their lensing features vary from Kerr like -- near the former -- to very non-Kerr like -- for some regions of the parameter space and for solutions with a large fraction of the hair in the Proca field. Concerning the latter, which are more exotic solutions, non-Kerr features such as cuspy, egg-like or ghost shadows emerge typically when the spacetime accommodates a more complex structure of FPOs than Kerr. As emphasised in the text above, it would be interesting to study the generality of some of the features unveiled here, such as: 1) the relation between cuspy shadows and the emergence of new pairs of light rings; or 2) the connection between ghost shadows and the existence of unstable FPOs not-related to the shadow. 

Furthermore, the (possibly) most intriguing result from our analysis arose when considering the comparison with observations. Let us contextualize it by recalling the analog result for Kerr BHs with synchronised scalar hair~\cite{Cunha_2019}. In this reference it was shown that the scalar hairy BHs were compatible with the EHT M87* measurement up to a fraction of the energy in the hair of the order of $10\%$. This seems an intuitive result, as this distribution of the energy between the BH and the surrounding environment seems to affect the shadow size by (roughly) $10\%$, which is of the order of the error in the measurement of the shadow size by the EHT for M87*. Naively, one could have expected a similar result for the KBHsPH discussed here. However, we found that in some regions of the parameter space -- and given sufficiently large $M\mu$ --, Proca hairy BHs with at least $29\%$\footnote{We quote this number since it is the thermodynamic limit of energy extraction from superradiance.} of their energy in the surrounding environment (rather than in the horizon) are still compatible with the EHT measurement of both M87* and Sgr A*. Our best guess, at the moment, is that the estimation of the variation of the shadow size by the fraction of energy outside the horizon can be misleading, as the shadow is an imprint not of the horizon but rather of the FPOs structure. For reasons that should be better understood this (naive estimate) works well for the scalar hairy BHs, but less for the Proca hairy ones. Thus, it would be interesting to understand if there is a correlation between the shadow size variation and the amount of hair that can be assigned to be outside the shell of FPOs around generic spinning BHs. An investigation on this aspect is underway, and we expect to report on it in the near future.

\section*{Acknowledgments}
This work was supported by the Center for Research
and Development in Mathematics and Applications
(CIDMA) through the Portuguese Foundation for
Science and Technology (FCT - Funda\c c\~ao para a
Ci\^encia e a Tecnologia), references UIDB/04106/2020
and UIDP/04106/2020, by national funds (OE),
through FCT, I.P., in the scope of the framework
contract foreseen in the numbers 4, 5 and 6 of the
article 23, of the Decree-Law 57/2016, of August
29, changed by Law 57/2017, of July 19 and by the
projects PTDC/FIS-OUT/28407/2017, CERN/FIS-
PAR/0027/2019, PTDC/FIS-AST/3041/2020 and
CERN/FIS-PAR/0024/2021. This work has further
been supported by the European Union’s Horizon 2020
research and innovation (RISE) programme H2020-
MSCA-RISE-2017 Grant No. FunFiCO-777740 and
by FCT through Project No. UIDB/00099/2020. PC
is supported by the Individual CEEC program 2020
funded by the FCT. IS is supported by the FCT
grant SFRH/BD/150788/2020 under the IDPASC Doctoral Program. Computations have been performed at
the Argus and Blafis cluster at the U. Aveiro.

\bibliographystyle{unsrt}
\bibliography{Refs}


\appendix
\section{\label{App1}Physical quantities of selected solutions }

\begin{table}[H]
	\centering
	\begin{tabular}{c|c|c|c|c|c|c|c|c|c}
		\hline
		\hline
		Label & $\omega$ & $M\mu$& $J \mu^2$& $M_{BH}\mu$& $J_{BH}\mu^2$& $M_{BH}/M$& $ J_{BH}/J$ & $ J/M^2$ & $J_{BH}/M^2_{BH} $ \\ \hline
		3.4 & 0.95 & 0.317 & 0.120 & 0.242 & 0.048 & 0.763 & 0.402 & 1.194 & 0.825   \\
		3.3 & 0.95 & 0.422 & 0.314 & 0.114 & 0.007 & 0.269 & 0.023 & 1.764 & 0.555   \\
		3.2 & 0.95 & 0.482 & 0.435 & 0.051 & 0.001 & 0.107 & 0.002 & 1.874 & 0.279   \\
		3.1 & 0.95 & 0.528 & 0.532 & 0.005 & 0.000 & 0.009 & 0.000 & 1.905 & 0.033   \\
		6.4 & 0.8 & 0.744 & 0.620 & 0.142 & 0.036 & 0.191 & 0.058 & 1.118 & 1.777   \\
		6.3 & 0.8 & 0.601 & 0.383 & 0.351 & 0.169 & 0.585 & 0.442 & 1.061 & 1.371   \\
		6.2 & 0.8 & 0.865 & 0.842 & 0.054 & 0.003 & 0.062 & 0.003 & 1.125 & 1.009   \\
		6.1 & 0.8 & 0.938 & 0.992 & 0.005 & 0.000 & 0.005 & 0.000 & 1.126 & 0.116   \\
		7.3 & 0.75 & 0.820 & 0.719 & 0.118 & 0.035 & 0.144 & 0.048 & 1.069 & 2.497   \\
		7.2 & 0.75 & 0.947 & 0.958 & 0.043 & 0.003 & 0.045 & 0.003 & 1.068 & 1.409   \\
		7.1 & 0.75 & 1.005 & 1.076 & 0.005 & 0.000 & 0.005 & 0.000 & 1.066 & 0.198   \\
		8.3 & 0.7 & 0.896 & 0.833 & 0.084 & 0.026 & 0.093 & 0.031 & 1.037 & 3.653   \\
		8.2 & 0.7 & 0.937 & 0.908 & 0.060 & 0.012 & 0.064 & 0.013 & 1.035 & 3.265   \\
		8.1 & 0.7 & 1.052 & 1.140 & 0.005 & 0.000 & 0.005 & 0.000 & 1.030 & 0.371   \\
		9.3 & 0.65 & 0.997 & 1.007 & 0.037 & 0.007 & 0.037 & 0.007 & 1.014 & 4.895   \\
		9.2 & 0.65 & 1.041 & 1.096 & 0.020 & 0.001 & 0.020 & 0.001 & 1.011 & 3.024   \\
		9.1 & 0.65 & 1.083 & 1.184 & 0.005 & 0.000 & 0.004 & 0.000 & 1.009 & 0.800   \\
		10.3 & 0.6 & 1.000 & 1.005 & 0.036 & 0.013 & 0.036 & 0.013 & 1.005 & 10.251   \\
		10.2 & 0.6 & 1.057 & 1.119 & 0.016 & 0.002 & 0.015 & 0.001 & 1.002 & 6.674   \\
		10.1 & 0.6 & 1.099 & 1.206 & 0.005 & 0.000 & 0.004 & 0.000 & 0.999 & 2.198   \\ \hline
	\end{tabular}
	\caption{\label{tab_data1} Physical quantities of the KBHsPH solutions highlighted in Figure \ref{fig1}.}
\end{table}

\begin{table}[H]
	\centering
	\begin{tabular}{c|c|c|c}
		\hline
		\hline
		Label & $\omega$ & $M\mu$& $J \mu^2$ \\ \hline
            1.0 & 0.99 & 0.246 & 0.247 \\
            2.0 & 0.97 & 0.420 & 0.424\\
            3.0 & 0.95 & 0.533 & 0.542\\
            4.0 & 0.9 & 0.725 & 0.749\\
            5.0 & 0.85 & 0.853 & 0.896\\
            6.0 & 0.8 & 0.946 & 1.007\\
            7.0 & 0.75 & 1.014 & 1.095\\
            8.0 & 0.7 & 1.063 & 1.162\\
            9.0 & 0.65 & 1.096 & 1.212\\
            10.0 & 0.6 & 1.117 & 1.245\\
            11.0 & 0.55 & 1.124 & 1.256\\
            12.0 & 0.5 & 1.112 & 1.234\\
            13.0 & 0.47 & 1.091 & 1.189	\\ \hline
            	\end{tabular}
	\caption{\label{tab_data2} Physical quantities of the Proca star solutions highlighted in Figure \ref{fig1}.}
\end{table}

\end{document}